\documentclass[preprint,authoryear,11pt]{elsarticle}

\usepackage{graphics}
\usepackage{graphicx}
\graphicspath{{Figures/}}
\usepackage{epsfig}
\usepackage{subfigure}
\usepackage{amssymb}
\usepackage{amsthm}
\usepackage{bm}
\usepackage{amsmath}
\usepackage{color}
\usepackage{geometry}
\usepackage{setspace}
\allowdisplaybreaks
\usepackage{enumerate}
\usepackage{booktabs}

\usepackage{hyperref}
\hypersetup{
	colorlinks=true,       
}

\biboptions{}

\journal{}

\begin{document}

\begin{frontmatter}



\title{Wrinkling of soft magneto-active plates}


\author[1]{Bin Wu\corref{cor1}}
\ead{bin.wu@nuigalway.ie}
\author[1]{Michel Destrade}

\cortext[cor1]{Corresponding author at: School of Mathematics, Statistics and Applied Mathematics, NUI Galway, University Road, Galway, Ireland.}
\address[1]{School of Mathematics, Statistics and Applied Mathematics,\\ NUI Galway, University Road, Galway, Ireland.}

\begin{abstract}

Coupled magneto-mechanical wrinkling has appeared in many scenarios of engineering and biology. Hence, soft magneto-active (SMA) plates  buckle when subject to critical uniform magnetic field normal to their wide surface. Here, we provide a systematic analysis of the wrinkling of SMA plates subject to an in-plane mechanical load and a transverse magnetic field. We consider two loading modes: plane-strain loading and uni-axial loading, and two models of  magneto-sensitive plates: the neo-Hookean ideal magneto-elastic model and the neo-Hookean magnetization saturation Langevin model. 
Our analysis relies on the theory of nonlinear magneto-elasticity and the associated linearized theory for superimposed perturbations. We derive  the Stroh formulation of the governing equations of wrinkling, and combine it with the surface impedance method to obtain explicitly the bifurcation equations identifying the onset of symmetric and antisymmetric wrinkles. We also obtain analytical expressions of instability in the thin- and thick-plate limits. For thin plates, we make the link with classical Euler buckling solutions. We also perform an exhaustive numerical analysis to elucidate the effects of loading mode, load amplitude, and saturation magnetization on the nonlinear static response and bifurcation diagrams. We find that antisymmetric wrinkling modes always occur before symmetric modes. Increasing the pre-compression or heightening the magnetic field has a destabilizing effect for SMA plates, while the saturation magnetization enhances their stability. We  show that the Euler buckling solutions are a good approximation to the exact bifurcation curves for thin plates.

\end{abstract}


\end{frontmatter}




\underline{\underline{}}
\section{Introduction}


Soft magneto-active (SMA) materials, such as magneto-active elastomers, are a particularly promising kind of smart materials that can respond to magnetic field excitation. In general, these SMA materials are prepared by mixing micron-sized magnetizable particles (such as carbonyl iron and neodymium-iron-boron) into a non-magnetic elastomeric matrix such as rubber or silicone \citep{rigbi1983response, ginder2002magnetostrictive, kim2018printing}. 
They can then deform significantly under the simple, remote, and reversible actuation of magnetic fields. Moreover, their overall magneto-mechanical properties can be altered actively by applying suitable magnetic fields, thus resulting in tunable vibration and wave characteristics. Owing to these superior magneto-mechanical coupling behaviors, SMA materials have recently attracted considerable academic and industrial interests, prefiguring various potential applications which include remote actuators and sensors \citep{lanotte2003potentiality, tian2011sensing, kim2018printing}, soft robotics and biomedical devices \citep{makarova2016tunable, luo2019magnetically, tang2019magnetic}, tunable vibration absorbers \citep{ginder2001magnetorheological, hoang2010adaptive}, and tunable wave devices \citep{yu2018magnetoactive, karami2019soft}.

A challenging problem that arises in the study of SMA materials is how to model  the strong nonlinearity and the magneto-mechanical coupling. Thus, a lot of academic interest has been devoted over the years to establish a general theoretical framework of nonlinear magneto-(visco)elasticity in order to describe appropriately the magneto-mechanical response of SMA materials \citep{tiersten1964coupled, brown1966magnetoelastic, yih1973linear, maugin2013continuum, brigadnov2003mathematical, dorfmann2004nonlinear, bustamante2010transversely, destrade2011magneto, saxena2013theory}. 
Comprehensive reviews regarding the theoretical development of nonlinear magneto-elasticity include those by  \citet{kankanala2004finitely} and \citet{dorfmann2014nonlinear}. 
Furthermore,  homogenization techniques based on micromechanical methods have also been developed to understand the connection between the magneto-active microstructures and the macroscopic physical or mechanical properties of SMA materials \citep{castaneda2011homogenization, galipeau2014magnetoactive}.

In a large variety of practical applications, the mechanical and magnetic loads (pre-stretch, magnetic field, etc.) influence the working performance of smart systems made of SMA materials and may lead to instability and even failure. In fact, it has long been observed that a magneto-elastic beam or plate in a uniform transverse magnetic field will buckle once the field reaches a critical value \citep{moon1968magnetoelastic, miya1978experimental, gerbal2015refined}. 
On the other hand, local buckling and other instability phenomena can be exploited to realize  active pattern switching devices and reconfigurable metamaterials. Hence \citet{psarra2017two, psarra2019wrinkling} made use of wrinkling and crinkling instabilities of a thin SMA film bonded on a soft non-magnetic substrate to achieve surface pattern control through a combined action of magnetic field and uni-axial pre-compression. \citet{goshkoderia2020instability} investigated experimentally  instability-induced pattern evolutions in SMA elastomer composites driven  by an applied magnetic field.

Thus, it is vital to theoretically study the stability of SMA structures and composites subject to coupled magneto-mechanical loads, so that we can  provide solid guidance for simulations and experiments. 
From a theoretical point of view, the classical buckling problem of a magneto-elastic beam-plate was first addressed by \citet{moon1968magnetoelastic}, based on the thin-plate theory and the assumption of a linear ferromagnetic material. \citet{yih1973linear} used a general theory of magneto-elasticity to re-examine this problem and to yield an identical antisymmetric buckling equation for thin plates. 
Following those works, many investigations looked at the same problem, trying to improve mathematical models to explain the discrepancy between experimental results and theoretical predictions \citep{wallerstein1972magnetoelastic, popelar1972postbuckling, dalrymple1974magnetoelastic, miya1978experimental, gerbal2015refined, singh2018magnetic}. Most of the above-mentioned works employed classical structural models to elucidate the magneto-mechanical coupling problem. 
More recently, using the theory of nonlinear magneto-elasticity, some researchers have explored the onset of instabilities of different SMA structures, including surface instabilities of isotropic SMA half-spaces \citep{ottenio2008incremental}, buckling modes of rectangular SMA blocks undergoing plane-strain loading \citep{kankanala2008magnetoelastic}, macroscopic instabilities of anisotropic SMA multilayered structures \citep{rudykh2013stability}, instabilities of a thin SMA layer resting on a soft non-magnetic substrate \citep{danas2014instability}, and instabilities of a cylindrical membrane \citep{reddy2018instabilities}.

\begin{figure}[htbp]
	\centering	
	\setlength{\abovecaptionskip}{5pt}
	\includegraphics[width=1.0\textwidth]{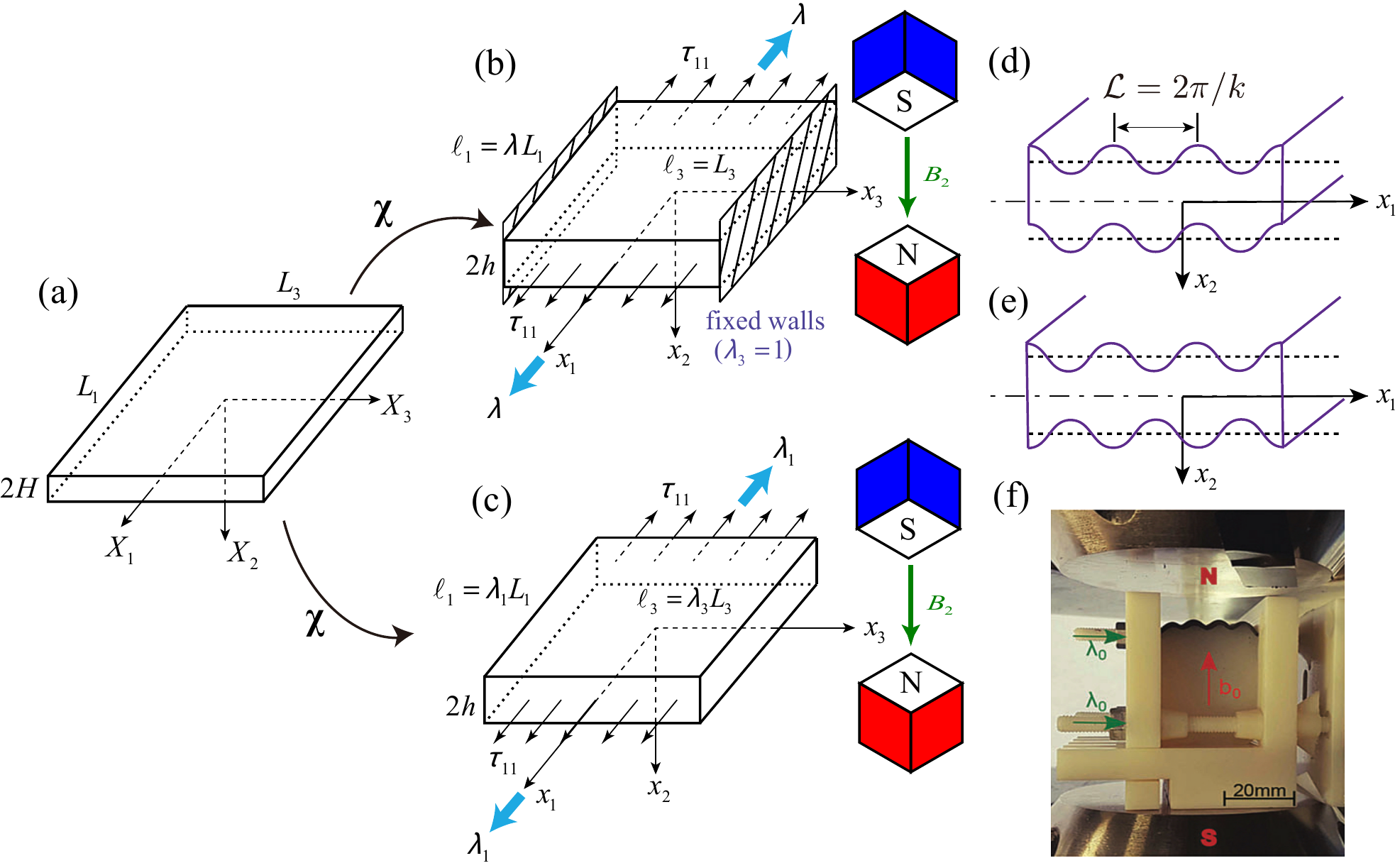}
	\caption{(a)-(e): Schematic diagram of a rectangular SMA plate with Cartesian coordinates and geometry. (a): Undeformed configuration before activation. (b): Deformed configuration subject to plane-strain loading ($\lambda_3=1$) with a transverse magnetic field $B_2$ and (c): Deformed configuration subject to uni-axial loading with a transverse magnetic field $B_2$  (the magnetic poles are assumed to be very close to the plate, to ensure quasi-uniformity of the mechanical and magnetic fields). (d)-(e): Antisymmetric and symmetric modes of wrinkling instability.
	(f): Antisymmetric wrinkling due to the application of an external transverse magnetic field to a pre-compressed SMA plate glued onto an inert substrate (taken from \cite{psarra2017two}).}
	\label{Fig1}
\end{figure}

The present work revisits the stability problem of SMA plates subject to an in-plane mechanical load and a uniform transverse magnetic field, and evaluates explicitly the onset of wrinkling instabilities. This work differs from previous works \citep{yih1973linear, kankanala2008magnetoelastic} in the following respects. (i) We adopt the nonlinear magneto-elasticity theory and the associated linearized theory developed by \citet{dorfmann2004nonlinear} and \citet{ottenio2008incremental} to derive the governing equations of nonlinear static response and the bifurcation equations of wrinkles. These theories introduce a total stress tensor and a modified (or total) energy density function to express constitutive relations in a simple and compact form. (ii) Two mechanical loading modes are considered: plane-strain loading (Fig.~\ref{Fig1}(b)) and uni-axial loading (Fig.~\ref{Fig1}(c)). (iii) To overcome the complexity of conventional displacement-based method, we employ the Stroh formulation and the surface impedance method \citep{su2018wrinkles} to obtain explicit expressions of the bifurcation equations of antisymmetric and symmetric wrinkling modes (Fig.~\ref{Fig1}(d) and \ref{Fig1}(e)). (iv) We also manage to derive the explicit bifurcation equations corresponding to the thin- and thick-plate limits, and to establish the thin-plate approximate formulas.

The paper is organized as follows. The equations of nonlinear magneto-elasticity are derived in Sec.~\ref{sec2}. The linearized stability analysis is conducted in Sec.~\ref{sec3} to obtain the exact bifurcation equations for symmetric and antisymmetric wrinkles. Section~\ref{sec4} specializes the analytical expressions to the neo-Hookean magnetization saturation material model, including the ideal magneto-elastic model. Numerical calculations are carried out in Sec.~\ref{sec5} to illustrate the effects of loading mode, load amplitude, and saturation magnetization on nonlinear static response and bifurcation diagrams. Section~\ref{sec6} concludes the work with a summary. Some mathematical expressions or derivations are provided in Appendices A-C.

\section{Equations of nonlinear magneto-elasticity} \label{sec2}

\subsection{Finite magneto-elasticity theory} \label{sec2.1}

We first present the basic equations governing the finite magneto-elastic deformations of an incompressible soft magneto-elastic body, as developed by \citet{dorfmann2004nonlinear}.

In the undeformed stress-free state, the body occupies in the Euclidean space a region $\mathcal{B}_r$ (the boundary being $\partial \mathcal{B}_r$, with the outward normal $\mathbf{N}$),
which is taken to be a \textit{reference configuration}. Any material point inside the body is then identified
with its position vector $\mathbf{X}$ in $\mathcal{B}_r$. The application of both mechanical load and magnetic field deforms quasistatically the body from $\mathcal{B}_r$ to the \textit{current configuration} $\mathcal{B}$ (the boundary being $\partial \mathcal{B}$, with the outward normal $\mathbf{n}$). The particle located at $\mathbf{X}$ in $\mathcal{B}_r$ now
occupies the position $\mathbf{x}=\bm{\chi}(\mathbf{X})$ in $\mathcal{B}$, where the vector function $\bm{\chi}$ is a one-to-one, orientation-preserving mapping with a sufficiently regular property. The associated deformation gradient tensor is $\mathbf{F}=\text{Grad} \, \bm{\chi}=\partial{\bm{\chi}}/\partial{\mathbf{X}}$ ($F_{i \alpha}=\partial x_i / \partial X_\alpha$ in Cartesian components) with Grad being the gradient operator in $\mathcal{B}_r$. Note that Greek indices are associated with
$\mathcal{B}_r$ and Roman indices with $\mathcal{B}$. 

The Jacobian of the deformation gradient is $J=\text{det}\, \mathbf{F}$, and it is equal to 1 at all times for incompressible materials. The left and right Cauchy-Green deformation tensors $\mathbf{b}=\mathbf{FF}^\text{T}$ and $\mathbf{C}=\mathbf{F}^\text{T}\mathbf{F}$ are used as the deformation measures, where the superscript $^\text{T}$ signifies the transpose.

In the Eulerian description, the equilibrium equations in the absence of mechanical body forces, and the Maxwell equations in the absence of time dependence, free charges and electric currents, are 
\begin{equation} \label{1}
\text{div}\,\bm{\tau }=\mathbf{0}, \qquad \text{div}\,\mathbf{B}=0,\qquad \text{curl}\,\mathbf{H}=\mathbf{0},
\end{equation}
where $\bm{\tau}$ is the \textit{total Cauchy stress tensor} including the contribution of magnetic body forces, $\mathbf{B}$ is the Eulerian magnetic induction vector and and $\mathbf{H}$ is the Eulerian magnetic field vector; curl and div are the curl and divergence operators in $\mathcal{B}$, respectively. The conservation of angular momentum implies that $\bm{\tau}$ is \textit{symmetric}.

The Eulerian magnetization vector $\mathbf{M}$ is defined by the standard relation
\begin{equation} \label{2}
\mathbf{B}={{\mu }_{0}}(\mathbf{H}+\mathbf{M}),
\end{equation}
where $\mu_0=4\pi \times 10^{-7}\,\text{N}\cdot \text{A}^{-2}$ is the magnetic permeability in vacuum.

In vacuum, there is no magnetization and Eq.~\eqref{2} reduces to ${{\mathbf{B}}^\star}={{\mu }_{0}}{{\mathbf{H}}^\star}$, where a star superscript identifies a quantity exterior to the material. The \textit{Maxwell stress tensor} $\bm{\tau}^\star$ in vacuum is
\begin{equation} \label{3}
{{\bm{\tau }}^\star}=\mu _{0}^{-1}\left[ {{\mathbf{B}}^\star}\otimes {{\mathbf{B}}^\star}- \tfrac{1}{2}\left( {{\mathbf{B}}^\star}\cdot {{\mathbf{B}}^\star} \right)\mathbf{I} \right],
\end{equation}
where $\mathbf{I}$ is the identity tensor. The external fields $\mathbf{B}^\star$ and $\mathbf{H}^\star$ satisfy $\text{div} \,{{\mathbf{B}}^\star}=0$ and $\text{curl}\, {{\mathbf{H}}^\star}=\mathbf{0}$, which leads to $\text{div}\,{{\bm{\tau }}^\star}=\mathbf{0}$.

The traction and magnetic boundary conditions are written in Eulerian form as
\begin{equation} \label{4}
\left( \bm{\tau }-{{\bm{\tau }}^\star} \right) \mathbf{n}={{\mathbf{t}}^{\text{a}}},\qquad \left( \mathbf{B}-{{\mathbf{B}}^\star} \right)\cdot \mathbf{n}=0, \qquad \left( \mathbf{H}-{{\mathbf{H}}^\star} \right)\times \mathbf{n}=\mathbf{0},
\end{equation}
where ${{\mathbf{t}}^{\text{a}}}$ is the applied mechanical traction vector per unit area of $\mathcal{\partial{B}}$. Note that the magnetic traction vector  ${{\mathbf{t}}^{\text{m}}}$ induced by
the external Maxwell stress ${{\bm{\tau }}^\star}$ is ${{\mathbf{t}}^{\text{m}}}={{\bm{\tau }}^\star} \mathbf{n}$. Using  Nanson's formula, we can transform the governing equations \eqref{1} and boundary conditions \eqref{4}
into the Lagrangian form.

Following the theory of nonlinear magneto-elasticity \citep{dorfmann2004nonlinear}, it is convenient to express the nonlinear constitutive relations for incompressible magneto-elastic materials in terms of a \textit{total energy function}, or \textit{modified free energy function}, $\Omega \left( \mathbf{F}, {\mathbf{B}_{\text{L}}} \right)$ per unit reference volume as
\begin{equation} \label{5}
\mathbf{T}=\frac{\partial \Omega }{\partial \mathbf{F}}-p{{\mathbf{F}}^{-1}},  \qquad {{\mathbf{H}}_{\text{L}}}=\frac{\partial \Omega }{\partial {{\mathbf{B}}_{\text{L}}}},
\end{equation}
where $\mathbf{T}={{\mathbf{F}}^{-1}}\bm{\tau}$ is the \textit{total} nominal stress tensor, the nominal magnetic field vector ${{\mathbf{B}}_{\text{L}}}={{\mathbf{F}}^{-1}}\mathbf{B}$ and the nominal magnetic induction vector  ${{\mathbf{H}}_{\text{L}}}={{\mathbf{F}}^{\text{T}}}\mathbf{H}$ are the Lagrangian counterparts of $\mathbf{B}$ and $\mathbf{H}$, respectively, and $p$ is a Lagrange multiplier related to the incompressibility constraint, which can be determined from the equilibrium equations and boundary conditions. 
The Eulerian counterparts of Eq.~\eqref{5} are
\begin{equation} \label{6}
\bm{\tau} =\mathbf{F}\frac{\partial \Omega }{\partial \mathbf{F}}-p\mathbf{I},\qquad
\mathbf{H}={{\mathbf{F}}^{-\text{T}}}\frac{\partial \Omega }{\partial {{\mathbf{B}}_{\text{L}}}}.
\end{equation}

For an incompressible isotropic material, the energy function $\Omega$ can be reduced to a function depending only on the following five invariants:
\begin{align} \label{7}
&{{I}_{1}}=\text{tr}\,\mathbf{C}, \qquad &&{{I}_{2}}=\tfrac{1}{2}\left[ {{\left( \text{tr}\,\mathbf{C} \right)}^{2}}-\text{tr}\left( {{\mathbf{C}}^{2}} \right) \right], && \notag\\ 
&{{I}_{4}}={{\mathbf{B}}_{\text{L}}}\cdot {{\mathbf{B}}_{\text{L}}},\qquad
&&{{I}_{5}}={{\mathbf{B}}_{\text{L}}}\cdot \left( \mathbf{C}{{\mathbf{B}}_{\text{L}}} \right),\qquad
&&{{I}_{6}}={{\mathbf{B}}_{\text{L}}}\cdot \left( {{\mathbf{C}}^{2}}{{\mathbf{B}}_{\text{L}}} \right).
\end{align}
According to Eqs.~\eqref{6} and \eqref{7}, the total Cauchy stress tensor $\bm{\tau}$ and the Eulerian magnetic field vector $\mathbf{H}$ are then derived as
\begin{align} \label{8}
& \bm{\tau} =2{{\Omega }_{1}}\mathbf{b}+2{{\Omega }_{2}}\left( {{I}_{1}}\mathbf{b}-{{\mathbf{b}}^{2}} \right)-p\mathbf{I}+2{{\Omega }_{5}}\mathbf{B}\otimes \mathbf{B}+2{{\Omega }_{6}}\left( \mathbf{B}\otimes \mathbf{bB}+\mathbf{bB}\otimes \mathbf{B} \right), \notag\\ 
& \mathbf{H}=2\left( {{\Omega }_{4}}{{\mathbf{b}}^{-1}}\mathbf{B}+{{\Omega }_{5}}\mathbf{B}+{{\Omega }_{6}}\mathbf{bB} \right),
\end{align}
where ${{\Omega }_{n}}=\partial \Omega /\partial {{I}_{n}}\, \left( n=1,2,4,5,6 \right)$.
%

\subsection{Pure homogeneous deformation of a plate} \label{sec2.2}

We now specialize the above equations of nonlinear magneto-elasticity to the homogeneous deformation of an SMA plate subject to an in-plane mechanical load and a uniform magnetic field along its thickness direction. In this work, we  consider two mechanical loading modes:  plane-strain loading (see Fig.~\ref{Fig1}(b)) and  uni-axial loading (see Fig.~\ref{Fig1}(c)).

We point out that to simplify the mathematical modelling and obtain explicit analytical solutions to the nonlinear response and the bifurcation criteria of SMA plates, we make the following assumptions. (i) The plate is entirely immersed in the external magnetic field and subjected to an applied uniform magnetic field in the thickness direction. (ii) The plate is of infinite lateral extent, i.e., the plate lateral dimensions are much larger than its thickness. (iii) The magnetic poles are very close to the plate, see Fig.~\ref{Fig1}(b) and \ref{Fig1}(c) for sketches and Fig.~\ref{Fig1}(f) for the experimental apparatus of \cite{psarra2017two}. These assumptions ensure that the field heterogeneity is only confined to the plate edges, while far away from the edges, the induced stress and magnetic field distributions can be envisioned as uniform, which is a good approximation in the sense of Saint-Venant's principle. A similar processing method has been adopted by \citet{su2020effect} to analyze the wrinkling of soft electro-active plates immersed in external electric fields.

Let $(X_1, X_2, X_3)$ and $(x_1, x_2, x_3)$ be the Cartesian coordinates in the reference and current configurations, respectively, along the $(\mathbf E_1, \mathbf E_2, \mathbf E_3)$ and $(\mathbf e_1, \mathbf e_2, \mathbf e_3$) unit vectors. In the undeformed configuration, the plate has uniform thickness $2H$ along the $X_2$ direction and lateral lengths $L_1, L_3$ in the ($X_1,X_3$) plane. The pure homogeneous deformations corresponding to the two loading modes are defined by ${{x}_{1}}={{\lambda }_{1}}{{X}_{1}}$, ${{x}_{2}}=\lambda_2 {{X}_{2}}$, ${{x}_{3}}={{\lambda }_{3}}{{X}_{3}}$, where $\lambda_i \, (i=1,2,3)$ is the principal stretch in the $X_i$ direction and $\lambda_2={{\left( {{\lambda }_{1}}{{\lambda }_{3}} \right)}^{-1}}$ by incompressibility. The resulting deformation gradient tensor is the diagonal matrix $\mathbf{F}= \text{diag}\left[{{\lambda }_{1}},{{({{\lambda }_{1}}{{\lambda }_{3}})}^{-1}}, {{\lambda }_{3}}\right]$ in the $\mathbf e_i \otimes \mathbf E_j$ basis. The current thickness and lateral lengths of the deformed plate are $2h$, $\ell_1$ and $\ell_3$, respectively.

We take the underlying Eulerian magnetic induction vector $\mathbf{B}$ to be in the $x_2$ direction, that is, $\mathbf{B}=[0,B_2,0]^{\text{T}}$. The associated Lagrangian field ${{\mathbf{B}}_{\text{L}}}={{\mathbf{F}}^{-1}}\mathbf{B}$ is $\mathbf{B}_{\text{L}}=[0,B_{\text{L}2},0]^{\text{T}}$ with 
${{B}_{\text{L}2}}={{\lambda }_{1}}{{\lambda }_{3}}{{B}_{2}}$. The five independent invariants in Eq.~\eqref{7} are written now as
\begin{align} \label{9}
&{{I}_{1}}=\lambda _{1}^{2}+{{\left( {{\lambda }_{1}}{{\lambda }_{3}} \right)}^{-2}}+\lambda _{3}^{2},\qquad
{{I}_{2}}=\lambda _{1}^{-2}+\lambda _{3}^{-2}+\lambda _{1}^{2}\lambda _{3}^{2}, \notag\\
&{{I}_{4}}=\lambda _{1}^{2}\lambda _{3}^{2}B_{2}^{2}, \qquad {{I}_{5}}={{\left( {{\lambda }_{1}}{{\lambda }_{3}} \right)}^{-2}}{{I}_{4}},\qquad
{{I}_{6}}={{\left( {{\lambda }_{1}}{{\lambda }_{3}} \right)}^{-4}}{{I}_{4}}.
\end{align}

Thus, we can define a reduced energy function $\omega$ as
$\omega \left( {{\lambda }_{1}},{{\lambda }_{3}},{{I}_{4}} \right)=\Omega \left({{I}_{1}},{{I}_{2}},{{I}_{4}},{{I}_{5}}, {{I}_{6}} \right)$. Based on Eq.~\eqref{9} and the chain  rule, the constitutive relations \eqref{8} are written as
\begin{equation} \label{10}
{{\tau }_{11}}-{{\tau }_{22}}={{\lambda }_{1}}{{\omega }_{{{\lambda }_{1}}}},\qquad {{\tau }_{33}}-{{\tau }_{22}}={{\lambda }_{3}}{{\omega }_{{{\lambda }_{3}}}},\qquad {{H}_{2}}=2{{\left( {{\lambda }_{1}}{{\lambda }_{3}} \right)}^{2}}{{B}_{2}}{{\omega }_{4}},
\end{equation}
where  ${{\omega }_{\lambda_1}}=\partial \omega /\partial {\lambda_1}$, ${{\omega }_{\lambda_3}}=\partial \omega /\partial {\lambda_3}$, and ${{\omega }_4}=\partial \omega /\partial {I_4}$. They satisfy
\begin{align} \label{11}
& {{\lambda }_{1}}{{\omega }_{{{\lambda }_{1}}}}=2\left[ \left( {{\Omega }_{1}}+{{\Omega }_{2}}\lambda _{3}^{2} \right)\left( \lambda _{1}^{2}-\lambda _{1}^{-2}\lambda _{3}^{-2} \right)-\left( {{\Omega }_{5}}+2{{\Omega }_{6}}\lambda _{1}^{-2}\lambda _{3}^{-2} \right)B_{2}^{2} \right], \notag\\ 
& {{\lambda }_{3}}{{\omega }_{{{\lambda }_{3}}}}=2\left[ \left( {{\Omega }_{1}}+{{\Omega }_{2}}\lambda _{1}^{2} \right)\left( \lambda _{3}^{2}-\lambda _{1}^{-2}\lambda _{3}^{-2} \right)-\left( {{\Omega }_{5}}+2{{\Omega }_{6}}\lambda _{1}^{-2}\lambda _{3}^{-2} \right)B_{2}^{2} \right], \\ 
& {{\left( {{\lambda }_{1}}{{\lambda }_{3}} \right)}^{2}}{{\omega }_{4}}={{\Omega }_{4}}{{\left( {{\lambda }_{1}}{{\lambda }_{3}} \right)}^{2}}+{{\Omega }_{5}}+{{\Omega }_{6}}{{\left( {{\lambda }_{1}}{{\lambda }_{3}} \right)}^{-2}}. \notag
\end{align}
Note that $H_1=H_3=0$ from Eq.~\eqref{8}$_2$. For constant $\lambda_1$, $\lambda_3$ and $B_2$, all the fields are uniform and hence satisfy the equilibrium equations and the Maxwell equations automatically.

It follows from the magnetic boundary conditions \eqref{4}$_{2,3}$ that the non-zero components of ${{\mathbf{B}}^\star}$ and ${\mathbf{H}}^\star$ are $B_{2}^\star={{B}_{2}}$ and $H_{2}^\star=\mu _{0}^{-1}{{B}_{2}}$, respectively. Thus, the non-zero components of the Maxwell stress \eqref{3} are 
\begin{equation} \label{12}
\tau _{11}^\star=\tau _{33}^\star=-\tau _{22}^\star=-\tfrac{1}{2}\mu _{0}^{-1}B_{2}^{2}.
\end{equation}
These Maxwell stress components generate the magnetic traction vector ${{\mathbf{t}}^{\text{m}}}={{\bm{\tau }}^\star} \mathbf{n}$. 

For \textit{uni-axial mechanical loading} in the $x_1$ direction (Fig.~\ref{Fig1}(c)), there are no mechanical tractions applied on the faces $x_2 = \pm h$ and $x_3 =\pm\ell_3/2$, only magnetic tractions. The traction boundary conditions \eqref{4}$_1$ yield
\begin{equation} \label{13}
{{\tau }_{22}}=\tau _{22}^\star,\qquad {{\tau }_{33}}=\tau _{33}^\star,\qquad {{\tau }_{11}}-\tau _{11}^\star=t_{1}^{\text{a}}={{\lambda }_{1}}{{s}_{1}},
\end{equation}
where ${{s}_{1}}$ is the nominal mechanical traction per unit area of $\mathcal{\partial{B}}_r$ applied on the faces $x_1 =\pm\ell_1/2$. Thus, we deduce from Eqs.~\eqref{10} and \eqref{13} that the governing equations of the nonlinear response for uni-axial loading are
\begin{equation} \label{14}
{{\lambda }_{1}}{{\omega }_{{{\lambda }_{1}}}}+\tau _{22}^\star-\tau _{11}^\star={{\lambda }_{1}}{{s}_{1}}, \qquad {{\lambda }_{3}}{{\omega }_{{{\lambda }_{3}}}}+\tau _{22}^\star-\tau _{33}^\star=0, \qquad {{H}_{2}}=2{{\left( {{\lambda }_{1}}{{\lambda }_{3}} \right)}^{2}}{{B}_{2}}{{\omega }_{4}},
\end{equation}
Note that Eq.~\eqref{14}$_2$ determines the induced principal stretch $\lambda_3$ in terms of the stretch $\lambda_1$ and magnetic field $B_2$. Then ${{s}_{1}}$ and  $H_2$ are calculated from Eqs.~\eqref{14}$_{1,3}$, respectively.

For \textit{plane-strain mechanical loading} in the $x_1$ direction (Fig.~\ref{Fig1}(b)), we have ${{\lambda }_{3}}=1$ and ${{\lambda }_{1}}=\lambda _{2}^{-1} \equiv {{\lambda }}$. There is  no mechanical traction applied on the faces $x_2 = \pm h$, but  lateral mechanical tractions, applied on the faces $x_1 =\pm\ell_1/2$ and $x_3 =\pm\ell_3/2$, are required to maintain the plane-strain deformation. As a result, Eq.~\eqref{9} becomes
\begin{equation} \label{15}
{{I}_{1}}={{I}_{2}}={{\lambda }^{2}}+{{\lambda }^{-2}}+1,\qquad {{I}_{4}}={{\lambda }^{2}}B_{2}^{2}, \qquad {{I}_{5}}={{\lambda }^{-2}}{{I}_{4}}, \qquad {{I}_{6}}={{\lambda }^{-4}}{{I}_{4}}.
\end{equation}
By introducing the reduced energy function $\tilde \omega$ as $\tilde{\omega }\left( \lambda ,{{I}_{4}} \right)=\Omega \left( {{I}_{1}},{{I}_{2}},{{I}_{4}},{{I}_{5}},{{I}_{6}} \right)$, it follows from the constitutive relations \eqref{8} that
\begin{align} \label{16}
&{{\tau }_{11}}-{{\tau }_{22}}=\lambda {{{\tilde{\omega }}}_{\lambda }}, \qquad {{H}_{2}}=2{{\lambda }^{2}}{{B}_{2}}{{{\tilde{\omega }}}_{4}}, \notag \\ 
&{{\tau }_{33}}-{{\tau }_{22}}=2\left[ \left( {{\lambda }^{-2}}{{\Omega }_{1}}+{{\Omega }_{2}} \right)\left( {{\lambda }^{2}}-1 \right)-\left( {{\Omega }_{5}}+2{{\lambda }^{-2}}{{\Omega }_{6}} \right)B_{2}^{2} \right], 
\end{align}
where ${\tilde{\omega }_{\lambda}}=\partial \tilde\omega /\partial {\lambda}$ and ${\tilde{\omega }_4}=\partial \tilde\omega /\partial {I_4}$. They are determined by
\begin{align} \label{17}
&\lambda {{\tilde{\omega }}_{\lambda }}=2\left[ \left( {{\lambda }^{2}}-{{\lambda }^{-2}} \right)\left( {{\Omega }_{1}}+{{\Omega }_{2}} \right)-\left( {{\Omega }_{5}}+2{{\Omega }_{6}}{{\lambda }^{-2}} \right)B_{2}^{2} \right], \notag\\
&{{\lambda }^{2}}{{\tilde{\omega }}_{4}}={{\lambda }^{2}}{{\Omega }_{4}}+{{\Omega }_{5}}+{{\lambda }^{-2}}{{\Omega }_{6}}.
\end{align}
For this loading mode, the traction boundary conditions \eqref{4}$_1$ read
\begin{equation} \label{18}
{{\tau }_{22}}=\tau _{22}^\star,\qquad {{\tau }_{11}}-\tau _{11}^\star=t_{1}^{\text{a}}=\lambda {{s}_{1}},\qquad {{\tau }_{33}}-\tau _{33}^\star=t_{3}^{\text{a}}={{s}_{3}},
\end{equation}
where ${{s}_{1}}$ and ${{s}_{3}}$ are the nominal mechanical tractions applied on the faces $x_1 =\pm\ell_1/2$ and $x_3 =\pm\ell_3/2$, respectively. In view of Eqs.~\eqref{16}$_{1,2}$ and \eqref{18}, the nonlinear static response for plane-strain loading is governed by
\begin{align} \label{19}
\lambda {{\tilde{\omega }}_{\lambda }}+\tau _{22}^\star-\tau _{11}^\star=\lambda {{s}_{1}}, \qquad {{\tau }_{33}}-{{\tau }_{22}}+\tau _{22}^\star-\tau _{33}^\star={{s}_{3}}, \qquad
{{H}_{2}}=2{{\lambda }^{2}}{{B}_{2}}{{{\tilde{\omega }}}_{4}},
\end{align}
where ${{\tau }_{33}}-{{\tau }_{22}}$ is given by Eq.~\eqref{16}$_3$. Thus, Eq.~\eqref{19} is used to calculate ${{s}_{1}}$, ${{s}_{3}}$ and $H_2$ in terms of the applied stretch $\lambda$ and magnetic field $B_2$.

\section{Linearized stability analysis} \label{sec3}

In this section we employ the linearized incremental theory of magneto-elasticity \citep{ottenio2008incremental, destrade2011magneto} and the Stroh formalism \citep{su2018wrinkles} to investigate the formation of small-amplitude wrinkles, signaling the onset of wrinkling instability of the SMA plate, for the two loading modes described in Sec.~\ref{sec2.2}.

\subsection{Incremental governing equations} \label{sec3.1}

Consider an infinitesimal incremental mechanical displacement $\mathbf{u}={\mathbf{\dot x}}$ along with an updated incremental magnetic induction vector $\mathbf{\dot{B}}_{\text{L0}}$, superimposed on the finitely deformed configuration reached via $\mathbf{x}=\bm{\chi}(\mathbf{X})$. Here and henceforth, a superposed dot indicates an increment.

The incremental balance laws and the incremental incompressibility condition are formulated in Eulerian, or updated Lagrangian, form as
\begin{equation} \label{20}
\text{div}\,{\mathbf{\dot{T}}_{0}}=\mathbf{0},\qquad
\text{div}\,{\mathbf{\dot{B}}_{\text{L0}}}=0,\qquad
\text{curl}\,{\mathbf{\dot{H}}_{\text{L0}}}=\mathbf{0},\qquad
\text{div}\,\mathbf{u}=\text{tr}\,\mathbf{L}=0,
\end{equation}
where ${\mathbf{\dot{T}}}_{0}=\mathbf{F}\mathbf{\dot{T}}$, ${\mathbf{\dot{B}}_{\text{L0}}}=\mathbf{F}{\mathbf{\dot{B}}_{\text{L}}}$ and ${\mathbf{\dot{H}}_{\text{L0}}}=\mathbf{F}^{-\text{T}}{\mathbf{\dot{H}}_{\text{L}}}$ are the push-forward versions of the corresponding Lagrangian increments $\mathbf{\dot{T}}$, ${\mathbf{\dot{B}}_{\text{L}}}$ and ${\mathbf{\dot{H}}_{\text{L}}}$, respectively, and $\mathbf{L}=\text{grad}\, \mathbf{u}$ is the incremental displacement gradient. The resulting push-forward quantities are identified with a subscript 0. 

The linearized incremental constitutive equations for incompressible SMA materials read
\begin{equation} \label{21}
{{\mathbf{\dot{T}}}_{0}}={\bm{\mathcal{A}}_{0}}\mathbf{L}+{{\mathbf{\Gamma }}_{0}} {{\mathbf{\dot{B}}}_{\text{L}0}} +p\mathbf{L}-\dot{p}\mathbf{I}, \qquad {{\mathbf{\dot{H}}}_{\text{L}0}}=\mathbf{\Gamma }_{0}^{\text{T}}\mathbf{L}+{{\mathbf{K }}_{0}} {{\mathbf{\dot{B}}}_{\text{L}0}},
\end{equation}
where ${\bm{\mathcal{A}}_{0}}$, ${{\mathbf{\Gamma }}_{0}}$ and ${{\mathbf{K}}_{0}}$ are, respectively, fourth-, third- and second-order tensors, which are referred to as \textit{instantaneous} magneto-elastic moduli tensors (see \citet{ottenio2008incremental} and \citet{destrade2011magneto} for their general expressions). In index notation, these magneto-elastic moduli tensors are given by
\begin{equation} \label{22}
{{\mathcal{A}}_{0piqj}}={{F}_{p\alpha }}{{F}_{q\beta }}{{\mathcal{A}}_{\alpha i\beta j}},\qquad
{{\Gamma }_{0piq}}={{F}_{p\alpha }}F_{\beta q}^{-1}{{\Gamma }_{\alpha i\beta }},\qquad
{{K}_{0ij}}=F_{\alpha i}^{-1}F_{\beta j}^{-1}{{K}_{\alpha \beta }},
\end{equation}
where $\bm{\mathcal{A}}$, ${\mathbf{\Gamma }}$ and ${\mathbf{K}}$ are the relevant \textit{referential} magneto-elastic moduli tensors, which are defined by ${\bm{\mathcal{A}}}={{{\partial }^{2}}\Omega}/({\partial {\mathbf{F}}\partial {\mathbf{F}}})$, ${\mathbf{\Gamma }}={{{\partial }^{2}}\Omega}/({\partial {\mathbf{F}}\partial {\mathbf{B}_{\text{L}}}})$, and $\mathbf{K}={{{\partial }^{2}}\Omega}/({\partial {\mathbf{B}_{\text{L}}}\partial {\mathbf{B}_{\text{L}}}})$. Note that the instantaneous moduli tensors have the symmetries ${{\mathcal{A}}_{0piqj}} ={{\mathcal{A}}_{0qjpi}}$, ${\Gamma }_{0piq}={\Gamma }_{0ipq}$, and ${{K}_{0ij}}={{K}_{0ji}}$.

Using the incremental form of the rotational balance condition $\mathbf{FT}=(\mathbf{FT})^{\text{T}}$, we have the following relation between ${\bm{\mathcal{A}}_{0}}$ and $\bm{\tau}$ for an incompressible material
\begin{equation} \label{23}
{{\mathcal{A}}_{0jisk}}-{{\mathcal{A}}_{0ijsk}}=\left( {{\tau }_{js}}+p{{\delta }_{js}} \right){{\delta }_{ik}}-\left( {{\tau }_{is}}+p{{\delta }_{is}} \right){{\delta }_{jk}}.
\end{equation}

The incremental fields exterior to the material read
\begin{equation} \label{24}
{{\dot{\mathbf{B}}}^\star}={{\mu }_{0}}{{\dot{\mathbf{H}}}^\star}, \qquad
{{\bm{\dot{\tau }}}^\star}=\mu _{0}^{-1}[ {{{\dot{\mathbf{B}}}}^\star}\otimes {\mathbf{B}^\star}+{{\mathbf{B}}^\star}\otimes {{{\dot{\mathbf{B}}}}^\star}-( {{\mathbf{B}}^\star}\cdot {{{\dot{\mathbf{B}}}}^\star})\mathbf{I}],
\end{equation}
where $\dot{\mathbf{B}}^\star$ and $\dot{\mathbf{H}}^\star$ are to satisfy $\text{div}\,{{\dot{\mathbf{B}}}^\star}=0$ and $\text{curl}\,{\dot{\mathbf{H}}^\star}=\mathbf{0}$, respectively, and hence $\bm{\dot{\tau }}^*$ is divergence-free, i.e., $\text{div}\,{{\bm{\dot{\tau }}}^\star}=\mathbf{0}$.

The mechanical and magnetic boundary conditions for the incremental fields can be expressed, in updated Lagrangian form, as
\begin{align} \label{25}
& \left[ \mathbf{\dot{T}}_{0}^{\text{T}}-{{{\bm{\dot{\tau }}}}^\star}+{{\bm{\tau }}^\star}{{\mathbf{L}}^{\text{T}}}-\left( \text{div}\,\mathbf{u} \right){{\bm{\tau }}^\star} \right] \mathbf{n}=\mathbf{\dot{t}}_{0}^{\text{A}}, \notag\\ 
& \left[ {{{\mathbf{\dot{B}}}}_{\text{L}0}}-{{{\mathbf{\dot{B}}}}^\star}+\mathbf{L}{{\mathbf{B}}^\star}-\left( \text{div}\,\mathbf{u} \right){{\mathbf{B}}^\star} \right]\cdot \mathbf{n}=0, \notag \\ 
& \left( {{{\mathbf{\dot{H}}}}_{\text{L}0}}-{{{\mathbf{\dot{H}}}}^{\mathbf{*}}}-{{\mathbf{L}}^{\text{T}}}{{\mathbf{H}}^{\mathbf{*}}} \right)\times \mathbf{n}=\mathbf{0}, 
\end{align}
where $\mathbf{\dot{t}}_{0}^{\text{A}}\text{d}a= {{\mathbf{\dot{t}}}^{\text{A}}}\text{d}A$, with $\mathbf{t}^{\text{A}}$ being the applied mechanical traction vector per unit area of $\mathcal{\partial{B}}_r$ (i.e., ${{\mathbf{t}}^{\text{A}}}\text{d}A={{\mathbf{t}}^{\text{a}}}\text{d}a$). Note that $\text{d}A$ and $\text{d}a$ are area elements of the reference and current configurations, respectively.

In this work we focus on incremental two-dimensional solutions independent of $x_3$, such that ${{u}_{3}}={{\dot{B}}_{\text{L}03}}={{\dot{H}}_{\text{L}03}}=0$, and hence $u_i=u_i({{x}_{1}},{{x}_{2}})$, ${\dot{B}_{\text{L}0i}}={\dot{B}_{\text{L}0i}}({{x}_{1}},{{x}_{2}})$ and ${\dot{H}_{\text{L}0i}}={\dot{H}_{\text{L}0i}}({{x}_{1}},{{x}_{2}})$ for $i=1,2$ and $\dot{p}=\dot{p}({{x}_{1}},{{x}_{2}})$.

From Eq.~\eqref{20}$_3$, the incremental magnetic field vector ${\mathbf{\dot{H}}_{\text{L0}}}$ is curl-free and thus an incremental magnetic scalar potential $\dot{\varphi}$ can be introduced such that ${\mathbf{\dot{H}}_{\text{L}0}} =-\text{grad}\dot{\varphi}$, with  components 
\begin{equation} \label{26}
{{\dot{H}}_{\text{L}01}}=-{{\dot{\varphi }}_{,1}},\qquad {{\dot{H}}_{\text{L}02}}=-{{\dot{\varphi }}_{,2}}.
\end{equation}
The incremental balance laws and incompressibility condition \eqref{20}$_{1,2,4}$ become
\begin{align} \label{27}
& {{{\dot{T}}}_{011,1}}+{{{\dot{T}}}_{021,2}}=0,&& {{{\dot{T}}}_{012,1}}+{{{\dot{T}}}_{022,2}}=0, \notag\\ 
&{{{\dot{B}}}_{\text{L}01,1}}+{{{\dot{B}}}_{\text{L}02,2}}=0, && {{u}_{1,1}}+{{u}_{2,2}}=0.
\end{align}

For pure homogeneous deformation of the SMA plate subject to the transverse magnetic field, we have $F_{ij}=0$ for $i \ne j$ and $B_1=B_3=0$. The magneto-elastic moduli tensors ${\bm{\mathcal{A}}_{0}}$, ${{\mathbf{\Gamma }}_{0}}$ and ${{\mathbf{K}}_{0}}$ satisfy \citep{ottenio2008incremental}
\begin{align} \label{28}
& {{\mathcal{A}}_{0iijk}}=0,\quad {{K}_{0jk}}=0, \quad \text{for} \quad j\ne k, \notag \\ 
& {{\Gamma }_{0ii1}}={{\Gamma }_{01ii}}={{\Gamma }_{0ii3}}={{\Gamma }_{03ii}}=0, \notag\\ 
& {{\Gamma }_{0ijk}}=0, \quad \text{for} \quad i\ne j\ne k\ne i. 
\end{align}
Consequently, using Eqs.~\eqref{26} and \eqref{28}, the incremental constitutive relations \eqref{21} are written in terms of $u_i\,(i=1,2)$ and $\dot{\varphi}$ as
\begin{align} \label{29}
& {{{\dot{T}}}_{011}}=\left( {{c}_{11}}+p \right){{u}_{1,1}}+{{c}_{12}}{{u}_{2,2}}+{{e}_{21}}{{{\dot{\varphi }}}_{,2}}-\dot{p}, \notag \\ 
& {{{\dot{T}}}_{012}}=\left( {{c}_{69}}+p \right){{u}_{1,2}}+{{c}_{66}}{{u}_{2,1}}+{{e}_{16}}{{{\dot{\varphi }}}_{,1}}, \notag \\ 
& {{{\dot{T}}}_{021}}=\left( {{c}_{69}}+p \right){{u}_{2,1}}+{{c}_{99}}{{u}_{1,2}}+{{e}_{16}}{{{\dot{\varphi }}}_{,1}}, \notag \\ 
& {{{\dot{T}}}_{022}}=\left( {{c}_{22}}+p \right){{u}_{2,2}}+{{c}_{12}}{{u}_{1,1}}+{{e}_{22}}{{{\dot{\varphi }}}_{,2}}-\dot{p},
\end{align}
and
\begin{align} \label{30}
&{{{\dot{B}}}_{\text{L}01}}={{e}_{16}}({{u}_{2,1}}+{{u}_{1,2}})-{{\mu }_{11}}{{{\dot{\varphi }}}_{,1}}, \notag \\ &{{{\dot{B}}}_{\text{L}02}}={{e}_{21}}{{u}_{1,1}}+{{e}_{22}}{{u}_{2,2}}-{{\mu }_{22}}{{{\dot{\varphi }}}_{,2}},  
\end{align}
where the effective material parameters $c_{ij}$, $e_{ij}$ and $\mu_{ij}$ are defined as
\begin{align} \label{31}
& {{\mu }_{11}}=K_{011}^{-1},\quad {{\mu }_{22}}=K_{022}^{-1},\quad {{e}_{16}}=-{{\Gamma }_{0211}}K_{011}^{-1},\quad {{e}_{21}}=-{{\Gamma }_{0112}}K_{022}^{-1},\quad {{e}_{22}}=-{{\Gamma }_{0222}}K_{022}^{-1}, \notag \\ 
& {{c}_{11}}={{\mathcal{A}}_{01111}}+{{\Gamma }_{0112}}{{e}_{21}},\quad {{c}_{12}}={{\mathcal{A}}_{01122}}+{{\Gamma }_{0112}}{{e}_{22}},\quad {{c}_{22}}={{\mathcal{A}}_{02222}}+{{\Gamma }_{0222}}{{e}_{22}}, \notag \\ 
& {{c}_{69}}={{\mathcal{A}}_{01221}}+{{\Gamma }_{0211}}{{e}_{16}},\quad {{c}_{66}}={{\mathcal{A}}_{01212}}+{{\Gamma }_{0211}}{{e}_{16}},\quad {{c}_{99}}={{\mathcal{A}}_{02121}}+{{\Gamma }_{0211}}{{e}_{16}}. 
\end{align}
%

\subsection{Stroh formulation and its resolution for plates} \label{sec3.2}

The two-dimensional incremental solutions, for the wrinkling instability with sinusoidal shape along the $x_1$ direction and amplitude variations along the $x_2$ direction (see Fig.~\ref{Fig1}(d) and \ref{Fig1}(e)), are sought in the form
\begin{multline} \label{32}
\left\{ {{u}_{1}},{{u}_{2}},\dot{\varphi},{{{\dot{T}}}_{021}}, {{{\dot{T}}}_{022}},{{{\dot{B}}}_{\text{L}02}} \right\}\\
=\Re\left\{ \left[ {{k}^{-1}}{{{\overline{U}}}_{1}}, {{k}^{-1}}{{{\overline{U}}}_{2}}, \sqrt{G/{{\mu }_{0}}}{{k}^{-1}}\overline{\Phi }, \text{i} G{{{\overline{\Sigma }}}_{21}}, \text{i} G{{{\overline{\Sigma }}}_{22}}, \text{i} \sqrt{G{{\mu }_{0}}}\overline{\Delta } \right] {{\text{e}}^{\text{i}k{{x}_{1}}}} \right\},
\end{multline}
where $\text{i}=\sqrt{-1}$, $\overline{U}_1$, $\overline{U}_2$, $\overline{\Phi}$, $\overline{\Sigma}_{21}$, $\overline{\Sigma}_{22}$ and $\overline{\Delta}$ are non-dimensional functions of $kx_2$ only, $k = 2\pi/\mathcal{L}$ is the wrinkling wavenumber with $\mathcal{L}$ being the wavelength of the wrinkles, and $G$ is the initial shear modulus  (in Pa) of the SMA plate in the absence of magnetic field.

Following a standard derivation procedure \citep{su2018wrinkles, su2019finite}, we can rewrite the incremental governing equations \eqref{27}, \eqref{29} and \eqref{30} in the following non-dimensional Stroh form:
\renewcommand{\arraystretch}{1.2}
\begin{equation} \label{33}
\bm{\overline{\eta}}'=\text{i}\mathbf{\overline{N}\bm{\overline{\eta}}} =\text{i}\left[ \begin{matrix}
{{{\mathbf{\overline{N}}}}_{1}} & {{{\mathbf{\overline{N}}}}_{2}}  \\
{{{\mathbf{\overline{N}}}}_{3}} & \mathbf{\overline{N}}_{1}^{\dagger }  \\
\end{matrix} \right]\bm{\overline{\eta}},
\end{equation}
where $\bm{\overline{\eta }}={{\left[ \begin{matrix}
{{{\overline{U}}}_{1}} & {{{\overline{U}}}_{2}} & {\overline{\Phi }} & {{{\overline{\Sigma }}}_{21}} & {{{\overline{\Sigma }}}_{22}} & {\overline{\Delta }}  \\
\end{matrix} \right]}^{\text{T}}}$ is the Stroh vector, $\mathbf{\overline{N}}$ is the $6\times6$ Stroh matrix, the prime denotes differentiation with respect to $kx_2$, and $^\dagger$ signifies the Hermitian operator. In what follows we use the generalized displacement and traction vectors $\mathbf{\overline{U}}={{\left[ \begin{matrix}
{{{\overline{U}}}_{1}} & {{{\overline{U}}}_{2}} & {\overline{\Phi }} \end{matrix} \right]}^{\text{T}}}$ and $\mathbf{\overline{S}}={{\left[ \begin{matrix}
{{{\overline{\Sigma }}}_{21}} & {{{\overline{\Sigma }}}_{22}} & {\overline{\Delta }} \end{matrix} \right]}^{\text{T}}}$ to express the Stroh vector as $\bm{\overline{\eta }}={{\left[ \begin{matrix}{\mathbf{\overline{U}}} & {\mathbf{\overline{S}}}
\end{matrix} \right]}^{\text{T}}}$. For reference, the $3\times3$ real sub-matrices $\overline{\mathbf{N}}_i\,(i=1,2,3)$ are presented in \ref{AppeA}.

For the constant Stroh matrix $\mathbf{\overline{N}}$ considered here, we seek solutions to Eq.~\eqref{33} in the form,
\begin{equation} \label{37}
\bm{\overline{\eta }}\left( k{{x}_{2}} \right)={{\bm{\overline{\eta }}}^{0}}{\text{e}^{\text{i}qk{{x}_{2}}}}.
\end{equation}
Substituting Eq.~\eqref{37} into Eq.~\eqref{33} yields an eigenvalue problem $\left( \mathbf{\overline{N}}-q\mathbf{I} \right){{\bm{\overline{\eta }}}^{0}}=\mathbf{0}$, with eigenvalue $q$ and eigenvector $\bm{\overline{\eta }}^{0}$. The requirement of non-trivial solutions results in $\text{det}\left(\mathbf{\overline{N}}-q\mathbf{I} \right)=0$, which gives the following bi-cubic characteristic equation in $q$:
\begin{equation}  \label{38}
\overline{c} \,\overline{g}\, {{q}^{6}}+\left[ 2\overline{b} \overline{g}+\overline{c}\overline{f}+\overline{d}\left( \overline{d}-2\overline{e} \right) \right]{{q}^{4}}+\left[ \overline{a}\,\overline{g}+2\overline{b}\,\overline{f}-2\bar{d}\left( \overline{d}-\overline{e} \right)-{{{\overline{e}}}^{2}}\overline{f}/\overline{g} \right]{{q}^{2}}+\overline{a}\overline{f}+{{\overline{d}}^{2}}=0,
\end{equation}
It is clear that the characteristic equation and hence the eigenvalues do not depend on $\tau_{22}$ (appearing in the Stroh matrix \eqref{A1}) for any choice of energy density function, despite the presence of external magnetic field. 

Thus, the two-dimensional wrinkling solutions to Eq.~\eqref{33} for an SMA plate are found as
\begin{equation} \label{39}
\bm{\overline{\eta }}\left( k{{x}_{2}} \right)=\left[ \begin{matrix}
\mathbf{\overline{U}}\left( k{{x}_{2}} \right)  \\
\mathbf{\overline{S}}\left( k{{x}_{2}} \right)  \\
\end{matrix} \right]=\sum\limits_{j=1}^{6}{{{A}_{j}}{{{\bm{\overline{\eta }}}}^{\left( j \right)}}{\text{e}^{\text{i}{{q}_{j}}k{{x}_{2}}}}}.
\end{equation}
where $A_j \,(j = 1,\ldots,6)$ are arbitrary constants to be determined, $q_j \,(j = 1,\ldots,6)$ are the eigenvalues, and $\bm{\overline{\eta}}^{(j)} \,(j = 1,\ldots,6)$ are the eigenvectors associated with $q_j$. The eigenvalues are complex conjugate pairs because of the real coefficients of Eq.~\eqref{38}.

So far, there is no restriction on the specific form of energy density function $\Omega$. 
To proceed further and demonstrate the possibility of obtaining concise and analytical eigen-equation and eigenvectors, we henceforth consider the magneto-elastic material to be characterized by the following \textit{Mooney-Rivlin magneto-elastic model},
\begin{equation} \label{40}
\Omega =\frac{G\left( 1-\beta  \right)}{2}\left( {{I}_{1}}-3 \right)+\frac{G\beta }{2}\left( {{I}_{2}}-3 \right)+F\left( {{I}_{5}} \right),
\end{equation}
where $F$ is an arbitrary function of $I_5$ only and $\beta \in [0,1]$ is a constant. Note that Eq.~\eqref{40} with $\beta=0$ reduces to the neo-Hookean magneto-elastic model, including, for example, the ideal model with no saturation and the magnetization saturation Langevin model, which we consider in Sec.~\ref{sec4}.

Using Eqs.~\eqref{40} and \eqref{A4}, the dimensionless parameters $\overline{a}-\overline{g}$ in Eq.~\eqref{A2} are calculated as
\begin{align} \label{41}
& \overline{a}=\lambda _{1}^{2}\left( 1-\beta  \right)+\lambda _{1}^{2}\lambda _{3}^{2}\beta -2\overline{B}_{2}^{2} \overline{{F}}_5, \qquad \overline{c}=\lambda _{1}^{-2}\lambda _{3}^{-2} \left(1-\beta  \right)+\lambda _{1}^{-2}\beta, \notag \\ 
& \overline{b}=\frac{1}{2}\left( \lambda _{1}^{-2}\lambda _{3}^{-2}+\lambda_{1}^{2} \right)\left[ \left( 1-\beta  \right)+\lambda_{3}^{2}\beta  \right]+\overline{B}_{2}^{2}
\left(3\overline{{F}}_5+2\overline{B}_{2}^{2}\overline{F}_{55} \right), \notag \\ 
& \overline{d}=-{{{\overline{B}}}_{2}},\qquad \overline{e}=-{{{\overline{B}}}_{2}}\left[ 1+\overline{{F}}_5/ \left(\overline{{F}}_5+2\overline{B}_{2}^{2} \overline{F}_{55} \right) \right], \notag \\
&\overline{f}=1/\left( 2\overline{{F}}_5 \right),\qquad \overline{g}=1/\left[ 2\left( \overline{{F}}_5+2\overline{B}_{2}^{2}\overline{F}_{55} \right) \right],
\end{align}
where ${{\overline{B}}_{2}}= {{B}_{2}}/\sqrt{G{{\mu }_{0}}}$ is the dimensionless applied magnetic induction; $\overline{F}_5$ and $\overline{F}_{55}$ are defined as
\begin{equation} \label{42}
\overline{F}_5={{\mu }_{0}}F_5={{\mu }_{0}} \partial F/\partial I_5, \qquad \overline{F}_{55}=G\mu _{0}^{2}F_{55}=G\mu _{0}^{2} \partial^2 F/\partial I_5^2.
\end{equation}
Then we find that the characteristic equation \eqref{38} factorizes as
\begin{equation}  \label{43}
\left( {{q}^{2}}+1 \right)\left( {{q}^{2}}+\lambda _{1}^{4}\lambda _{3}^{2} \right)\left( \overline{F}_5 {{q}^{2}}+\overline{{F}}_5+2\overline{B}_{2}^{2}\overline{F}_{55} \right)=0,
\end{equation}
which yields six pure imaginary eigenvalues as
\begin{equation} \label{44}
{{q}_{1}}=-{{q}_{4}}=\text{i}{{p}_{1}},\qquad {{q}_{2}}=-{{q}_{5}}=\text{i}{{p}_{2}},\qquad {{q}_{3}}=-{{q}_{6}}=\text{i}{{p}_{3}},
\end{equation}
where the real numbers $p_1$, $p_2$, and $p_3$ are
\begin{equation} \label{45}
{{p}_{1}}=1,\qquad {{p}_{2}}=\lambda _{1}^{2}{{\lambda }_{3}},\qquad {{p}_{3}}=\sqrt{1+2\overline{B}_{2}^{2}\overline{F}_{55}/\overline{{F}}_5}.
\end{equation}
The associated eigenvectors are derived as
\begin{align} \label{46}
& {{{\bm{\overline{\eta }}}}^{\left( 1 \right)}}={{\left[ 1,\,\text{i},\, \text{i}\overline{d}/\overline{f},\, \text{i} 
(2\overline{c}+{{{\overline{d}}}^{2}}/\overline{f} -{{{\overline{\tau }}}_{22}}), \, {{{\overline{\tau }}}_{22}}-( \overline{a}+\overline{c}+ {{{\overline{d}}}^{2}}/\overline{f}), \, -\overline{d} \right]}^{\text{T}}}, \notag \\ 
& {{{\bm{\overline{\eta }}}}^{\left( 2 \right)}}={{\left[ {{p}_{2}}, \, \text{i},\, \text{i}\overline{d}/\overline{f},\, \text{i}(\overline{a}+\overline{c}+2{{{\overline{d}}}^{2}}/ \overline{f}-{{{\overline{\tau }}}_{22}}),\, ( {{{\overline{\tau }}}_{22}}-2\overline{c}){{p}_{2}}, \, -\overline{d}{{p}_{2}} \right]}^{\text{T}}}, \notag \\ 
& {{{\bm{\overline{\eta }}}}^{\left( 3 \right)}}={{\left[ 0,\, 0,\, \text{i}{{p}_{3}},\, \text{i}{{p}_{3}}\overline{d},\, -\overline{d},\, \overline{f} \right]}^{\text{T}}}, \notag \\ 
& {{{\bm{\overline{\eta }}}}^{\left( 4 \right)}}={{({{{\bm{\overline{\eta }}}}^{(1)}})}^{*}},\qquad {{{\bm{\overline{\eta }}}}^{\left( 5 \right)}}={{({{{\bm{\overline{\eta }}}}^{(2)}})}^{*}},\qquad {{{\bm{\overline{\eta }}}}^{\left( 6 \right)}}={{({{{\bm{\overline{\eta }}}}^{(3)}})}^{*}},
\end{align}
where the asterisk $^*$ denotes the complex conjugate. We observe from Eq.~\eqref{46} that
\begin{align} \label{47}
& \overline{\eta }_{1}^{\left( j+3 \right)}=\overline{\eta }_{1}^{\left( j \right)}, && \overline{\eta }_{2}^{\left( j+3 \right)}=-\overline{\eta }_{2}^{\left( j \right)}, && \overline{\eta }_{3}^{\left( j+3 \right)}=-\overline{\eta }_{3}^{\left( j \right)}, \notag\\ 
& \overline{\eta }_{4}^{\left( j+3 \right)}=-\overline{\eta }_{4}^{\left( j \right)}, && \overline{\eta }_{5}^{\left( j+3 \right)}=\overline{\eta }_{5}^{\left( j \right)}, && \overline{\eta }_{6}^{\left( j+3 \right)}=\overline{\eta }_{6}^{\left( j \right)}, \qquad \left( j=1,2,3 \right),
\end{align}
where $\overline{\eta}_{i}^{\left( j \right)}$ is the $i$-th component of the eigenvector $\bm{\overline{\eta}}^{(j)}$.

\subsection{Incremental boundary conditions} \label{sec3.3}

We take the applied mechanical traction $\mathbf{t}^{\text{A}}$ as a \textit{dead load} (i.e., $\mathbf{\dot{t}}_{0}^{\text{A}}=\mathbf{\dot{t}}^{\text{A}}=\mathbf{0}$), so that the general incremental boundary conditions \eqref{25} are specialized to the two-dimensional problem at $x_2=\pm h$, as
\begin{align}   \label{50}
& {{{\dot{T}}}_{021}}=-\tau _{11}^\star{{u}_{2,1}} +\dot{\tau }_{21}^\star,
&&{{{\dot{T}}}_{022}}=-\tau _{22}^\star{{u}_{2,2}} +\dot{\tau }_{22}^\star, \notag \\ 
& {{{\dot{B}}}_{\text{L}02}} =\dot{B}_{2}^\star-B_{2}^\star{{u}_{2,2}},
&&{{{\dot{H}}}_{\text{L}01}}=\dot{H}_{1}^\star+H_{2}^\star{{u}_{2,1}}.
\end{align}

From $\text{curl}\,{\dot{\mathbf{H}}^\star}=\mathbf{0}$, we deduce the existence of an incremental magnetic scalar potential $\dot{\varphi}^\star=\dot{\varphi} ^\star(x_1,x_2)$ in vacuum such that
\begin{equation}   \label{51}
\dot{H}_{1}^\star=-\dot{\varphi }_{,1}^\star, \qquad
\dot{H}_{2}^\star=-\dot{\varphi }_{,2}^\star, \qquad
\dot{B}_{1}^\star=-{{\mu }_{0}}\dot{\varphi }_{,1}^\star, \qquad
\dot{B}_{2}^\star=-{{\mu }_{0}}\dot{\varphi }_{,2}^\star.
\end{equation}
Substituting Eq.~\eqref{51}$_{3,4}$ into $\text{div}\,{{\dot{\mathbf{B}}}^\star}=0$ results in the Laplace equation for $\dot{\varphi }^\star$,
\begin{equation}   \label{52}
\dot{\varphi }_{,11}^\star+\dot{\varphi }_{,22}^\star=0.
\end{equation}
To satisfy the decay condition $\dot{\varphi}^\star \to 0$ as $x_2 \to \pm \infty$, we take the solutions to Eq.~\eqref{52} which are localized near the interfaces $x_2 = \pm h$, as
\begin{align}    \label{53}
& \dot{\varphi }_{+}^\star ={{k}^{-1}}A_{+}^\star{{\text{e}}^{-k{{x}_{2}}}}{{\text{e}}^{\text{i}k{{x}_{1}}}},\quad ({{x}_{2}}>h), \notag \\ 
& \dot{\varphi }_{-}^\star ={{k}^{-1}}A_{-}^\star{{\text{e}}^{k{{x}_{2}}}}{{\text{e}}^{\text{i}k{{x}_{1}}}},\quad ({{x}_{2}}<-h),
\end{align}
where $A_{+}^\star$ and $A_{-}^\star$ are arbitrary constants. Thus, the associated incremental Maxwell stress tensor \eqref{24}$_2$ has non-zero components
\begin{equation}   \label{54}
\dot{\tau }_{11}^\star=\dot{\tau }_{33}^\star=-\dot{\tau }_{22}^\star=B_{2}^\star\dot{\varphi }_{+,2}^\star,\qquad
\dot{\tau }_{12}^\star=\dot{\tau }_{21}^\star=-B_{2}^\star\dot{\varphi }_{+,1}^\star
\end{equation}
and 
\begin{equation}    \label{55}
\dot{\tau }_{11}^\star=\dot{\tau }_{33}^\star=-\dot{\tau }_{22}^\star=B_{2}^\star\dot{\varphi }_{-,2}^\star,\qquad
\dot{\tau }_{12}^\star=\dot{\tau }_{21}^\star=-B_{2}^\star\dot{\varphi }_{-,1}^\star
\end{equation}
for ${{x}_{2}}>h$ and  ${{x}_{2}}<-h$, respectively.

Inserting Eqs.~\eqref{26}$_1$, \eqref{32}, \eqref{51}$_1$ and \eqref{53}$_1$ into the incremental magnetic boundary condition \eqref{50}$_4$ at the plate top surface $x_2=+h$ leads to the relation
\begin{equation}    \label{56}
\overline{A}_{+}^\star{{\text{e}}^{-kh}}=\overline{\Phi }\left( kh \right) +{{\overline{B}}_{2}}{{\overline{U}}_{2}}\left( kh \right),
\end{equation}
where $\overline{A}_{+}^\star=A_{+}^\star\sqrt{{{\mu }_{0}}/G}$. Using Eqs.~\eqref{12}, \eqref{32}, \eqref{51}$_4$, \eqref{53}$_1$, \eqref{54} and \eqref{56}, we write the incremental boundary conditions \eqref{50}$_{1-3}$ at $x_2=+h$ in an impedance form, as
\begin{equation}    \label{57}
\mathbf{\overline{S}}(kh)=\text{i}\mathbf{\overline{Z}}_{+}^\star\mathbf{\overline{U}}(kh),
\end{equation}
where
\begin{equation}    \label{58}
\mathbf{\overline{Z}}_{+}^\star=\left[ \begin{matrix}
0 & \text{i}\overline{B}_{2}^{2}/2 & \text{i}{{{\overline{B}}}_{2}}  \\
-\text{i}\overline{B}_{2}^{2}/2 & -\overline{B}_{2}^{2} & -{{{\overline{B}}}_{2}}  \\
-\text{i}{{{\overline{B}}}_{2}} & -{{{\overline{B}}}_{2}} & -1  \\
\end{matrix} \right]
\end{equation}
is a surface impedance matrix connecting the vectors $\mathbf{\overline{S}}(kh)$ and $\mathbf{\overline{U}}(kh)$ at the face $x_2=+h$.

Similarly, at the plate bottom surface $x_2=-h$, we obtain 
\begin{equation}    \label{59}
\overline{A}_{-}^\star{{\text{e}}^{-kh}}=\overline{\Phi }\left( -kh \right) +{{\overline{B}}_{2}}{{\overline{U}}_{2}}\left( -kh \right),\qquad \overline{A}_{-}^\star=A_{-}^\star\sqrt{{{\mu }_{0}}/G},
\end{equation}
and
\begin{align}    \label{60}
\mathbf{\overline{S}}(-kh)=\text{i}\mathbf{\overline{Z}}_{-}^\star\mathbf{\overline{U}}(-kh),
\end{align}
where the surface impedance matrix in the lower half-space is
\begin{equation}    \label{61}
\mathbf{\overline{Z}}_{-}^\star=\left[ \begin{matrix}
0 & \text{i}\overline{B}_{2}^{2}/2 & \text{i}{{{\overline{B}}}_{2}}  \\
-\text{i}\overline{B}_{2}^{2}/2 & \overline{B}_{2}^{2} & {{{\overline{B}}}_{2}}  \\
-\text{i}{{{\overline{B}}}_{2}} & {{{\overline{B}}}_{2}} & 1  \\
\end{matrix} \right].
\end{equation}
%

\subsection{Bifurcation equations for wrinkling instabilities} \label{sec3.4}

Using Eq.~\eqref{A6}, we rewrite Eqs.~\eqref{57} and \eqref{60} as
\begin{multline}    \label{62}
\left[ \begin{matrix}
\mathbf{\overline{S}}\left( kh \right)  \\
\mathbf{\overline{S}}\left( -kh \right)  \\
\end{matrix} \right]=\text{i}\left[ \begin{matrix}
\mathbf{\overline{Z}}_{+}^\star & \mathbf{0}  \\
\mathbf{0} & \mathbf{\overline{Z}}_{-}^\star  \\
\end{matrix} \right]\left[ \begin{matrix}
\mathbf{\overline{U}}(kh)  \\
\mathbf{\overline{U}}(-kh)  \\
\end{matrix} \right] \\ 
=\text{i} \left[ \begin{matrix}
\mathbf{\overline{Z}}_{+}^\star & \mathbf{0}  \\
\mathbf{0} & \mathbf{\overline{Z}}_{-}^\star  \\
\end{matrix} \right]\left[ \begin{matrix}
\overline{\eta }_{1}^{(1)}E_{1}^{+} & \overline{\eta }_{1}^{(2)}E_{2}^{+} & \overline{\eta }_{1}^{(3)}E_{3}^{+} & \overline{\eta }_{1}^{(4)}E_{1}^{-} & \overline{\eta }_{1}^{(5)}E_{2}^{-} & \overline{\eta }_{1}^{(6)}E_{3}^{-}  \\
\overline{\eta }_{2}^{(1)}E_{1}^{+} & \overline{\eta }_{2}^{(2)}E_{2}^{+} & \overline{\eta }_{2}^{(3)}E_{3}^{+} & \overline{\eta }_{2}^{(4)}E_{1}^{-} & \overline{\eta }_{2}^{(5)}E_{2}^{-} & \overline{\eta }_{2}^{(6)}E_{3}^{-}  \\
\overline{\eta }_{3}^{(1)}E_{1}^{+} & \overline{\eta }_{3}^{(2)}E_{2}^{+} & \overline{\eta }_{3}^{(3)}E_{3}^{+} & \overline{\eta }_{3}^{(4)}E_{1}^{-} & \overline{\eta }_{3}^{(5)}E_{2}^{-} & \overline{\eta }_{3}^{(6)}E_{3}^{-}  \\
\overline{\eta }_{1}^{(1)}E_{1}^{-} & \overline{\eta }_{1}^{(2)}E_{2}^{-} & \overline{\eta }_{1}^{(3)}E_{3}^{-} & \overline{\eta }_{1}^{(4)}E_{1}^{+} & \overline{\eta }_{1}^{(5)}E_{2}^{+} & \overline{\eta }_{1}^{(6)}E_{3}^{+}  \\
\overline{\eta }_{2}^{(1)}E_{1}^{-} & \overline{\eta }_{2}^{(2)}E_{2}^{-} & \overline{\eta }_{2}^{(3)}E_{3}^{-} & \overline{\eta }_{2}^{(4)}E_{1}^{+} & \overline{\eta }_{2}^{(5)}E_{2}^{+} & \overline{\eta }_{2}^{(6)}E_{3}^{+}  \\
\overline{\eta }_{3}^{(1)}E_{1}^{-} & \overline{\eta }_{3}^{(2)}E_{2}^{-} & \overline{\eta }_{3}^{(3)}E_{3}^{-} & \overline{\eta }_{3}^{(4)}E_{1}^{+} & \overline{\eta }_{3}^{(5)}E_{2}^{+} & \overline{\eta }_{3}^{(6)}E_{3}^{+}  \\
\end{matrix} \right]\left[ \begin{matrix}
{{A}_{1}}  \\ {{A}_{2}}  \\ {{A}_{3}}  \\
{{A}_{4}}  \\ {{A}_{5}}  \\ {{A}_{6}}  \\
\end{matrix} \right].
\end{multline}
Substituting Eqs.~\eqref{47} and \eqref{A7} into Eq.~\eqref{62} yields
\begin{equation}     \label{63}
\left[ \begin{matrix}
{{D}_{1}}E_{1}^{+} & {{D}_{2}}E_{2}^{+} & {{D}_{3}}E_{3}^{+} & -{{D}_{1}}E_{1}^{-} & -{{D}_{2}}E_{2}^{-} & -{{D}_{3}}E_{3}^{-}  \\
F_{1}^{+}E_{1}^{+} & F_{2}^{+}E_{2}^{+} & F_{3}^{+}E_{3}^{+} & F_{1}^{-}E_{1}^{-} & F_{2}^{-}E_{2}^{-} & F_{3}^{-}E_{3}^{-}  \\
G_{1}^{+}E_{1}^{+} & G_{2}^{+}E_{2}^{+} & G_{3}^{+}E_{3}^{+} & G_{1}^{-}E_{1}^{-} & G_{2}^{-}E_{2}^{-} & G_{3}^{-}E_{3}^{-}  \\
{{D}_{1}}E_{1}^{-} & {{D}_{2}}E_{2}^{-} & {{D}_{3}}E_{3}^{-} & -{{D}_{1}}E_{1}^{+} & -{{D}_{2}}E_{2}^{+} & -{{D}_{3}}E_{3}^{+}  \\
F_{1}^{-}E_{1}^{-} & F_{2}^{-}E_{2}^{-} & F_{3}^{-}E_{3}^{-} & F_{1}^{+}E_{1}^{+} & F_{2}^{+}E_{2}^{+} & F_{3}^{+}E_{3}^{+}  \\
G_{1}^{-}E_{1}^{-} & G_{2}^{-}E_{2}^{-} & G_{3}^{-}E_{3}^{-} & G_{1}^{+}E_{1}^{+} & G_{2}^{+}E_{2}^{+} & G_{3}^{+}E_{3}^{+}  \\
\end{matrix} \right]\left[ \begin{matrix}
{{A}_{1}}  \\
{{A}_{2}}  \\
{{A}_{3}}  \\
{{A}_{4}}  \\
{{A}_{5}}  \\
{{A}_{6}}  \\
\end{matrix} \right]=0,
\end{equation}
where
\begin{align}     \label{64}
& {{D}_{j}}=\overline{\eta }_{4}^{\left( j \right)}+{{{\overline{B}}}_{2}}\left( {{{\overline{B}}}_{2}}\overline{\eta }_{2}^{\left( j \right)}+2\overline{\eta }_{3}^{\left( j \right)} \right)/2, \notag\\ 
& F_{j}^{\pm }=\overline{\eta }_{5}^{\left( j \right)}-\overline{B}_{2}^{2}\overline{\eta }_{1}^{\left( j \right)}/2\pm \text{i}{{{\overline{B}}}_{2}}\left( {{{\overline{B}}}_{2}}\overline{\eta }_{2}^{\left( j \right)}+\overline{\eta }_{3}^{\left( j \right)} \right), \notag\\ 
& G_{j}^{\pm }=\overline{\eta }_{6}^{\left( j \right)}-{{{\overline{B}}}_{2}}\overline{\eta }_{1}^{\left( j \right)}\pm \text{i}\left( {{{\overline{B}}}_{2}}\overline{\eta }_{2}^{\left( j \right)}+\overline{\eta }_{3}^{\left( j \right)} \right),\quad \left( j=1,2,3 \right).
\end{align}

By conducting some simple linear matrix manipulations of Eq.~\eqref{63} and using the relations $E_{j}^{\text{+}}\text{+}E_{j}^{-}=2\cosh \left( {{p}_{j}}kh \right)$ and $E_{j}^{\text{+}}-E_{j}^{-}=-2\sinh \left( {{p}_{j}}kh \right)$, we obtain two
sets of independent linear algebraic equations,
\begin{equation}     \label{65}
{{\mathbf{P}}^{\text{sym}}}\left[ \begin{matrix}
{{A}_{1}}+{{A}_{4}}  \\
{{A}_{2}}+{{A}_{5}}  \\
{{A}_{3}}+{{A}_{6}}  \\
\end{matrix} \right]=0, \qquad {{\mathbf{P}}^{\text{anti}}}\left[ \begin{matrix}
{{A}_{1}}-{{A}_{4}}  \\
{{A}_{2}}-{{A}_{5}}  \\
{{A}_{3}}-{{A}_{6}}  \\
\end{matrix} \right]=0,
\end{equation}
where the $3\times3$ coefficient matrices $\mathbf{P}^{\text{sym}}$ and $\mathbf{P}^{\text{anti}}$ have non-zero components
\begin{align}      \label{66}
& P_{1j}^{\text{sym}}=\left[ \overline{\eta }_{4}^{\left( j \right)}+{{{\overline{B}}}_{2}}\left( {{{\overline{B}}}_{2}}\overline{\eta }_{2}^{\left( j \right)}+2\overline{\eta }_{3}^{\left( j \right)} \right)/2 \right]\tanh \left( {{p}_{j}}kh \right), \notag \\ 
& P_{2j}^{\text{sym}}=\left( \overline{\eta }_{5}^{\left( j \right)}-\overline{B}_{2}^{2}\overline{\eta }_{1}^{\left( j \right)}/2 \right)-\text{i}{{{\overline{B}}}_{2}}\left( {{{\overline{B}}}_{2}}\overline{\eta }_{2}^{\left( j \right)}+\overline{\eta }_{3}^{\left( j \right)} \right)\tanh \left( {{p}_{j}}kh \right), \notag \\ 
& P_{3j}^{\text{sym}}=\left( \overline{\eta }_{6}^{\left( j \right)}-{{{\overline{B}}}_{2}}\overline{\eta }_{1}^{\left( j \right)} \right)-\text{i}\left( {{{\overline{B}}}_{2}}\overline{\eta }_{2}^{\left( j \right)}+\overline{\eta }_{3}^{\left( j \right)} \right)\tanh \left( {{p}_{j}}kh \right),
\end{align}
and
\begin{align}    \label{67}
& P_{1j}^{\text{anti}}=\overline{\eta }_{4}^{\left( j \right)}+{{{\overline{B}}}_{2}}\left( {{{\overline{B}}}_{2}}\overline{\eta }_{2}^{\left( j \right)}+2\overline{\eta }_{3}^{\left( j \right)} \right)/2, \notag \\ 
& P_{2j}^{\text{anti}}=\left( \overline{\eta }_{5}^{\left( j \right)}-\overline{B}_{2}^{2}\overline{\eta }_{1}^{\left( j \right)}/2 \right)\tanh \left( {{p}_{j}}kh \right)-\text{i}{{{\overline{B}}}_{2}}\left( {{{\overline{B}}}_{2}}\overline{\eta }_{2}^{\left( j \right)}+\overline{\eta }_{3}^{\left( j \right)} \right), \notag \\ 
& P_{3j}^{\text{anti}}=\left( \overline{\eta }_{6}^{\left( j \right)}-{{{\overline{B}}}_{2}}\overline{\eta }_{1}^{\left( j \right)} \right)\tanh \left( {{p}_{j}}kh \right)-\text{i}\left( {{{\overline{B}}}_{2}}\overline{\eta }_{2}^{\left( j \right)}+\overline{\eta }_{3}^{\left( j \right)} \right), 
\end{align}
for $j=1,2,3$. For non-trivial solutions of Eq.~\eqref{65}, the determinants of coefficient matrices must vanish, i.e.,
\begin{equation} \label{68}
\text{det}\left({{\mathbf{P}}^{\text{sym}}} \right) =0,\qquad
\text{det}\left( {{\mathbf{P}}^{\text{anti}}}\right)=0,
\end{equation}
which identify, respectively, possible symmetric and antisymmetric wrinkling modes.

Substituting Eqs.~\eqref{tau22}, \eqref{41}, \eqref{45}$_{1,2}$ and \eqref{46}$_{1,2,3}$ into Eqs.~\eqref{66}-\eqref{68} and with the help of Mathematica (Wolfram Research, Inc., 2013), we are able to obtain the \emph{explicit bifurcation equation for antisymmetric modes} as
\begin{multline} \label{69}
\left[ 1+\beta \left( \lambda _{3}^{2}-1 \right) \right]\left[ 2{\overline{F}_5}{{p}_{3}}+\tanh \left( {{p}_{3}}kh \right) \right] \times \\
\left[ {{\left( 1+\lambda _{1}^{4}\lambda _{3}^{2} \right)}^{2}}\tanh \left( kh \right)-4\lambda _{1}^{2}{{\lambda }_{3}}\tanh \left( \lambda _{1}^{2}{{\lambda }_{3}}kh \right) \right] \\
=\lambda _{1}^{2}\lambda _{3}^{2}\left( \lambda _{1}^{4}\lambda _{3}^{2}-1 \right) \overline{B}_{2}^{2}{{\left( 1-2{\overline{F}_5} \right)}^{2}}\tanh \left( {{p}_{3}}kh \right),
\end{multline}
where $kh=\lambda _{1}^{-1}\lambda _{3}^{-1}kH$. The bifurcation equation for symmetric modes is the same as Eq.~\eqref{69} except that tanh is replaced by coth everywhere. Note that Eq.~\eqref{69} is applicable to both the uni-axial and plane-strain loading considered in Sec.~\ref{sec2.2}.


In the \textit{thick-plate} or \textit{short-wave} limit ($kH\to \infty$), the tanh functions in Eq.~\eqref{69} are replaced by 1 and the bifurcation criteria for both symmetric and antisymmetric modes reduce to
\begin{multline}  \label{70}
\left[ 1+\beta \left( \lambda _{3}^{2}-1 \right) \right]\left( 1+2{\overline{F}_5}{{p}_{3}} \right) \left[ {{\left( \lambda _{1}^{2}{{\lambda }_{3}} \right)}^{3}}+{{\left( \lambda _{1}^{2}{{\lambda }_{3}} \right)}^{2}}+3\lambda _{1}^{2}{{\lambda }_{3}}-1 \right] \\
=\lambda _{1}^{2}\lambda _{3}^{2}\left( \lambda _{1}^{2}{{\lambda }_{3}}+1 \right) \overline{B}_{2}^{2}{{\left( 1-2{\overline{F}_5} \right)}^{2}}.
\end{multline}
Note that the bifurcation equation \eqref{70} identifies the surface wrinkling instability for the magneto-elastic half-space, which can also be derived based on the surface impedance method shown in \ref{AppeB}. For ${{\overline{B}}_{2}}=0$, Eq.~\eqref{70} reduces to ${{\left( \lambda _{1}^{2}{{\lambda }_{3}} \right)}^{3}}+{{\left( \lambda _{1}^{2}{{\lambda }_{3}} \right)}^{2}}+3\lambda _{1}^{2}{{\lambda }_{3}}-1=0$, corresponding to the surface instability of a purely elastic half-space \citep{flavin1963surface}.

In the \textit{thin-plate} or \textit{long-wave} limit ($kH\to 0$), we find that the antisymmetric bifurcation equation \eqref{69} can be rearranged as
\begin{equation} \label{71}
2{\overline{F}_5}\left( \lambda _{1}^{2}-\lambda _{1}^{-2}\lambda _{3}^{-2} \right)\left[ 1+\beta \left( \lambda _{3}^{2}-1 \right) \right]=\overline{B}_{2}^{2}{{\left( 1-2{\overline{F}_5} \right)}^{2}}.
\end{equation}
For ${{\overline{B}}_{2}}=0$, Eq.~\eqref{71} recovers the critical buckling condition $\lambda _{1}^{2}{{\lambda }_{3}}=1$ for the purely elastic case, which results in $\lambda_1^{\text{cr}}=1.0$ for both the uni-axial and plane-strain loading. This indicates that in the absence of magnetic field, the infinite-long plate or thin-plate buckles immediately when subject to an extremely small in-plane compression.

\section{Specialization to the neo-Hookean magneto-elastic solid} \label{sec4}

For definiteness, we now specialize the previous
results to the neo-Hookean ($\beta=0$) ideal magneto-elastic model and the neo-Hookean magnetization saturation Langevin model, which are characterized by the following energy functions, respectively,
\begin{equation} \label{72}
\Omega =\frac{G}{2}\left( {{I}_{1}}-3 \right)
+\frac{1-\chi }{2{{\mu }_{0}}}{{I}_{5}},
\end{equation}
and
\begin{equation} \label{73}
\Omega =\frac{G}{2}\left( {{I}_{1}}-3 \right)
+\frac{{{I}_{5}}}{2{{\mu }_{0}}}+\frac{{{\mu }_{0}}{{({{m}^{s}})}^{2}}}{3\chi }\left\{ \ln \left( \frac{3\chi \sqrt{{{I}_{5}}}}{{{\mu }_{0}}{{m}^{s}}} \right)-\ln \left[ \sinh \left( \frac{3\chi \sqrt{{{I}_{5}}}}{{{\mu }_{0}}{{m}^{s}}} \right) \right] \right\},
\end{equation}
where $m^s$ is the saturation magnetization and $\chi$ is the magnetic susceptibility that is associated with the relative magnetic permeability $\mu_r$ by ${{\mu }_{r}}=1/\left( 1-\chi  \right)$.

The neo-Hookean ideal model \eqref{72} has been used to study the stability of anisotropic magnetorheological elastomers in finite deformations \citep{rudykh2013stability}. The compressible counterpart of the neo-Hookean magnetization saturation Langevin model \eqref{73} has been adopted to investigate instability-induced pattern evolutions of the heterogeneous materials and structures \citep{psarra2017two, psarra2019wrinkling, goshkoderia2020instability}.

Note that the magneto-elastic material models \eqref{72} and \eqref{73} take no account of  particle-particle interactions and thus neglect magneto-mechanical coupling in terms of pure material magnetostriction. However, it is emphasized that the neo-Hookean magnetization saturation model \eqref{73} allows for a satisfactory quantitative and very good qualitative agreement with the experimental data presented in previous works \citep{psarra2017two, psarra2019wrinkling, goshkoderia2020instability}, although it is anticipated to be less accurate in the post-bifurcation regime, especially when wrinkles develop substantially due to the large shear strains. As a result, more elaborate magneto-mechanical models such as the ones proposed recently by \citet{mukherjee2020microstructurally} are required to explore the effects of the strain-stiffening and the constituent phase properties (such as particle volume fraction, particle-particle interactions, etc) on the wrinkling instability of SMA structures, but such models are out of the scope of this paper.

Using Eqs.~\eqref{2}, \eqref{8}$_2$ and \eqref{72} or \eqref{73}, we get the governing equations of the magnetization response
\begin{equation} \label{74}
\mathbf{M}=\mu _{0}^{-1}\chi \mathbf{B}
\end{equation}
for the ideal model, and
\begin{align} \label{75}
\mathbf{M}=\left[ \frac{{{m}^{s}}}{\left| \mathbf{B} \right|}\coth \left( \frac{3\chi \left| \mathbf{B} \right|}{{{\mu }_{0}}{{m}^{s}}} \right)-\frac{{{\mu }_{0}}{{({{m}^{s}})}^{2}}}{3\chi {{\left| \mathbf{B} \right|}^{2}}} \right]\mathbf{B}
\end{align}
for the saturation Langevin model. In the limit of small magnetic field ($\mathbf{B} \to \mathbf{0}$), one can verify that the saturation magnetization response \eqref{75} is compatible with the linear magnetization response \eqref{74}.

Now introduce the following dimensionless quantities in terms of the initial shear modulus $G$ and the vacuum magnetic permeability $\mu_0$:
\begin{align} \label{76}
& {{{\overline{I}}}_{5}}=\frac{{{I}_{5}}}{G{{\mu }_{0}}}=\overline{B}_{2}^{2},\qquad {{{\overline{M}}}_{2}}={{M}_{2}}\sqrt{\frac{{{\mu }_{0}}}{G}},\qquad {{{\overline{m}}}^{s}}={{m}^{s}}\sqrt{\frac{{{\mu }_{0}}}{G}}, \notag \\ 
& {{{\overline{s}}}_{1}}=\frac{{{s}_{1}}}{G},\qquad {{{\overline{s}}}_{3}}=\frac{{{s}_{3}}}{G},\qquad {{{\overline{\tau }}}_{11}}=\frac{{{\tau }_{11}}}{G}, \qquad {{{\overline{\tau }}}_{33}} =\frac{{{\tau}_{33}}}{G}.
\end{align}
Inserting Eqs.~\eqref{72} and \eqref{73} into Eq.~\eqref{42} gives
\begin{equation} \label{77}
{\overline{F}_5}=(1-\chi)/2,\qquad \overline{F}_{55}=0
\end{equation}
for the ideal model, and
\begin{align} \label{78}
& {\overline{F}_5}=\frac{1}{2}\left[ \text{1+}\frac{{{({{{\overline{m}}}^{s}})}^{2}}}{3\chi {{{\overline{I}}}_{5}}}-\frac{{{{\overline{m}}}^{s}}}{\sqrt{{{{\overline{I}}}_{5}}}}\coth \left( \frac{3\chi }{{{{\overline{m}}}^{s}}}\sqrt{{{{\overline{I}}}_{5}}} \right) \right], \notag \\ 
& \overline{{F}}_{55}=\frac{1}{2}\left[ -\frac{{{({{{\overline{m}}}^{s}})}^{2}}}{3\chi \overline{I}_{5}^{2}}+\frac{{{{\overline{m}}}^{s}}}{2\overline{I}_{5}^{3/2}}\coth \left( \frac{3\chi \sqrt{{{{\overline{I}}}_{5}}}}{{{{\overline{m}}}^{s}}} \right)+\frac{3\chi } {2{{{\overline{I}}}_{5}}}\sinh^{-2}\left( \frac{3\chi \sqrt{{{{\overline{I}}}_{5}}}}{{{{\overline{m}}}^{s}}} \right) \right]
\end{align}
for the saturation Langevin model.

Thus, the magnetization responses \eqref{74} and \eqref{75} to the applied transverse magnetic field are written in non-dimensional form, as
\begin{equation}  \label{79}
M_{2}^\star=\left( 1-2{\overline{F}_5}\right)B_{2}^\star,
\end{equation}
where $B_{2}^\star \equiv {{{B}_{2}}}/({{{\mu }_{0}}{{m}^{s}}})=\overline{B}_2/\overline{m}^s$ and $M_{2}^\star\equiv{{{M}_{2}}}/{{{m}^{s}}}={{{{\overline{M}}}_{2}}}/{{{{\overline{m}}}^{s}}}$.

Substituting Eqs.~\eqref{72} and \eqref{73} into Eqs.~\eqref{11}$_{1,2}$, \eqref{16}$_{3}$ and \eqref{17}$_{1}$, we  obtain  the nonlinear mechanical response from Eqs.~\eqref{14}$_{1,2}$ and \eqref{19}$_{1,2}$ as
\begin{equation}  \label{80}
\lambda _{1}^{2}-\lambda _{1}^{-2}\lambda _{3}^{-2}+\left( 1-2{\overline{F}_5} \right)\overline{B}_{2}^{2}={{\lambda }_{1}}{{\overline{s}}_{1}},\qquad \lambda _{3}^{2}-\lambda _{1}^{-2}\lambda _{3}^{-2}+\left( 1-2{\overline{F}_5}\right)\overline{B}_{2}^{2}=0
\end{equation}
for uni-axial loading, and 
\begin{equation}  \label{81}
{{\lambda }^{2}}-{{\lambda }^{-2}}+\left( 1-2{\overline{F}_5} \right)\overline{B}_{2}^{2}=\lambda {{\overline{s}}_{1}},\qquad {{\overline{s}}_{3}}=1-{{\lambda }^{-2}}+\left( 1-2{\overline{F}_5} \right)\overline{B}_{2}^{2}
\end{equation}
for plane-strain loading.
Practically, we determine the response by prescribing the pre-stretch $\lambda_1$ or $\lambda$ and solving \eqref{80} for $\overline s_1$ and $\lambda_3$ (uni-axial loading) or solving \eqref{81} for $\overline s_1$ and $\overline s_3$ (plane-strain loading).

For the neo-Hookean saturation Langevin model, the bifurcation or buckling equations \eqref{69}-\eqref{71} are the same except that $\beta=0$ and ${\overline{F}_5}$ and ${\overline{F}_{55}}$ are given in Eq.~\eqref{78}.

For the neo-Hookean ideal model, we find from Eqs.~\eqref{45}$_3$ and \eqref{77} that ${{p}_{3}}=1$. As a result, the bifurcation equation \eqref{69} reduces to
\begin{multline}  \label{82}
\left[ 1-\chi +\tanh \left( kh \right) \right]\left[ {{\left( 1+\lambda _{1}^{4}\lambda _{3}^{2} \right)}^{2}}\tanh \left( kh \right)-4\lambda _{1}^{2}{{\lambda }_{3}}\tanh \left( \lambda _{1}^{2}{{\lambda }_{3}}kh \right) \right] \\
=\overline{B}_{2}^{2}{{\chi }^{2}}\lambda _{1}^{2}\lambda _{3}^{2}\left( \lambda _{1}^{4}\lambda _{3}^{2}-1 \right)\tanh \left( kh \right).
\end{multline}
Equations \eqref{70} and \eqref{71} become
\begin{equation}   \label{83}
\left( 2-\chi  \right)\left[ {{\left( \lambda _{1}^{2}{{\lambda }_{3}} \right)}^{3}}+{{\left( \lambda _{1}^{2}{{\lambda }_{3}} \right)}^{2}}+3\lambda _{1}^{2}{{\lambda }_{3}}-1 \right]=\overline{B}_{2}^{2}{{\chi }^{2}}\lambda _{1}^{2}\lambda _{3}^{2}\left( \lambda _{1}^{2}{{\lambda }_{3}}+1 \right)
\end{equation}
for the thick-plate limit ($kH\to \infty$), and
\begin{equation}  \label{84}
\left( 1-\chi  \right)\left( \lambda _{1}^{2}-\lambda _{1}^{-2}\lambda _{3}^{-2} \right) =\overline{B}_{2}^{2}{{\chi }^{2}}
\end{equation}
for the thin-plate limit ($kH\to 0$).


\section{Results and discussion}\label{sec5}


We first conduct numerical calculations in Sec.~\ref{Sec5.1} to  investigate quantitatively the nonlinear static response of incompressible SMA plates subject to  mechanical and magnetic loads. Section~\ref{Sec5.2} then focuses on the bifurcation analysis to calculate the critical mechanical and/or magnetic field generating the wrinkling instability.

We consider two different loading modes (plane-strain and uni-axial loading) and two neo-Hookean magneto-elastic models (\eqref{72} and \eqref{73}). The material properties used in the numerical computations are taken as $G=10$~kPa, $\chi=0.4$ and $\mu_0 m^s=0.5$~T, which are obtained from experiments with a class of magnetorheological elastomers \citep{psarra2017two, psarra2019wrinkling} consisting of a soft silicone mixed with iron particles at a volume fraction of 20\%. More details about the fabrication technique can be found in the paper by \citet{psarra2017two}.


\subsection{Nonlinear static response} \label{Sec5.1}


The nonlinear static response of an SMA plate subject to mechanical and magnetic loads is calculated from Eqs.~\eqref{79}-\eqref{81}.

\begin{figure}[htbp]
	\centering
	\setlength{\abovecaptionskip}{5pt}
	\includegraphics[width=1.0\textwidth]{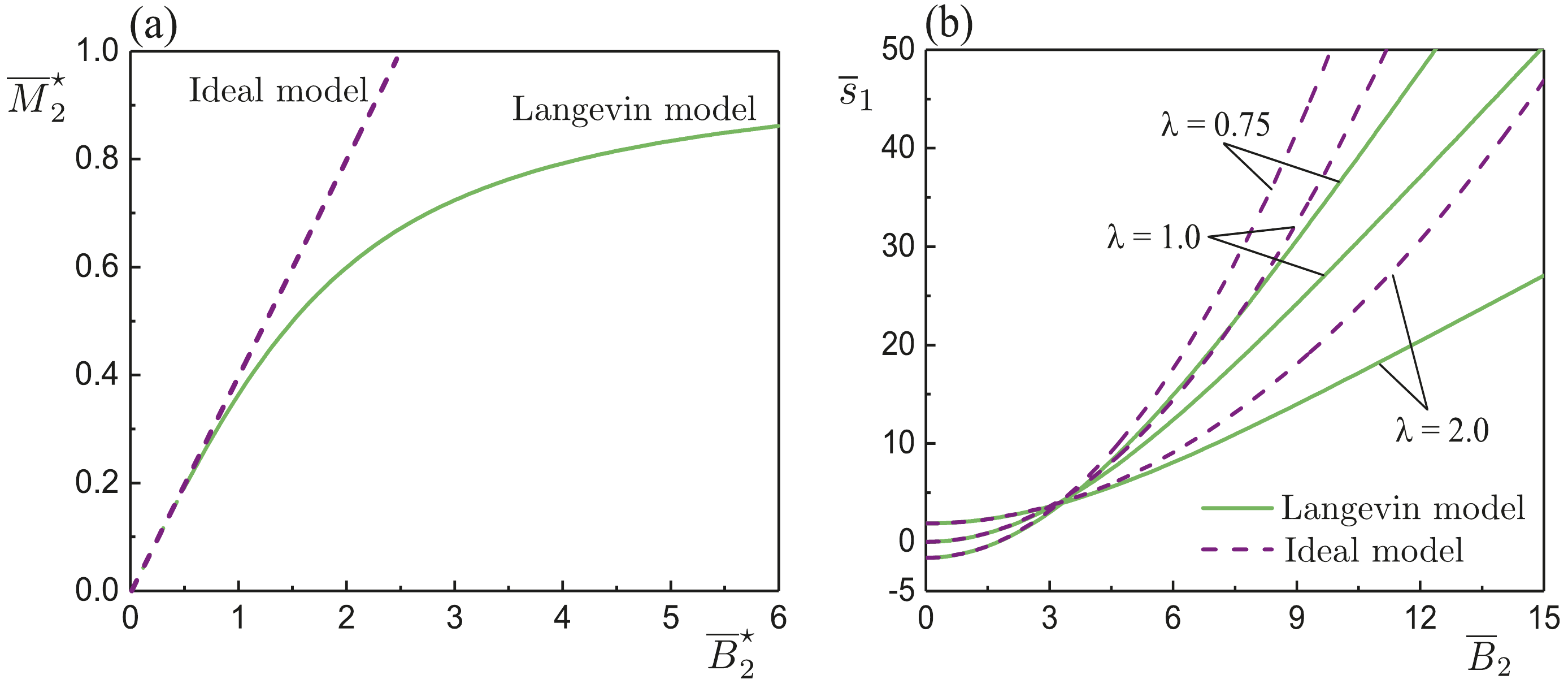}
	\caption{Plane-strain loading: (a) magnetization response of the dimensionless magnetization $M_{2}^\star={{{{\overline{M}}}_{2}}}/{{{{\overline{m}}}^{s}}}$ versus the dimensionless magnetic induction field $B_{2}^\star=\overline{B}_2/\overline{m}^s$ from Eq.~\eqref{79}; (b) response of the dimensionless nominal mechanical traction $\overline{s}_{1}$ versus the dimensionless magnetic induction field $\overline{B}_2$ from Eq.~\eqref{81}$_1$ for three different values of pre-stretch $\lambda=0.75, 1.0, 2.0$. Solid lines correspond to the ideal magneto-elastic model while dashed lines represent the saturation Langevin model.}
	\label{Fig2}
\end{figure}

For plane-strain loading, Fig.~\ref{Fig2}(a) shows the magnetization response of the dimensionless magnetization $M_{2}^\star={{{{\overline{M}}}_{2}}}/{{{{\overline{m}}}^{s}}}$ versus the dimensionless magnetic induction field $B_{2}^\star=\overline{B}_2/\overline{m}^s$ (see Eq.~\eqref{79}) for the two material models \eqref{72} and \eqref{73}. We see from Fig.~\ref{Fig2}(a) that a neo-Hookean ideal magneto-elastic plate exhibits a linear magnetization response, whereas the response of a plate with magnetization saturation is nonlinear. The magnetization responses of the two material models are essentially identical for $\overline{B}_2^\star \le 1.0$ (i.e., $\overline{B}_2 \le 4.46$). The magnetization begins to saturate at $\overline{B}_2^\star \simeq 6.0$ (i.e., $\overline{B}_2 \simeq 26.76$). 
We note from Eqs.~\eqref{78}$_1$ and \eqref{79} that the magnetization response of each material model is independent of the mechanical stretch ratio.

For plane-strain loading, the effect of the magnetic induction field $\overline{B}_2$ on the nominal mechanical traction $\overline{s}_{1}$ applied to the SMA plate is plotted in Fig.~\ref{Fig2}(b) for three different values of pre-stretch $\lambda=0.75, 1.0, 2.0$. Clearly, when $\overline{B}_2$ increases, the required mechanical traction increases monotonically, which indicates that the SMA plate has an in-plane contraction trend because of the increasing external Maxwell stress. For the three pre-stretches $\lambda=0.75, 1.0, 2.0$, the mechanical tractions corresponding to the two material models overlap up to $\overline{B}_2 \simeq 3.8, 4.2, 5.0$, respectively. However, the difference predicted by the two material models becomes more evident with subsequent increases in $\overline{B}_2$. Specifically, at the same level of $\overline{B}_2$, the plate with saturation magnetization effect requires a smaller mechanical traction as compared to the ideal magneto-elastic plate. This is because the presence of saturation magnetization will induce a smaller in-plane contraction trend.

\begin{figure}[htbp]
	\centering
	\setlength{\abovecaptionskip}{5pt}
	\includegraphics[width=0.48\textwidth]{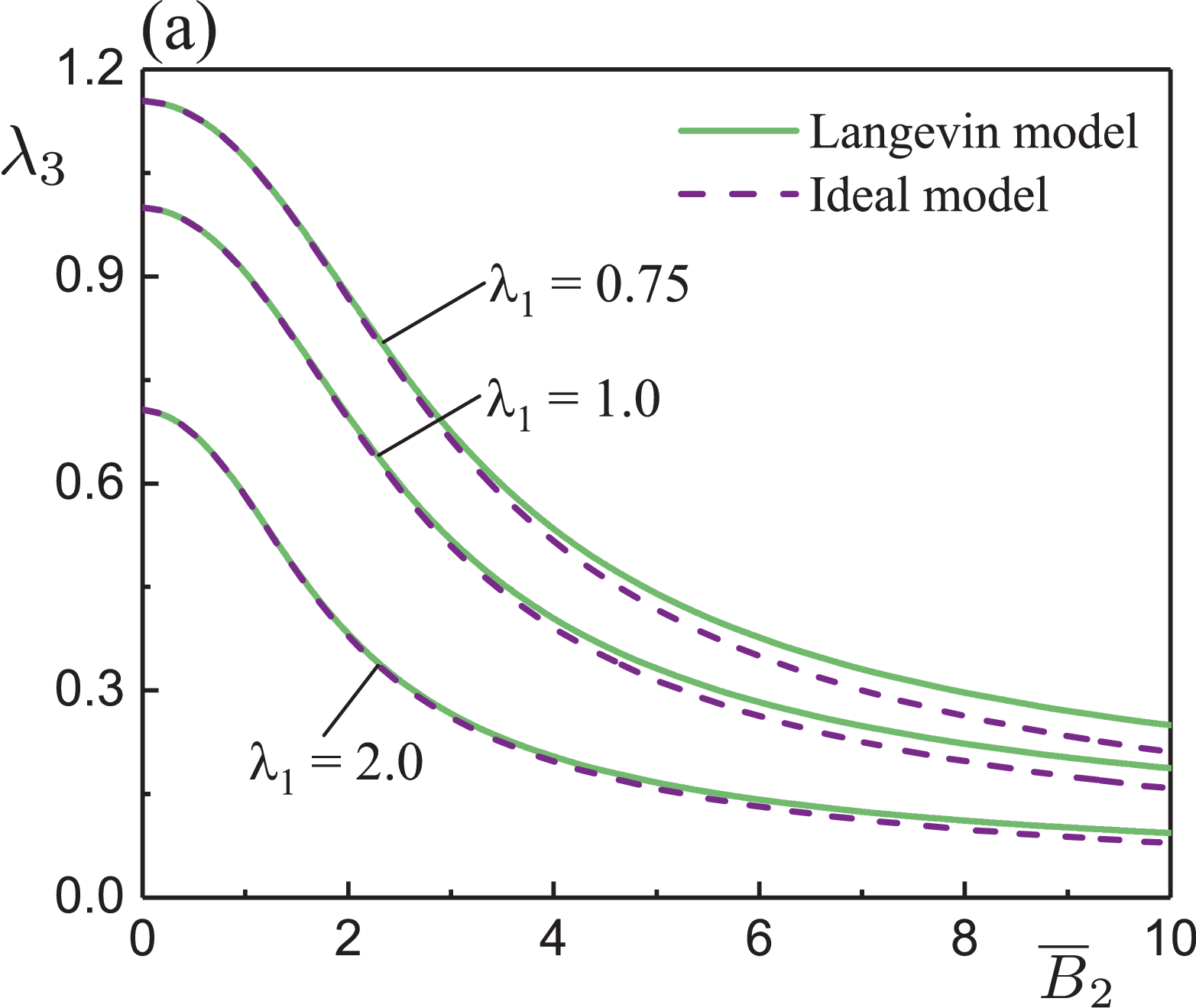}
	\hspace{0.005\textwidth}
	\includegraphics[width=0.48\textwidth]{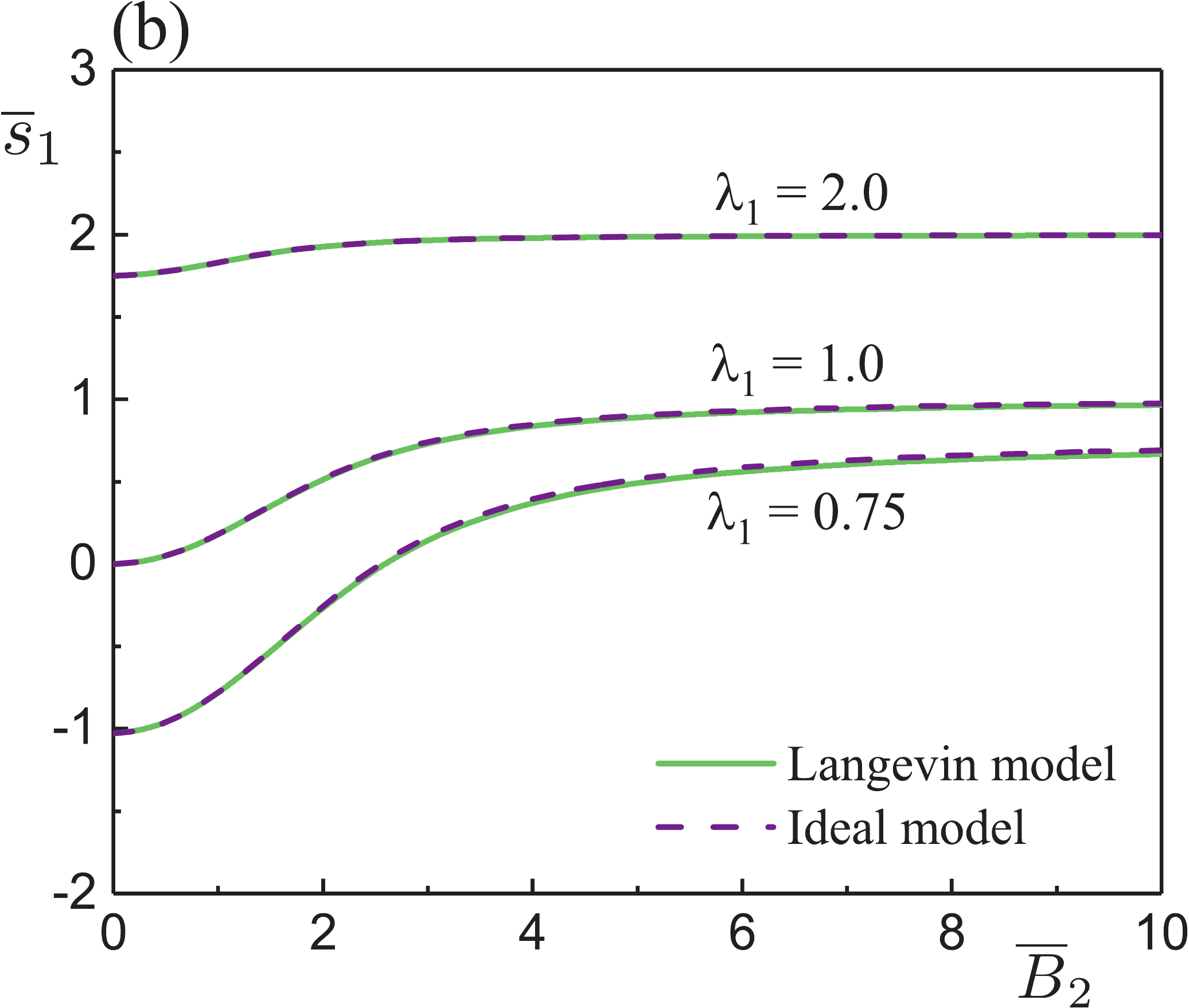}
	\caption{Uni-axial loading for three different values of pre-stretch $\lambda_1=0.75, 1.0, 2.0$: (a) response of the stretch ratio $\lambda_3$ versus the dimensionless magnetic induction field $\overline{B}_2$ from Eq.~\eqref{80}$_2$; (b) response of the dimensionless nominal mechanical traction $\overline{s}_{1}$ versus $\overline{B}_2$ from Eq.~\eqref{80}$_1$. Solid lines correspond to the ideal magneto-elastic model while dashed lines represent the saturation Langevin model.}
	\label{Fig3}
\end{figure}

For uni-axial loading, the magnetization responses of the two material models are essentially the same as those for plane-strain loading, as shown in Fig.~\ref{Fig2}(a). For three different values of pre-stretch $\lambda_1=0.75, 1.0, 2.0$, Fig.~\ref{Fig3}(a) and \ref{Fig3}(b) display the effect of the magnetic induction field $\overline{B}_2$ on the stretch ratio $\lambda_3$ and the nominal mechanical traction $\overline{s}_{1}$, respectively, for the two material models.
Again, we point out that the induced mechanical and magnetic field distributions are assumed to be uniform when solving the nonlinear static response, as explained previously, and that the material models \eqref{72} and \eqref{73} neglect pure material magnetostriction.
We observe from Fig.~\ref{Fig3}(a) that the stretch ratio $\lambda_3$ decreases notably with increasing $\overline{B}_2$ \emph{only} due to the magnetic traction induced by the external Maxwell stress. The stretch $\lambda_3$ for the ideal model is slightly lower than that predicted by the saturation Langevin model, because the saturation magnetization generates a smaller stretch $\lambda_2$ in the thickness direction. It is clear from Fig.~\ref{Fig3}(b) that the mechanical traction $\overline{s}_{1}$ increases gradually with $\overline{B}_2$. After $\overline{B}_2$ reaches a certain value, $\overline{s}_{1}$ remains unchanged. This is because the in-plane elongation trend due to the compression in the $x_3$ direction counteracts the in-plane contraction tendency due to the external Maxwell stress in the $x_1$ direction. Furthermore, in the whole $\overline{B}_2$ range of interest, the mechanical tractions based on the two material models are almost identical,  because the saturation magnetization does not alter the contraction trend in the $x_1$ direction and  just affects the compression amount in the $x_3$ direction.


\subsection{Bifurcation analysis} \label{Sec5.2}


We now examine the critical values of stretch in the $x_1$ direction and of transverse magnetic induction field for which antisymmetric (see Fig.~\ref{Fig1}(d)) and symmetric (see Fig.~\ref{Fig1}(e)) modes of wrinkling instability appear. 

For \textit{plane-strain} loading, antisymmetric modes are identified by bifurcation equations \eqref{69} and \eqref{82} with $\lambda_1=\lambda, \lambda_3=1$ for the two material models. The critical buckling fields of antisymmetric modes for \textit{uni-axial} loading are calculated by solving bifurcation equations \eqref{69} and \eqref{82} together with nonlinear static response \eqref{80}$_2$. The corresponding critical fields of symmetric modes are calculated by using coth to replace tanh in Eqs.~\eqref{69} and \eqref{82}. The wrinkling criteria for thin- and thick-plate limits are obtained by evaluating Eqs.~\eqref{70} and \eqref{71} or Eqs.~\eqref{83} and \eqref{84} for the two material models.

\begin{figure}[htbp]
	\centering
	\setlength{\abovecaptionskip}{5pt}
	\includegraphics[width=\textwidth]{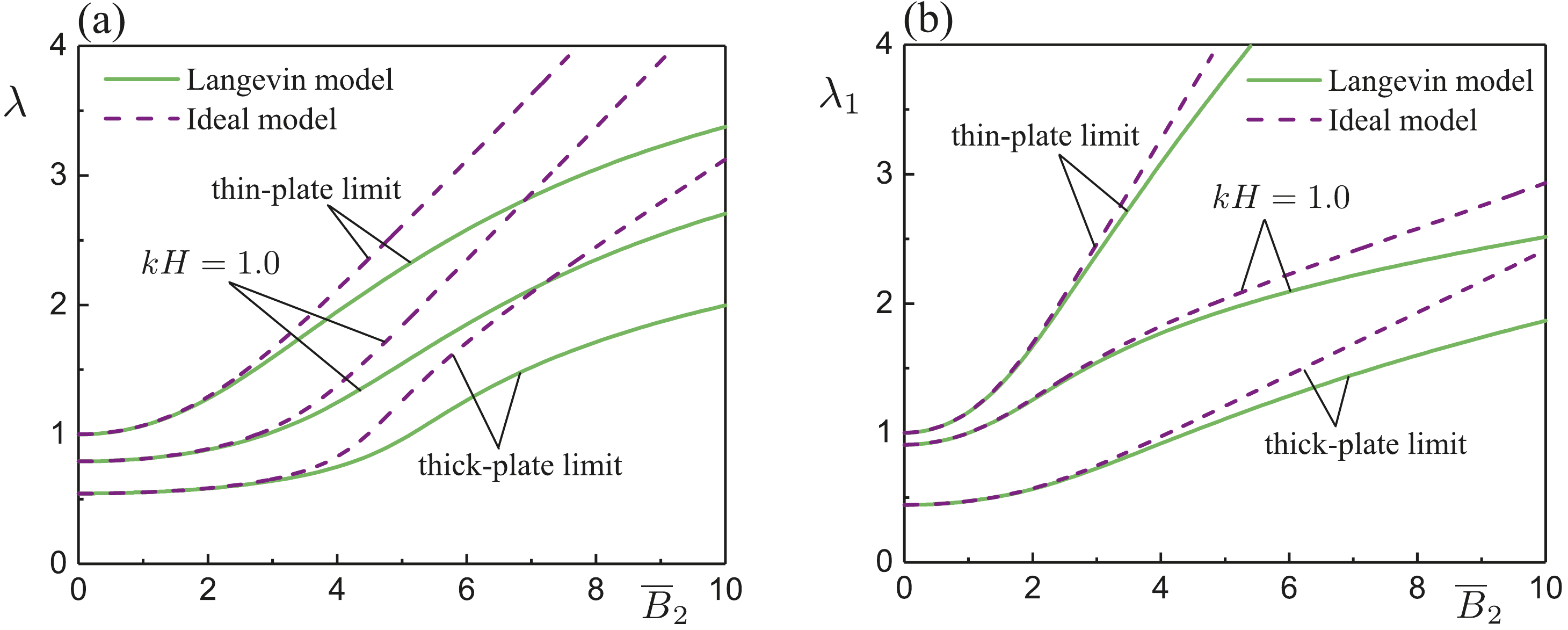}
	\caption{Bifurcation curves (or critical combinations of stretch ratio and dimensionless magnetic induction field) of antisymmetric wrinkling modes for the thin- and thick-plate limits and $kH =1.0$: (a) plane-strain loading ($\lambda$ versus $\overline{B}_2$); (b) uni-axial loading ($\lambda_1$ versus $\overline{B}_2$). Solid lines correspond to the ideal magneto-elastic model while dashed lines represent the saturation Langevin model.}
	\label{Fig4}
\end{figure}

For plane-strain loading and uni-axial loading, Fig.~\ref{Fig4}(a) and \ref{Fig4}(b) illustrate the critical combinations of stretch ratio $\lambda$ or $\lambda_1$ and dimensionless magnetic induction field $\overline{B}_2$ of antisymmetric wrinkling modes for the two material models. Specifically, Fig.~\ref{Fig4} shows the results corresponding to the thin- and thick-plate limits and a representative value of $kH=1$. Since it is found that antisymmetric modes always occur first, Fig.~\ref{Fig4} does not display the results for symmetric modes, which are addressed below.

We first focus on the results for the ideal magneto-elastic model in Fig.~\ref{Fig4}. In the absence of magnetic field ($\overline{B}_2=0$, purely elastic case), a thin plate with $kH \to 0$ is unstable for $\lambda<1$ ($\lambda_1<1$) and is stable for any $\lambda>1$ ($\lambda_1>1$) for plane-strain (uni-axial) loading. For both loading modes, a larger value of the parameter $kH$ requires an increasing compression to induce instability. 
In the thick-plate limit $kH\to \infty$, we recover the well-known critical compression stretches for surface instability in the purely elastic case, namely $\lambda^{\text{cr}}=0.544$ and $\lambda_1^{\text{cr}}=0.444$ for plane-strain and uni-axial loading, respectively \citep{beatty1998stability}. 
For a fixed non-zero $\overline{B}_2$, the variation trends of the critical stretch $\lambda^{\text{cr}}$ or $\lambda_1^{\text{cr}}$ with increasing $kH$ are qualitatively the same as that for $\overline{B}_2=0$. Besides, the critical magnetic field $\overline{B}_2^{\text{cr}}$, for a given stretch $\lambda$ or $\lambda_1$, increases monotonically with $kH$, resulting in enhanced stability. 

Furthermore, for a given $kH$, Fig.~\ref{Fig4}(a) shows that the critical stretch $\lambda^{\text{cr}}$ exhibits a monotonous increase when $\overline{B}_2$ goes up, which means that the plate become more and more unstable and the magnetic field has a destabilizing effect. In particular,  thin plates with $kH \to 0$ are unstable in tension ($\lambda^{\text{cr}}>1$) for non-zero $\overline{B}_2$, while the plate with $kH=1$ has a wrinkling instability in tension for $\overline{B}_2 \gtrsim 2.8$. Similar phenomena are observed in Fig.~\ref{Fig4}(b) for uni-axial loading.

We now evaluate the effect of saturation magnetization on the stability. Fig.~\ref{Fig4}(a) shows that for plane-strain loading, the critical stretch $\lambda^{\text{cr}}$ predicted by the saturation Langevin model coincides with that based on the ideal model for small and moderate values of $\overline{B}_2$, the range of which depends on $kH$. For example, the results based on the two material models are almost the same when $\overline{B}_2\lesssim 3.0, 3.5, 4.0$ for $kH=0,1,\infty$, respectively. However, the presence of saturation magnetization reduces remarkably the critical stretch $\lambda^{\text{cr}}$ for a large value of $\overline{B}_2$. These phenomena are also found in Fig.~\ref{Fig4}(b) for uni-axial loading. Nevertheless, compared with plane-strain loading, the effect of saturation magnetization on the critical fields is weaker for uni-axial loading because there is no constraint in the $x_3$ direction.


\subsubsection{Critical stretch for a prescribed magnetic load} \label{Sec5.2.1}

\begin{figure}[h!]
	\centering	
	\setlength{\abovecaptionskip}{5pt}
	\includegraphics[width=0.48\textwidth]{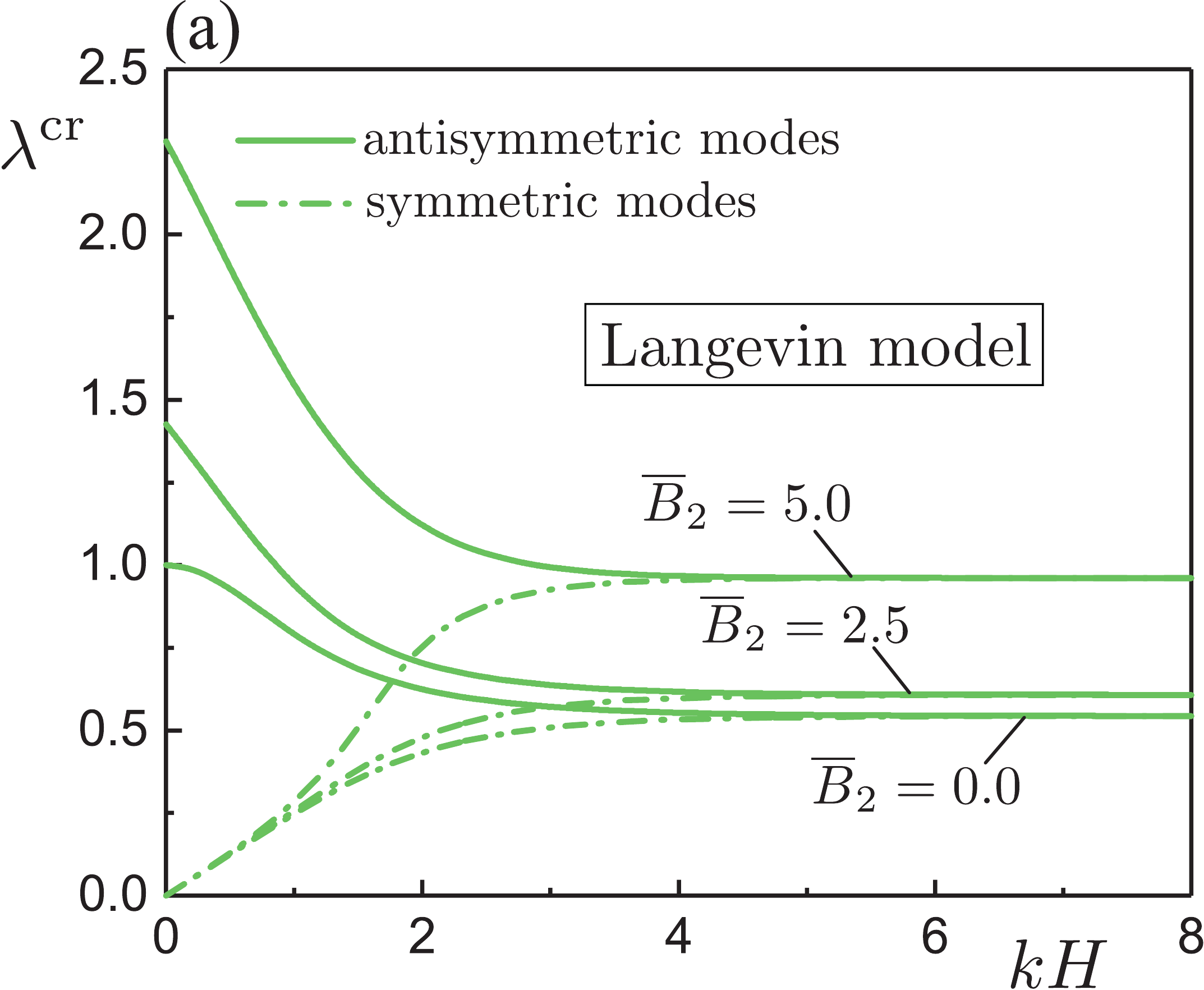}
	\hspace{0.005\textwidth}
	\includegraphics[width=0.48\textwidth]{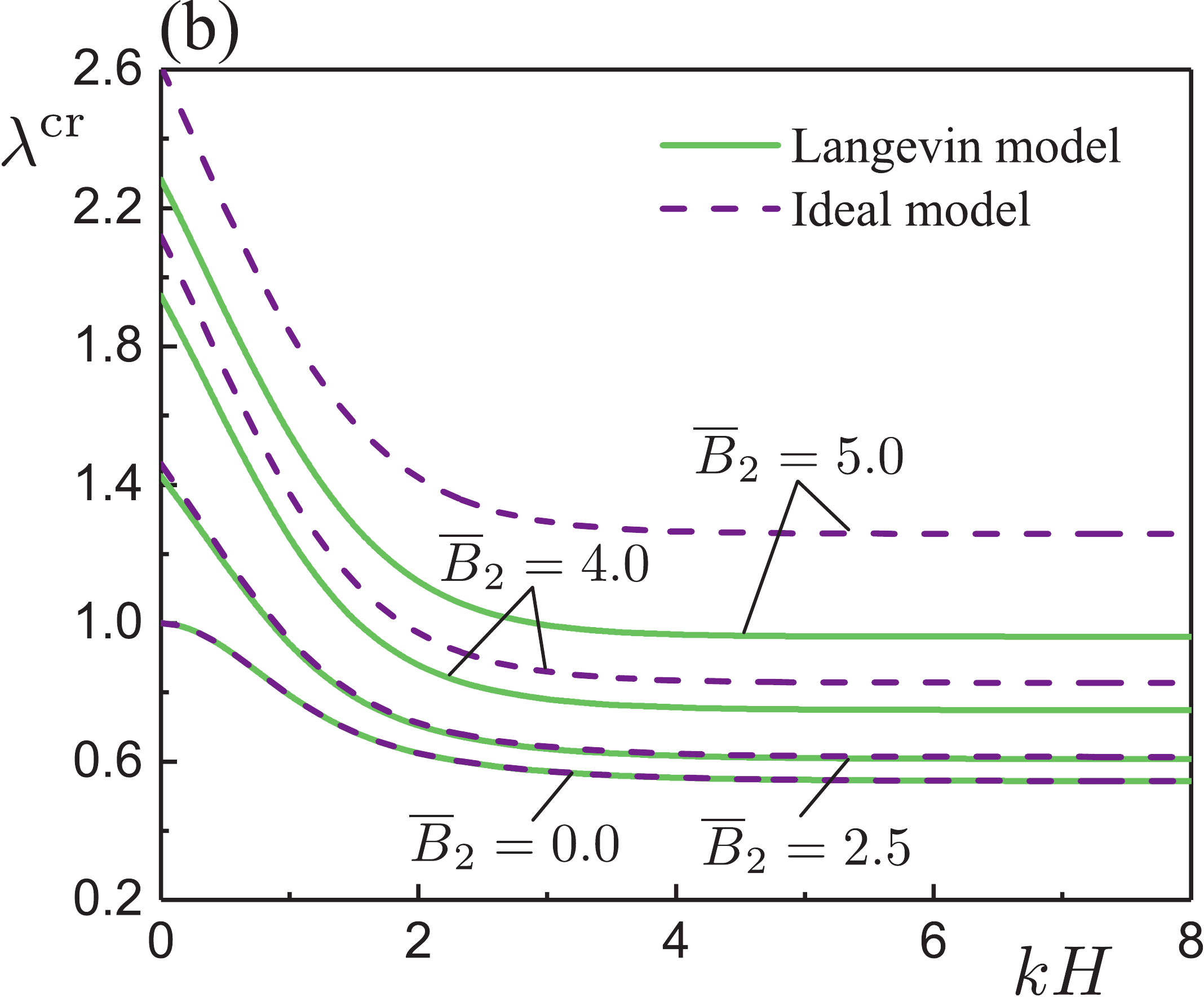}
	\caption{Plane-strain loading: (a) critical stretch $\lambda^{\text{cr}}$ as a function of $kH$ for antisymmetric (solid lines) and symmetric (dashed-dotted lines) modes of the neo-Hookean saturation Langevin plates subject to three prescribed values of $\overline{B}_2=0.0, 2.5, 5.0$; (b) $\lambda^{\text{cr}}$ as a function of $kH$ for antisymmetric modes of the neo-Hookean ideal (dashed lines) and saturation Langevin (solid lines) magneto-elastic plates subject to four fixed values of $\overline{B}_2=0.0, 2.5, 4.0, 5.0$.}
	\label{Fig5}
\end{figure}
\begin{figure}[htbp]
	\centering	
	\setlength{\abovecaptionskip}{5pt}
	\includegraphics[width=\textwidth]{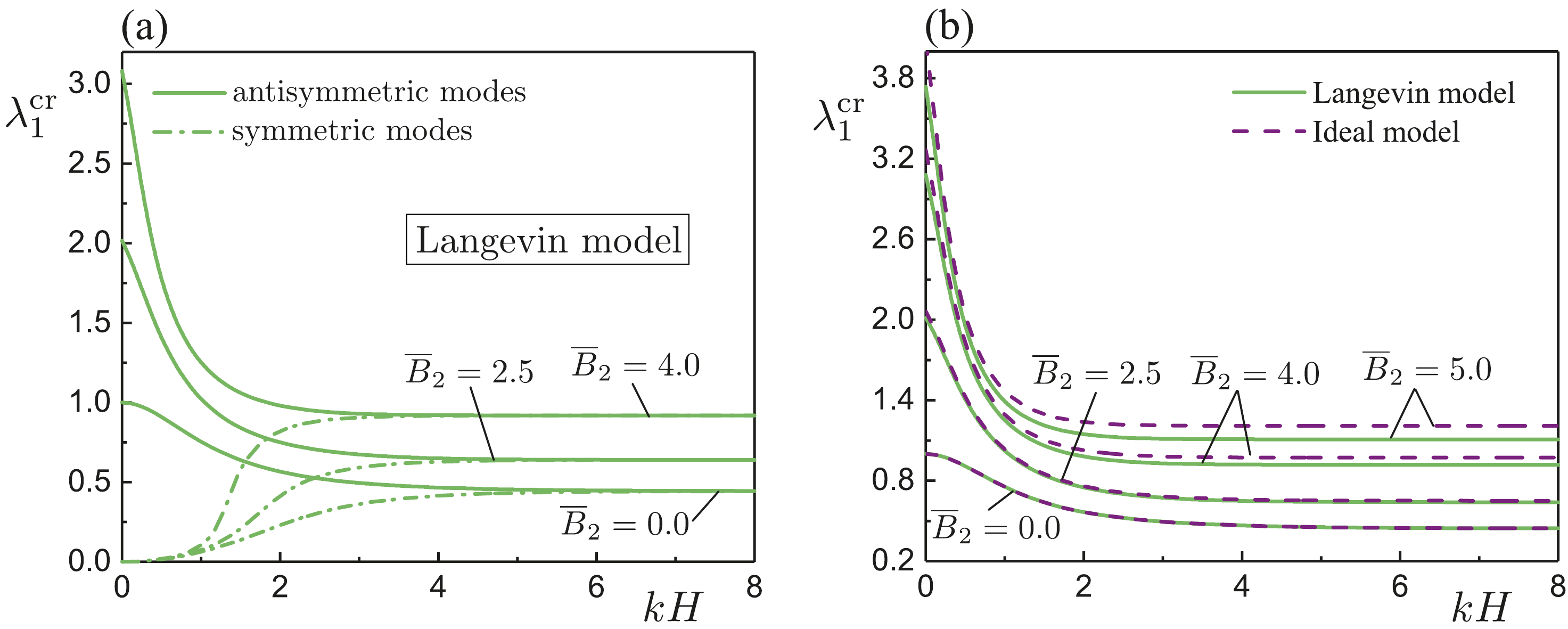}
	\caption{Uni-axial loading: (a) critical stretch $\lambda_1^{\text{cr}}$ as a function of $kH$ for anti-symmetric (solid lines) and symmetric (dashed-dotted lines) modes of the neo-Hookean saturation Langevin plates subject to three prescribed values of $\overline{B}_2=0.0, 2.5, 4.0$; (b) $\lambda_1^{\text{cr}}$ as a function of $kH$ for anti-symmetric modes of the neo-Hookean ideal (dashed lines) and saturation Langevin (solid lines) magneto-elastic plates subject to four fixed values of $\overline{B}_2=0.0, 2.5, 4.0, 5.0$.}
	\label{Fig6}
\end{figure}

For a prescribed magnetic induction field, we first determine the critical stretch of the underlying deformed configuration for which antisymmetric and symmetric wrinkling modes are induced. Specifically, for plane-strain (uni-axial) loading, Fig.~\ref{Fig5}(a) (Fig.~\ref{Fig6}(a)) displays the variation of the critical stretch $\lambda^{\text{cr}}$ ($\lambda_1^{\text{cr}}$) with $kH$ for the neo-Hookean magnetization saturation SMA plates subject to three representative values of $\overline{B}_2=0, 2.5, 5.0$ ($\overline{B}_2=0, 2.5, 4.0$), wherein the antisymmetric and symmetric solutions are represented by the solid and dashed-dotted lines, respectively.

We find from Figs.~\ref{Fig5}(a) and \ref{Fig6}(a) that antisymmetric modes always occur before symmetric modes become possible for any value of $kH$. Therefore, to realize a symmetric buckling mode we must in principle suppress the appearance of the antisymmetric mode. 

For a fixed $\overline{B}_2$, the critical stretch $\lambda^{\text{cr}}$ (or $\lambda_1^{\text{cr}}$) required to initiate the antisymmetric instability decreases monotonically with increasing $kH$, asymptotically approaching the surface instability of the thick-plate limit when $kH \to \infty$. However, the symmetric bifurcation curves exhibit an opposite trend. Thus, the stable range of combinations of $\lambda$ (or $\lambda_1$) and $kH$ is determined by the region above the solid line for a fixed $\overline{B}_2$. Moreover, the critical stretch $\lambda^{\text{cr}}$ (or $\lambda_1^{\text{cr}}$) for any value of $kH$ is shifted upwards when raising $\overline{B}_2$, thus indicating that the SMA plate is destabilized by the application of an increasing magnetic field. In particular, for plane-strain loading with $\overline{B}_2=0, 2.5, 5.0$, the critical stretches $\lambda^{\text{cr}}$ of a plate with $kH \to 0$ are 1.000, 1.426, 2.282, while those with $kH \to \infty$ are 0.544, 0.608, 0.961, respectively. For uni-axial loading with $\overline{B}_2=0, 2.5, 4.0$, the critical stretches $\lambda_1^{\text{cr}}$ of the thin-plate limit are equal to 1.000, 2.016, 3.080, while those of the thick-plate limit are 0.444, 0.639, 0.918, respectively. The effect of small magnetic field on the stability  is more significant for uni-axial loading than for plane-strain loading.

For antisymmetric modes, Figs.~\ref{Fig5}(b) and \ref{Fig6}(b) illustrate how the saturation magnetization affects the bifurcation curves ($\lambda^{\text{cr}}$ versus $kH$ and $\lambda_1^{\text{cr}}$ versus $kH$) for plane-strain and uni-axial loading, respectively. We see that the bifurcation curves based on the two material models overlap in the entire range of $kH$ for $\overline{B}_2 \le 2.5$. This means that the critical stretch is hardly affected by the saturation magnetization for small to moderate values of the magnetic field. 
However, the saturation magnetization plays  an important role in determining the critical stretch $\lambda^{\text{cr}}$ or $\lambda_1^{\text{cr}}$ for large values of $\overline{B}_2$: the figures show that it then stabilizes the SMA plate, and that its effect is much stronger for plane-strain loading than for uni-axial loading.


\subsubsection{Critical magnetic induction field for a fixed pre-stretch} \label{Sec5.2.2}

\begin{figure}[h!]
	\centering
	\setlength{\abovecaptionskip}{5pt}	
	\includegraphics[width=0.48\textwidth]{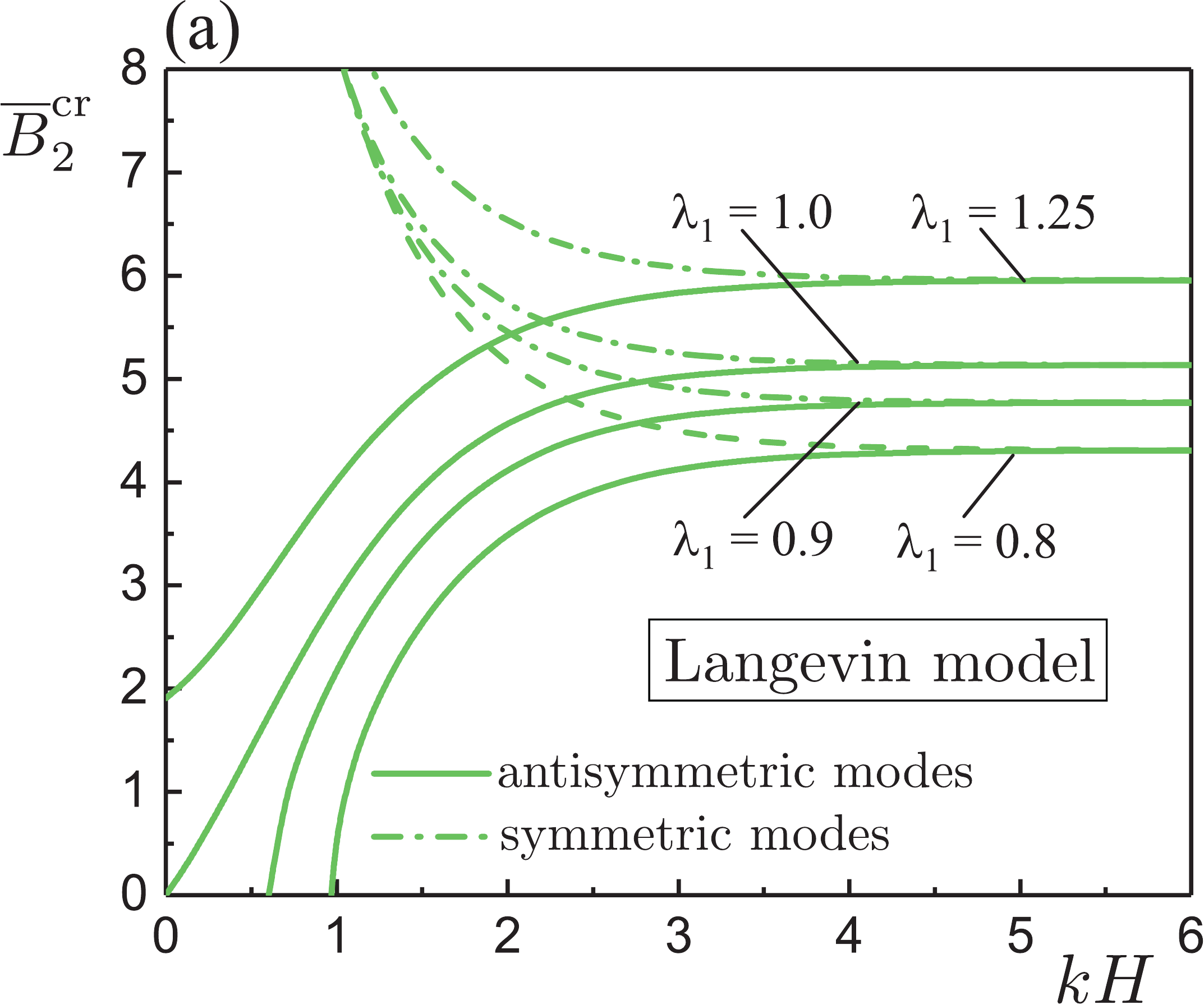}
	\hspace{0.005\textwidth}
	\includegraphics[width=0.48\textwidth]{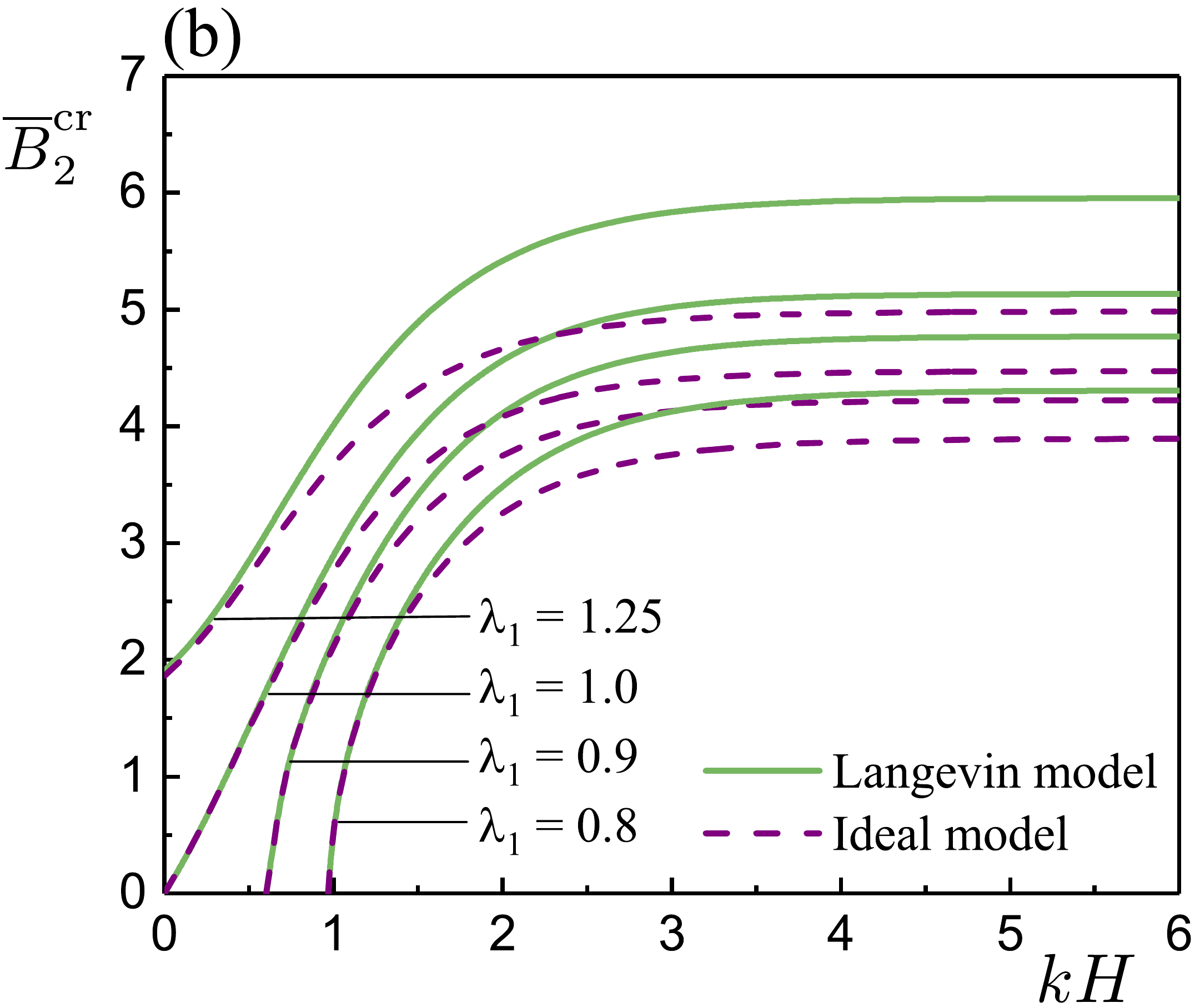}
	\caption{Plane-strain loading for four fixed values of pre-stretch $\lambda=0.8, 0.9, 1.0, 1.25$: (a) critical magnetic induction field $\overline{B}_2^{\text{cr}}$ as a function of $kH$ for antisymmetric (solid lines) and symmetric (dashed-dotted lines) modes of the neo-Hookean saturation Langevin plates; (b) $\overline{B}_2^{\text{cr}}$ as a function of $kH$ for antisymmetric modes of the neo-Hookean ideal (dashed lines) and saturation Langevin (solid lines) magneto-elastic plates.}
	\label{Fig7}
\end{figure}
\begin{figure}[h!]
	\centering
	\setlength{\abovecaptionskip}{5pt}	
	\includegraphics[width=0.48\textwidth]{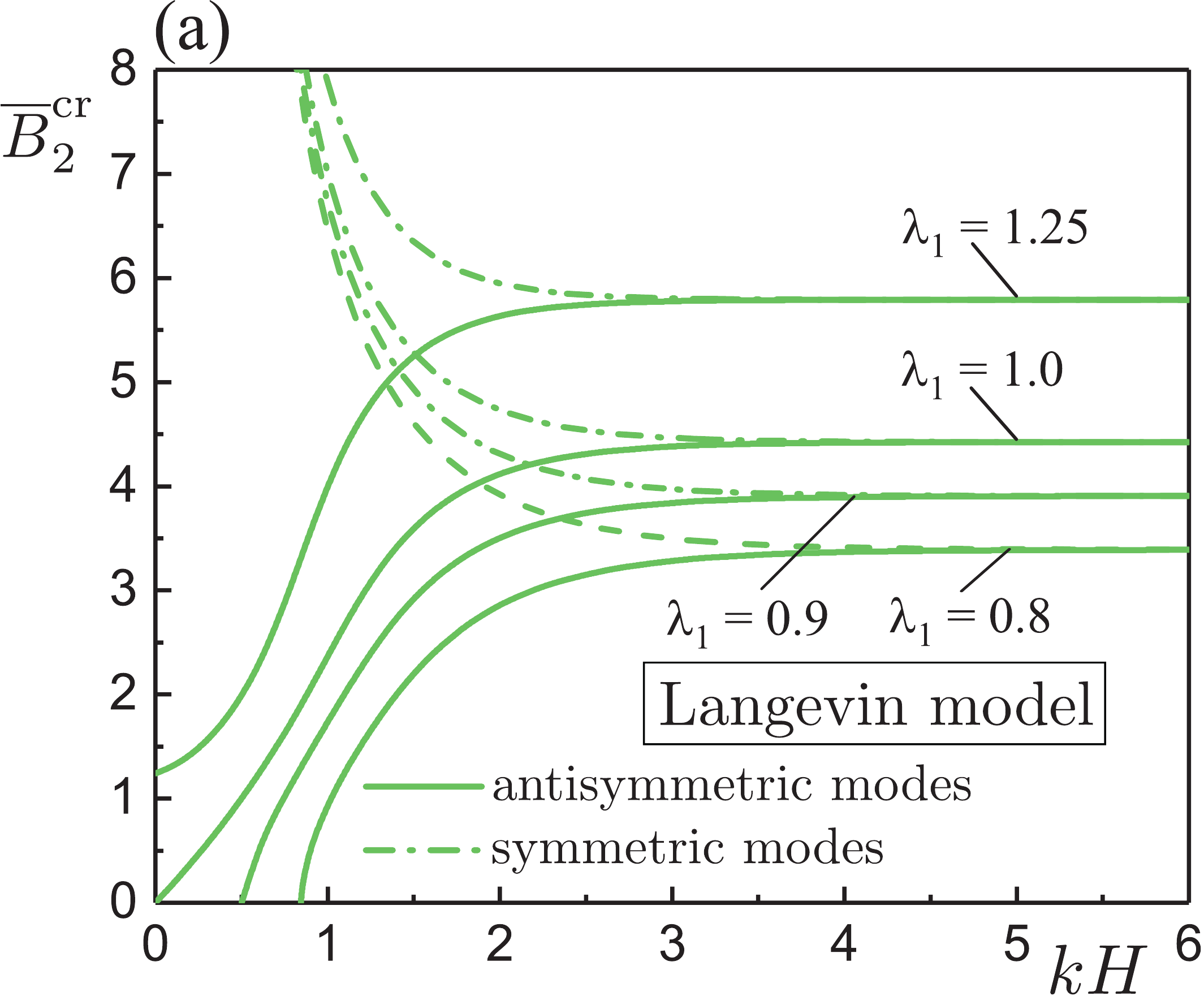}
	\hspace{0.005\textwidth}
	\includegraphics[width=0.48\textwidth]{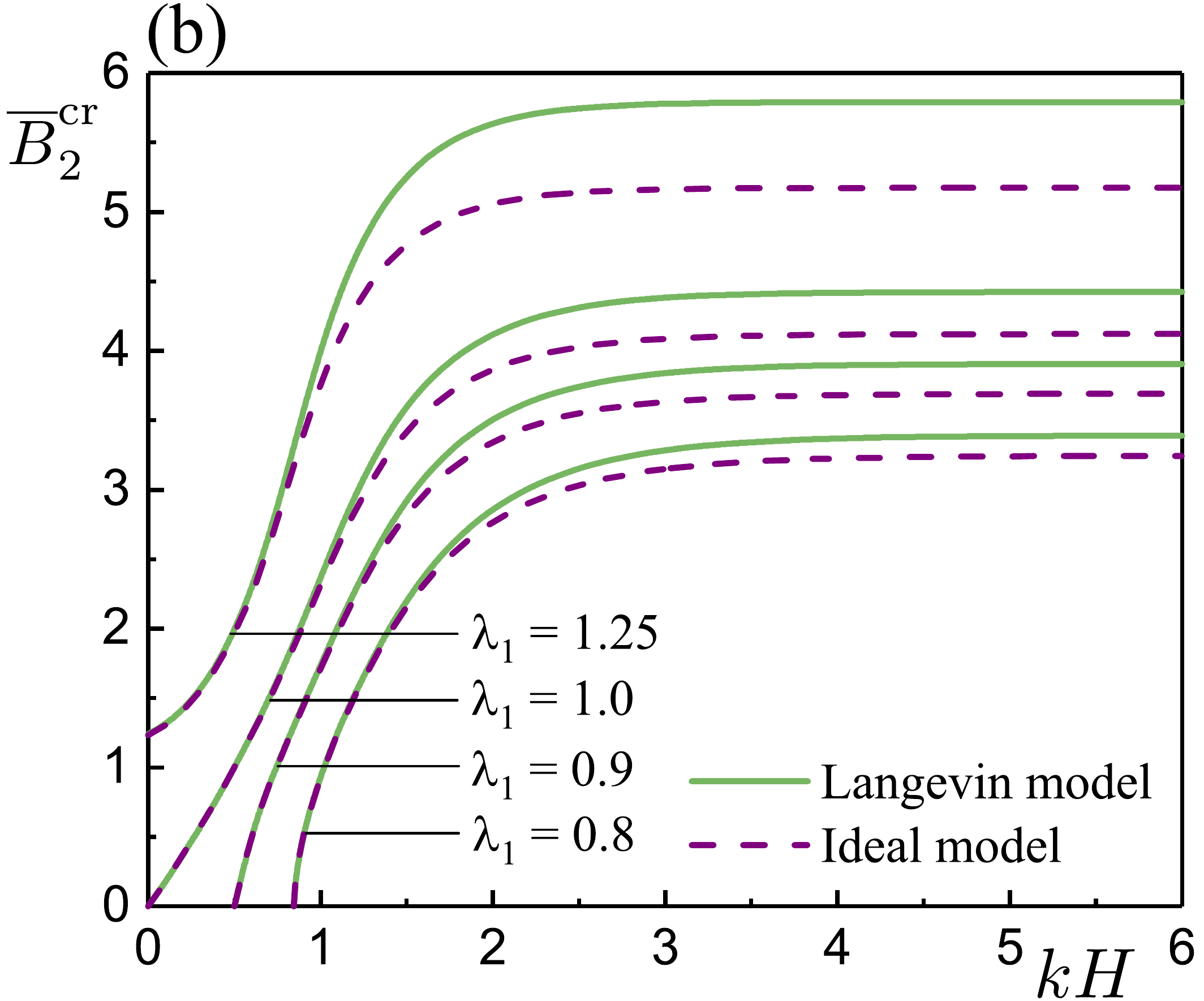}
	\caption{Uni-axial loading for four fixed values of pre-stretch $\lambda_1=0.8, 0.9, 1.0, 1.25$: (a) critical magnetic induction field $\overline{B}_2^{\text{cr}}$ as a function of $kH$ for antisymmetric (solid lines) and symmetric (dashed-dotted lines) modes of the neo-Hookean saturation Langevin plates; (b) $\overline{B}_2^{\text{cr}}$ as a function of $kH$ for antisymmetric modes of the neo-Hookean ideal (dashed lines) and saturation Langevin (solid lines) magneto-elastic plates.}
	\label{Fig8}
\end{figure}

We now evaluate the effect of pre-stretch on the bifurcation curves ($\overline{B}_2^\text{cr}$ versus $kH$) of antisymmetric and symmetric wrinkling modes. For plane-strain (uni-axial) loading, Fig.~\ref{Fig7}(a) (Fig.~\ref{Fig8}(a)) shows the critical magnetic induction field $\overline{B}_2^{\text{cr}}$ as a function of $kH$ for the neo-Hookean magnetization saturation SMA plates for four different levels of pre-stretch $\lambda$ or $\lambda_1=0.8, 0.9, 1.0, 1.25$. Note that the solid and dashed-dotted lines denote the antisymmetric and symmetric solutions, respectively.

We observe from Figs.~\ref{Fig7}(a) and \ref{Fig8}(a) that antisymmetric wrinkling modes are always triggered before the symmetric modes in the entire range of $kH$, which is analogous to what Figs.~\ref{Fig5}(a) and \ref{Fig6}(a) show. As $kH$ increases, the critical magnetic field $\overline{B}_2^{\text{cr}}$ for a given pre-stretch increases gradually for antisymmetric modes and decreases monotonically for symmetric modes, both asymptotically tending to the surface instability of the thick-plate limit. The stable region of $\overline{B}_2$ and $kH$ for a fixed pre-stretch is enclosed below the solid line for antisymmetric modes. Furthermore, an increase in the pre-stretch results in a larger value of $\overline{B}_2^{\text{cr}}$ for a given $kH$. This means that increasing the pre-stretch enhances the stability of SMA plates. 

Besides, we see from Figs.~\ref{Fig7}(a) and \ref{Fig8}(a) that in the absence of pre-stretch (i.e., $\lambda=1$ or $\lambda_1=1$), a thin plate ($kH\to 0$) buckles immediately when subject to an extremely small transverse magnetic field.
By contrast, for a thin plate with a pre-stretch (say $\lambda=1.25$ or $\lambda_1=1.25$), a non-zero magnetic induction field $\overline{B}_2^{\text{cr}} \simeq 2.0$ (plane-strain loading) or $\overline{B}_2^{\text{cr}} \simeq 1.25$ (uni-axial loading) is required to trigger the buckling instability. Interestingly, if an SMA plate with small values of $kH$ is subject to a pre-compression (i.e., the pre-stretch is less than 1), the underlying configuration is unstable for any applied $\overline{B}_2$. For example, a plate with $\lambda=0.8$ or $\lambda_1=0.8$ is unstable for $kH \lesssim 0.97$ under plane-strain loading and for $kH \lesssim 0.85$ under uni-axial loading. 
Physically, this means  that plates with small values of $kH$ do not support  pre-compression even if there is no magnetic field.

For antisymmetric modes, the effect of saturation magnetization on the bifurcation curves ($\overline{B}_2^{\text{cr}}$ versus $kH$) is highlighted in Figs.~\ref{Fig7}(b) and \ref{Fig8}(b) for plane-strain and uni-axial loadings, respectively. We observe that for a given pre-stretch, the two material models predict an identical critical magnetic field $\overline{B}_2^{\text{cr}}$ for small and moderate values of $kH$. For example, the bifurcation curves of a plate without pre-stretch ($\lambda=1$ or $\lambda_1=1$) overlap for $kH \lesssim 1.0$ and $kH \lesssim 1.4$ under plane-strain and uni-axial loading, respectively. For a large value of $kH$ with a fixed pre-stretch, the saturation Langevin model leads to a higher $\overline{B}_2^{\text{cr}}$ compared with the prediction of the ideal magneto-elastic model. On the other hand, for a large value of $kH$, the predicted difference based on the two material models become larger and larger when increasing the pre-stretch.


\subsubsection{Euler's buckling approximations} \label{Sec5.2.3}


For the magneto-elastic coupling case, it is  useful to establish thin-plate approximate equations (i.e., the Euler buckling solutions) of the antisymmetric wrinkling modes, as they always occurs first.
The derivation procedure is provided in \ref{AppeC} in detail. 
We specialize the analysis to the neo-Hookean ideal magneto-elastic model \eqref{72} because its predicted bifurcation curves coincide with those based on the magnetization saturation model for small and moderate values of $kH$, as shown in Figs.~\ref{Fig5}-\ref{Fig8}.

For plane-strain loading, we find that the critical stretch is approximated as 
\begin{equation} 
\lambda^\text{cr} ={{\lambda }_{0}} + \left[\frac{1-\lambda _{0}^{4}}{2\left( 1+\lambda _{0}^{4} \right)\left( 1-\chi  \right)}\right]kH-\left[\frac{2\lambda _{0}^{3}}{3\left( 1+\lambda _{0}^{4} \right)}-\frac{\lambda _{0}^{12}+11\lambda _{0}^{8}-9\lambda _{0}^{4}-3}{8{{\lambda }_{0}}{{\left( 1+\lambda _{0}^{4} \right)}^{3}}{{\left( 1-\chi  \right)}^{2}}}\right]{{\left( kH \right)}^{2}},
\end{equation}
where
\begin{equation}
\lambda _{0} = \sqrt{\frac{\chi^2 \overline B_2^2}{2(1-\chi)} + \sqrt{\frac{\chi^4 \overline B_2^4}{4(1-\chi)^2}+1}}.
\end{equation}
Note that if there is no magnetic field ($\overline{B}_{2}=0$), then $\lambda_0=1$, the correction of order one  vanishes, and $\lambda^\text{cr} =1-{{\left( kH \right)}^{2}}/3$, in agreement with the classical Euler solution for the buckling of a slender plate under plane-strain loading \citep{beatty1998stability}.
In \ref{AppeC} we also establish the quadratic expansions of the critical magnetic field and the corresponding expansions for the uni-axial loading mode.

\begin{figure}[htbp]
	\centering
	\setlength{\abovecaptionskip}{5pt}	
	\includegraphics[width=\textwidth]{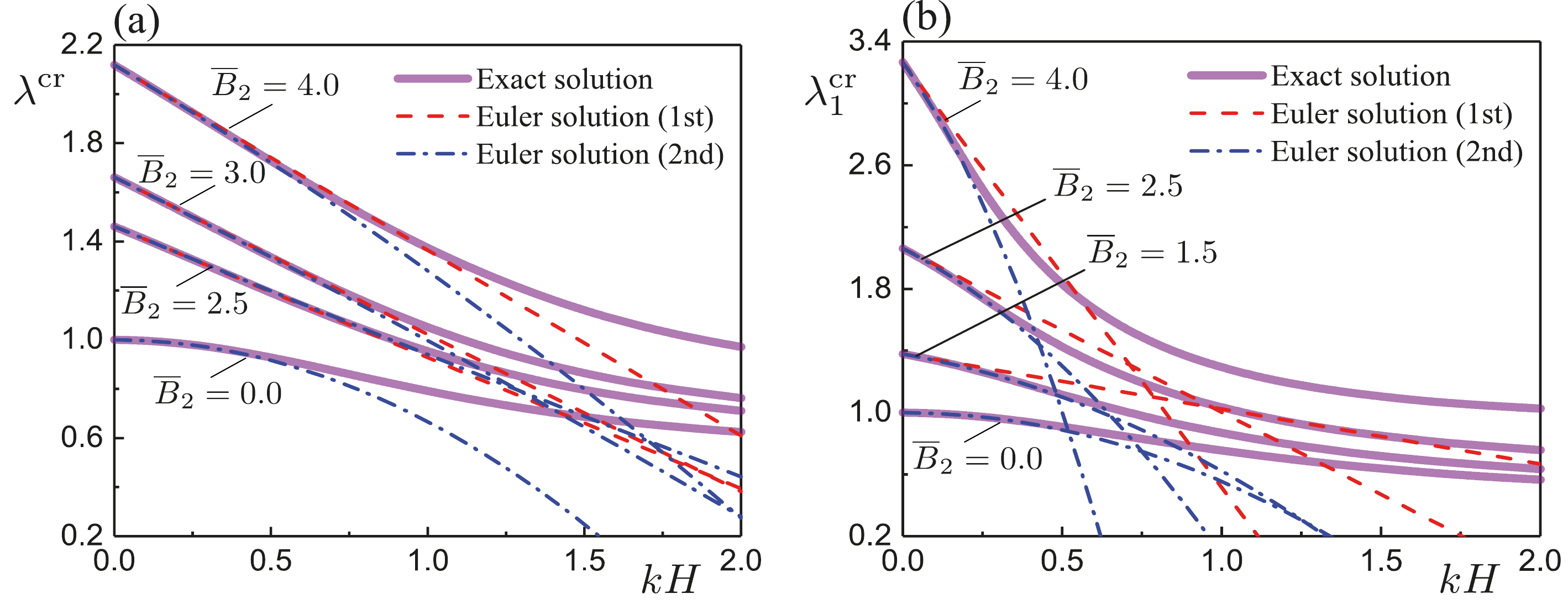}
	\caption{Critical stretch as a function of $kH$ for antisymmetric modes of a neo-Hookean ideal magneto-elastic plate subject to different fixed values of $\overline{B}_2$: (a) plane-strain loading ($\lambda^{\text{cr}}$ versus $kH$) with $\overline{B}_2=0.0, 2.5, 3.0, 4.0$; (b) uni-axial loading ($\lambda_1^{\text{cr}}$ versus $kH$) with $\overline{B}_2=0.0, 1.5, 2.5, 4.0$. The solid lines, dashed lines, and dashed-dotted lines represent, respectively, the exact solutions, the first-order and second-order Euler buckling solutions.}
	\label{Fig9}
\end{figure}

\begin{figure}[htbp]
	\centering
	\setlength{\abovecaptionskip}{5pt}	
	\includegraphics[width=0.48\textwidth]{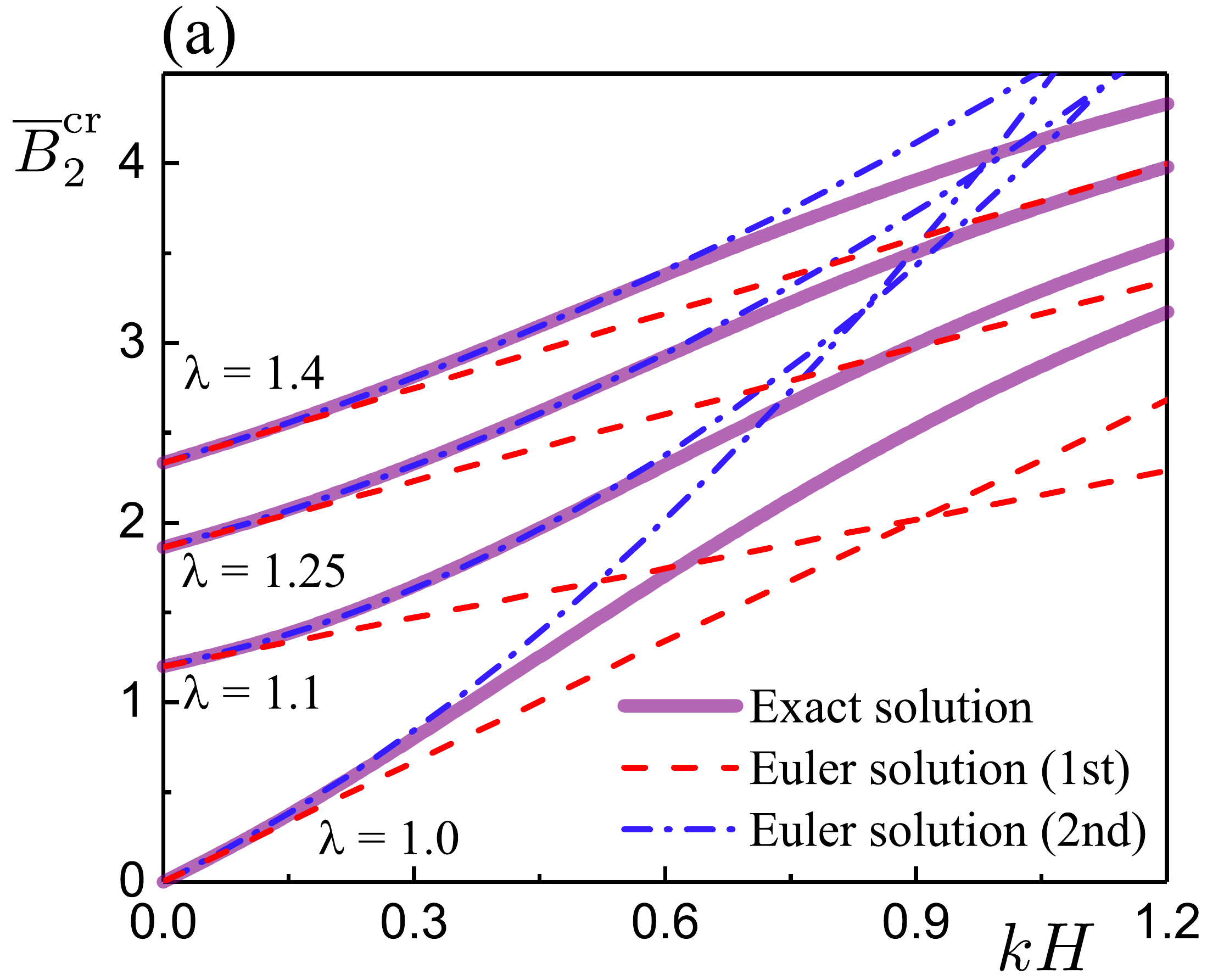}
	\hspace{0.005\textwidth}
	\includegraphics[width=0.48\textwidth]{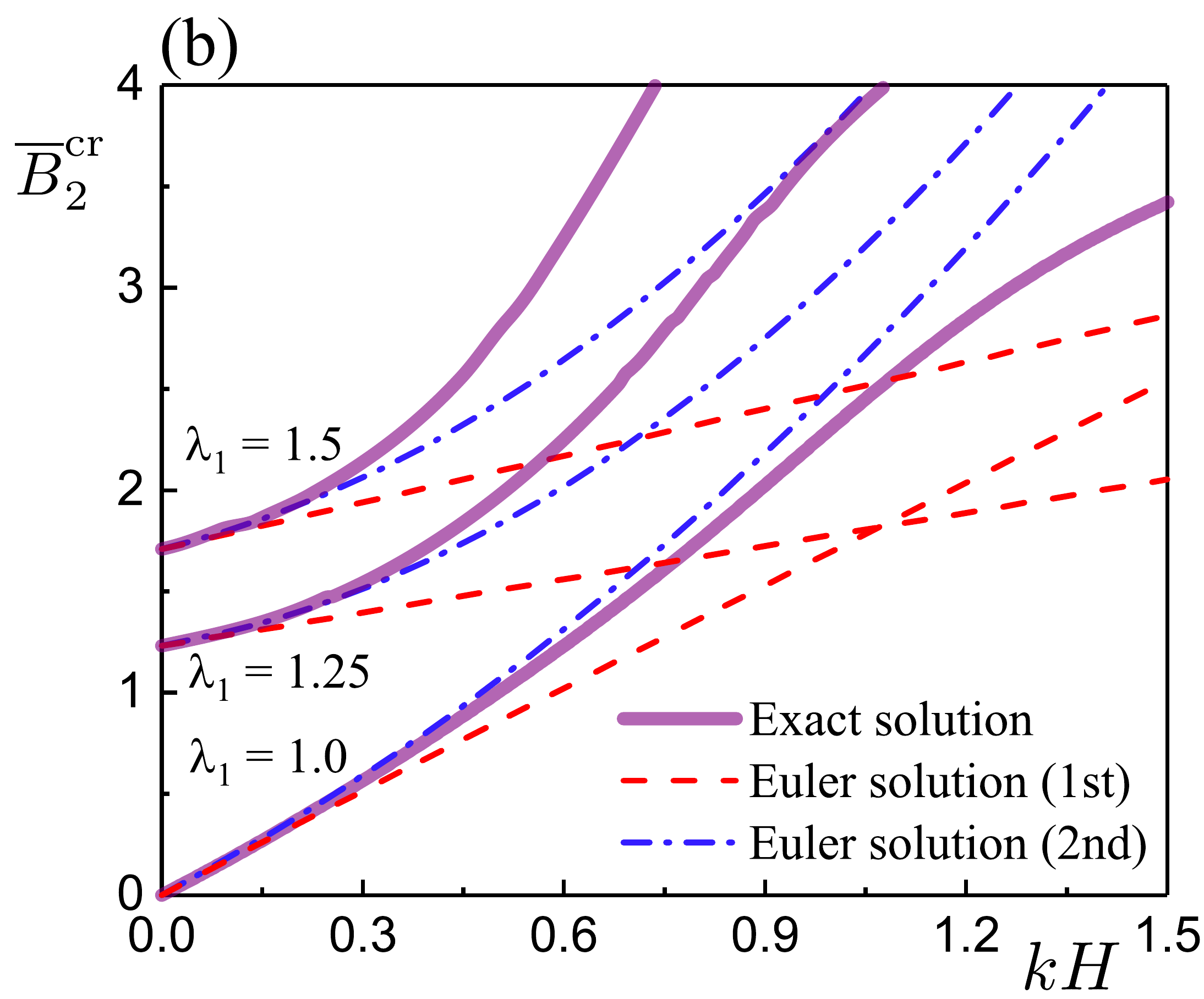}
	\caption{Critical magnetic induction field $\overline{B}_2^\text{cr}$ as a function of $kH$ for antisymmetric modes of a neo-Hookean ideal magneto-elastic plate subject to different fixed pre-stretch: (a) plane-strain loading with $\lambda=1.0, 1.1, 1.25, 1.4$; (b) uni-axial loading with $\lambda_1=1.0, 1.25, 1.5$. The solid lines, dashed lines, and dashed-dotted lines represent, respectively, the exact solutions, the first-order and second-order Euler buckling solutions.}
	\label{Fig10}
\end{figure}

Fig.~\ref{Fig9} compares the critical stretch $\lambda^{\text{cr}}$ or $\lambda_1^{\text{cr}}$ versus $kH$ based on the exact solutions to that calculated by the thin-plate buckling approximations. Fig.~\ref{Fig10} illustrates the bifurcation curves of the critical magnetic induction field $\overline{B}_2^{\text{cr}}$ versus $kH$ obtained by the exact solutions and the Euler buckling solutions. The results for plane-strain loading are shown in Figs.~\ref{Fig9}(a) and \ref{Fig10}(a) while those for uni-axial loading are depicted in Figs.~\ref{Fig9}(b) and \ref{Fig10}(b). The solid, dashed, and dashed-dotted lines represent, respectively, the exact solutions, the first-order and second-order Euler solutions.

We see from Fig.~\ref{Fig9} that in the absence of magnetic field ($\overline{B}_2=0$), the $\lambda^{\text{cr}}-kH$ curve and the $\lambda_1^{\text{cr}}-kH$ curve for the thin plate should be approximated quadratically, as in the purely elastic case \citep{beatty1998stability}. For $\overline{B}_2 \neq 0$, the earliest correction for the stretch is of the first order in $kH$. For plane-strain loading, the first-order Euler solutions provide enough accuracy to approximate the exact solutions without having to resort to the second-order correction. However, for uni-axial loading, the linear approximations are not great and the quadratic corrections are required to approximate the exact bifurcation curves.

Fig.~\ref{Fig10} shows that in the absence of pre-stretch ($\lambda=1$ or $\lambda_1=1$), the first-order Euler buckling solutions can approximate well the $\overline{B}_2^{\text{cr}}-kH$ curve for the thin plate under the two loading modes, as described by \citet{kankanala2008magnetoelastic}. But for a pre-stretch larger than 1, the $\overline{B}_2^{\text{cr}}-kH$ curve should be approximated quadratically for small values of $kH$ to enlarge the effective range of Euler's buckling solutions.


\section{Conclusions}\label{sec6}


In the framework of nonlinear magneto-elasticity theory and its associated incremental formulation, we presented a comprehensive theoretical analysis of the wrinkling instability of SMA plates under the combined action of transverse magnetic field and in-plane mechanical loading. We discussed two loading modes (plane-strain loading and uni-axial loading) and two types of neo-Hookean magneto-elastic material models (ideal model and magnetization saturation model). Employing the Stroh formulation and the surface impedance method, we derived explicit bifurcation equations of symmetric and antisymmetric modes and obtained their corresponding thin- and thick-plate limits analytically. Finally, we conducted detailed calculations to demonstrate the dependence of the nonlinear static response and bifurcation diagrams on the loading mode, load amplitude, and saturation magnetization. Our main observations are summarized below:

\begin{enumerate}[(1)]
	\item In contrast to the ideal model, introducting saturation magnetization results in a nonlinear magnetization response. Its effect on nonlinear mechanical response and critical buckling fields is more significant for  plane-strain loading than for uni-axial loading.
	
	\item  Antisymmetric wrinkling modes always appear before symmetric modes are expressed.
	
	\item Increasing the pre-compression and the magnetic field weakens the stability of SMA plates. However, the saturation magnetization effect strengthens their stability, especially for large pre-stretch or high magnetic field.	
	
	\item The thin-plate approximate formulas agree well with the exact bifurcation curves for thin plates.
\end{enumerate}

Note that we made the assumption of uniform magneto-mechanical biasing fields in this work to simplify the mathematical modelling of infinite SMA plates and to derive explicit analytical solutions. One factor that influences the uniformity of biasing fields is the shape and geometrical size of the SMA specimen. The mechanical and magnetic field distributions are usually non-uniform for ellipsoidal and cylindrical SMA specimens (see, for example, \citet{martin2006magnetostriction, rambausek2018analytical, lefevre2017general, lefevre2020two}). In addition, the SMA plate slenderness may affect the magneto-mechanical field distributions. If the biasing fields are not uniform (i.e., they vary with spatial coordinates), then the Stroh formulation for wrinkling instabilities becomes a set of first-order differential equations with \emph{variable} coefficients, which are difficult to solve analytically. Nonetheless, as shown by the work of \citet{shuvalov2004general} for example, the Stroh formalism is still useful and forms the basis of a robust numerical resolution with the surface impedance matrix method. Non-uniform biasing fields are beyond the scope of this preliminary paper and are worthy of further research.


\section*{Acknowledgement}


This work was supported by the Government of Ireland Postdoctoral Fellowship from the Irish Research Council (No. GOIPD/2019/65). We thank the reviewers for raising insightful remarks on a previous version of the paper. We are grateful to Danas Kostas (CNRS, Ecole Polytechnique) for fruitful discussions.
  
\appendix


\section{Magneto-elastic moduli components} \label{AppeA}


The three  $3\times3$ real sub-matrices $\overline{\mathbf{N}}_i\,(i=1,2,3)$ of the Stroh matrix $\mathbf{\overline{N}}$ in the Stroh formulation \eqref{33} are 
\begin{align} \label{A1}
& {{{\mathbf{\overline{N}}}}_{1}}=\left[ \begin{matrix}
0 & -1+{{{\overline{\tau }}}_{22}}/\overline{c} & -\overline{d}/ \overline{c}  \\-1 & 0 & 0  \\ -\overline{e}/\overline{g} & 0 & 0  \\ \end{matrix} \right],
\qquad
{{{\mathbf{\overline{N}}}}_{2}}=\left[ \begin{matrix}
1/\overline{c} & 0 & 0  \\  0 & 0 & 0  \\ 0 & 0 & -1/\overline{g}  \\  \end{matrix} \right], \notag \\ 
& {{{\mathbf{\overline{N}}}}_{3}}=\left[ \begin{matrix}
-2\left( \overline{b}+\overline{c}-{{{\overline{\tau }}}_{22}} \right) & 0 & 0  \\
0 & -\overline{a}+{{\left( \overline{c}-{{{\overline{\tau }}}_{22}} \right)}^{2}}/\overline{c} & -\overline{d}{{{\overline{\tau }}}_{22}}/\overline{c}  \\
0 & -\overline{d}{{{\overline{\tau }}}_{22}}/\overline{c} & \overline{f}+{{{\overline{d}}}^{2}}/\overline{c}  \\
\end{matrix} \right],
\end{align}
where the dimensionless parameters $\overline{a}-\overline{g}$ and $\overline{\tau}_{22}$ are defined as
\begin{align} \label{A2}
& \overline{a}=a/G, &&\overline{b}=b/G, &&\overline{c}=c/G, &&{{{\overline{\tau }}}_{22}}={{\tau }_{22}}/G, \notag \\ 
& \overline{d}=d/\sqrt{G{{\mu }_{0}}}, &&\overline{e}=e/\sqrt{G{{\mu }_{0}}}, &&\overline{f}=f/{{\mu }_{0}}, && \overline{g}=g/{{\mu }_{0}},
\end{align}
with the magneto-elastic material parameters $a-g$ being
\begin{align} \label{A3}
& a={{c}_{66}}={{\mathcal{A}}_{01212}}-\frac{\Gamma _{0211}^{2}}{{{K}_{011}}}, \quad c={{c}_{99}}={{\mathcal{A}}_{02121}}-\frac{\Gamma _{0211}^{2}}{{{K}_{011}}}, \quad d={{e}_{16}}=-\frac{{{\Gamma }_{0211}}}{{{K}_{011}}}, \notag \\ 
& e={{e}_{22}}-{{e}_{21}}=\frac{{{\Gamma }_{0112}}-{{\Gamma }_{0222}}}{{{K}_{022}}}, \quad f={{\mu }_{11}}=\frac{1}{{{K}_{011}}}, \quad g={{\mu }_{22}}=\frac{1}{{{K}_{022}}}, \notag \\ 
& 2b={{c}_{11}}+{{c}_{22}}-2{{c}_{12}}-2{{c}_{69}}+\frac{{{\left( {{e}_{22}}-{{e}_{21}} \right)}^{2}}}{{{\mu }_{22}}}=2\left( {{b}_{0}}+\frac{\Gamma _{0211}^{2}}{{{K}_{011}}} \right), \notag \\ 
& 2{{b}_{0}}={{\mathcal{A}}_{01111}}+{{\mathcal{A}}_{02222}}-2{{\mathcal{A}}_{01122}}-2{{\mathcal{A}}_{01221}},
\end{align}
where ${\bm{\mathcal{A}}_{0}}$, ${{\mathbf{\Gamma }}_{0}}$ and ${{\mathbf{K}}_{0}}$ are the magneto-elastic moduli tensors. Note that we have used Eq.~\eqref{31} and the relation ${\mathcal{A}}_{01221} +p={{\mathcal{A}}_{02121}}-{{\tau }_{22}}$ from Eq.~\eqref{23} to derive the Stroh matrix \eqref{A1}.

The total Cauchy stress tensor used in the Stroh matrix \eqref{A1} is obtained from Eqs.~\eqref{12} and \eqref{13}$_1$ as
\begin{equation} \label{tau22}
{{\overline{\tau }}_{22}}=\overline{\tau }_{22}^\star=\overline{B}_{2}^{2}/2,
\end{equation}
where ${{\overline{B}}_{2}}={{B}_{2}}/\sqrt{G{{\mu }_{0}}}$ is the dimensionless applied magnetic induction.

Using the incremental theory of magneto-elasticity \citep{ottenio2008incremental, destrade2011magneto}, we compute the instantaneous magneto-elastic moduli appearing in Eq.~\eqref{A3} for the applied bi-axial deformation $\lambda_1,\lambda_3$ and the transverse magnetic field $B_2$, as follows
\begin{align} \label{A4}
& {{\mathcal{A}}_{01212}}=2\lambda _{1}^{2}\left( {{\Omega }_{1}}+\lambda _{3}^{2}{{\Omega }_{2}}+B_{2}^{2}{{\Omega }_{6}} \right), \notag \\[4pt]
& {{\mathcal{A}}_{02121}}=2\lambda _{1}^{-2}\lambda _{3}^{-2}\left\{ \left( {{\Omega }_{1}}+\lambda _{3}^{2}{{\Omega }_{2}} \right)+B_{2}^{2}\left[ \lambda _{1}^{2}\lambda _{3}^{2}{{\Omega }_{5}}+\left( 2+\lambda _{1}^{4}\lambda _{3}^{2} \right){{\Omega }_{6}} \right] \right\}, \notag \\[4pt] 
& {{b}_{0}}=\left( \lambda _{1}^{-2}\lambda _{3}^{-2}+\lambda _{1}^{2} \right)\left( {{\Omega }_{1}}+\lambda _{3}^{2}{{\Omega }_{2}} \right)+2{{\left( \lambda _{1}^{2}-\lambda _{1}^{-2}\lambda _{3}^{-2} \right)}^{2}}\left( {{\Omega }_{11}}+2\lambda _{3}^{2}{{\Omega }_{12}}+\lambda _{3}^{4}{{\Omega }_{22}} \right) \notag \\ 
& \phantom{K_022} +B_{2}^{2}\left\{ 4\left( \lambda _{1}^{-2}\lambda _{3}^{-2}-\lambda _{1}^{2} \right)\left[ \left( {{\Omega }_{15}}+\lambda _{3}^{2}{{\Omega }_{25}} \right)+2\lambda _{1}^{-2}\lambda _{3}^{-2}\left( {{\Omega }_{16}}+\lambda _{3}^{2}{{\Omega }_{26}} \right) \right] \right. \notag \\ 
& \phantom{K_022} \left.  +\,{{\Omega }_{5}}+2\left( 3\lambda _{1}^{-2}\lambda _{3}^{-2}-\lambda _{1}^{2} \right){{\Omega }_{6}} \right\}+2B_{2}^{4}\left( {{\Omega }_{55}}+4\lambda _{1}^{-2}\lambda _{3}^{-2}{{\Omega }_{56}}+4\lambda _{1}^{-4}\lambda _{3}^{-4}{{\Omega }_{66}} \right), \notag \\[4pt] 
& {{\Gamma }_{0211}}=2{{B}_{2}}\left[ {{\Omega }_{5}}+\left( \lambda _{1}^{2}+\lambda _{1}^{-2}\lambda _{3}^{-2} \right){{\Omega }_{6}} \right], \notag \\[4pt] 
& {{\Gamma }_{0112}}-{{\Gamma }_{0222}}=4{{B}_{2}}\left\{ \left( \lambda _{1}^{4}\lambda _{3}^{2}-1 \right)\left[ \left( {{\Omega }_{14}}+\lambda _{3}^{2}{{\Omega }_{24}} \right)+\lambda _{1}^{-2}\lambda _{3}^{-2}\left( {{\Omega }_{15}}+\lambda _{3}^{2}{{\Omega }_{25}} \right) \right. \right. \notag \\ 
& \phantom{K_022} \left. +\lambda _{1}^{-4}\lambda _{3}^{-4}\left( {{\Omega }_{16}}+\lambda _{3}^{2}{{\Omega }_{26}} \right) \right]-{{\Omega }_{5}}-2\lambda _{1}^{-2}\lambda _{3}^{-2}{{\Omega }_{6}} \notag \\ 
& \phantom{K_022} \left. -B_{2}^{2}\left( \lambda _{1}^{2}\lambda _{3}^{2}{{\Omega }_{45}}+2{{\Omega }_{46}}+{{\Omega }_{55}}+3\lambda _{1}^{-2}\lambda _{3}^{-2}{{\Omega }_{56}}+2\lambda _{1}^{-4}\lambda _{3}^{-4}{{\Omega }_{66}} \right) \right\}, \notag \\[4pt] 
& {{K }_{011}}=2\left( \lambda _{1}^{-2}{{\Omega }_{4}}+{{\Omega }_{5}}+\lambda _{1}^{2}{{\Omega }_{6}} \right), \notag \\[4pt] 
& {{K }_{022}}=2\left( \lambda _{1}^{2}\lambda _{3}^{2}{{\Omega }_{4}}+{{\Omega }_{5}}+\lambda _{1}^{-2}\lambda _{3}^{-2}{{\Omega }_{6}} \right)+4B_{2}^{2}\left( \lambda _{1}^{4}\lambda _{3}^{4}{{\Omega }_{44}}+2\lambda _{1}^{2}\lambda _{3}^{2}{{\Omega }_{45}} \right. \notag\\ 
& \left.\phantom{{{K }_{022}}} +2{{\Omega }_{46}}+{{\Omega }_{55}}+2\lambda _{1}^{-2}\lambda _{3}^{-2}{{\Omega }_{56}}+\lambda _{1}^{-4}\lambda _{3}^{-4}{{\Omega }_{66}} \right).
\end{align}

After obtaining the eigenvalues $q_j$ and eigenvectors $\bm{\overline{\eta}}^{(j)}$, the generalized displacement and traction vectors at the faces $x_2=\pm h$ are expressed, according to Eq.~\eqref{39}, as
\begin{equation}  \label{A6}
\left[ \begin{matrix}
\mathbf{\overline{U}}\left( kh \right)  \\
\mathbf{\overline{U}}\left( -kh \right)  \\
\end{matrix} \right]=\left[ \begin{matrix}
\overline{\eta }_{1}^{\left( 1 \right)}E_{1}^{+} & \overline{\eta }_{1}^{\left( 2 \right)}E_{2}^{+} & \overline{\eta }_{1}^{\left( 3 \right)}E_{3}^{+} & \overline{\eta }_{1}^{\left( 4 \right)}E_{1}^{-} & \overline{\eta }_{1}^{\left( 5 \right)}E_{2}^{-} & \overline{\eta }_{1}^{\left( 6 \right)}E_{3}^{-}  \\
\overline{\eta }_{2}^{\left( 1 \right)}E_{1}^{+} & \overline{\eta }_{2}^{\left( 2 \right)}E_{2}^{+} & \overline{\eta }_{2}^{\left( 3 \right)}E_{3}^{+} & \overline{\eta }_{2}^{\left( 4 \right)}E_{1}^{-} & \overline{\eta }_{2}^{\left( 5 \right)}E_{2}^{-} & \overline{\eta }_{2}^{\left( 6 \right)}E_{3}^{-}  \\
\overline{\eta }_{3}^{\left( 1 \right)}E_{1}^{+} & \overline{\eta }_{3}^{\left( 2 \right)}E_{2}^{+} & \overline{\eta }_{3}^{\left( 3 \right)}E_{3}^{+} & \overline{\eta }_{3}^{\left( 4 \right)}E_{1}^{-} & \overline{\eta }_{3}^{\left( 5 \right)}E_{2}^{-} & \overline{\eta }_{3}^{\left( 6 \right)}E_{3}^{-}  \\
\overline{\eta }_{1}^{\left( 1 \right)}E_{1}^{-} & \overline{\eta }_{1}^{\left( 2 \right)}E_{2}^{-} & \overline{\eta }_{1}^{\left( 3 \right)}E_{3}^{-} & \overline{\eta }_{1}^{\left( 4 \right)}E_{1}^{+} & \overline{\eta }_{1}^{\left( 5 \right)}E_{2}^{+} & \overline{\eta }_{1}^{\left( 6 \right)}E_{3}^{+}  \\
\overline{\eta }_{2}^{\left( 1 \right)}E_{1}^{-} & \overline{\eta }_{2}^{\left( 2 \right)}E_{2}^{-} & \overline{\eta }_{2}^{\left( 3 \right)}E_{3}^{-} & \overline{\eta }_{2}^{\left( 4 \right)}E_{1}^{+} & \overline{\eta }_{2}^{\left( 5 \right)}E_{2}^{+} & \overline{\eta }_{2}^{\left( 6 \right)}E_{3}^{+}  \\
\overline{\eta }_{3}^{\left( 1 \right)}E_{1}^{-} & \overline{\eta }_{3}^{\left( 2 \right)}E_{2}^{-} & \overline{\eta }_{3}^{\left( 3 \right)}E_{3}^{-} & \overline{\eta }_{3}^{\left( 4 \right)}E_{1}^{+} & \overline{\eta }_{3}^{\left( 5 \right)}E_{2}^{+} & \overline{\eta }_{3}^{\left( 6 \right)}E_{3}^{+}  \\
\end{matrix} \right]\left[ \begin{matrix}
{{A}_{1}}  \\ {{A}_{2}}  \\ {{A}_{3}}  \\
{{A}_{4}}  \\ {{A}_{5}}  \\ {{A}_{6}}  \\
\end{matrix} \right],
\end{equation}
and
\begin{equation}  \label{A7}
\left[ \begin{matrix}
\mathbf{\overline{S}}\left( kh \right)  \\
\mathbf{\overline{S}}\left( -kh \right)  \\
\end{matrix} \right]=\left[ \begin{matrix}
\overline{\eta }_{4}^{\left( 1 \right)}E_{1}^{+} & \overline{\eta }_{4}^{\left( 2 \right)}E_{2}^{+} & \overline{\eta }_{4}^{\left( 3 \right)}E_{3}^{+} & \overline{\eta }_{4}^{\left( 4 \right)}E_{1}^{-} & \overline{\eta }_{4}^{\left( 5 \right)}E_{2}^{-} & \overline{\eta }_{4}^{\left( 6 \right)}E_{3}^{-}  \\
\overline{\eta }_{5}^{\left( 1 \right)}E_{1}^{+} & \overline{\eta }_{5}^{\left( 2 \right)}E_{2}^{+} & \overline{\eta }_{5}^{\left( 3 \right)}E_{3}^{+} & \overline{\eta }_{5}^{\left( 4 \right)}E_{1}^{-} & \overline{\eta }_{5}^{\left( 5 \right)}E_{2}^{-} & \overline{\eta }_{5}^{\left( 6 \right)}E_{3}^{-}  \\
\overline{\eta }_{6}^{\left( 1 \right)}E_{1}^{+} & \overline{\eta }_{6}^{\left( 2 \right)}E_{2}^{+} & \overline{\eta }_{6}^{\left( 3 \right)}E_{3}^{+} & \overline{\eta }_{6}^{\left( 4 \right)}E_{1}^{-} & \overline{\eta }_{6}^{\left( 5 \right)}E_{2}^{-} & \overline{\eta }_{6}^{\left( 6 \right)}E_{3}^{-}  \\
\overline{\eta }_{4}^{\left( 1 \right)}E_{1}^{-} & \overline{\eta }_{4}^{\left( 2 \right)}E_{2}^{-} & \overline{\eta }_{4}^{\left( 3 \right)}E_{3}^{-} & \overline{\eta }_{4}^{\left( 4 \right)}E_{1}^{+} & \overline{\eta }_{4}^{\left( 5 \right)}E_{2}^{+} & \overline{\eta }_{4}^{\left( 6 \right)}E_{3}^{+}  \\
\overline{\eta }_{5}^{\left( 1 \right)}E_{1}^{-} & \overline{\eta }_{5}^{\left( 2 \right)}E_{2}^{-} & \overline{\eta }_{5}^{\left( 3 \right)}E_{3}^{-} & \overline{\eta }_{5}^{\left( 4 \right)}E_{1}^{+} & \overline{\eta }_{5}^{\left( 5 \right)}E_{2}^{+} & \overline{\eta }_{5}^{\left( 6 \right)}E_{3}^{+}  \\
\overline{\eta }_{6}^{\left( 1 \right)}E_{1}^{-} & \overline{\eta }_{6}^{\left( 2 \right)}E_{2}^{-} & \overline{\eta }_{6}^{\left( 3 \right)}E_{3}^{-} & \overline{\eta }_{6}^{\left( 4 \right)}E_{1}^{+} & \overline{\eta }_{6}^{\left( 5 \right)}E_{2}^{+} & \overline{\eta }_{6}^{\left( 6 \right)}E_{3}^{+}  \\
\end{matrix} \right]\left[ \begin{matrix}
{{A}_{1}}  \\ {{A}_{2}}  \\ {{A}_{3}}  \\
{{A}_{4}}  \\ {{A}_{5}}  \\ {{A}_{6}}  \\
\end{matrix} \right],
\end{equation}
where $E_{j}^{\pm }={\text{e}^{\pm \text{i}{{q}_{j}}kh}}\left( j=1,\ldots, 6 \right)$. Thus, we note from Eq.~\eqref{44} that $E_{j}^{\pm }={\text{e}^{\mp {{p}_{j}}kh}}$ for $j=1,2,3$ since $q_j=\text{i}p_j\left( j=1,2,3 \right)$, and that $E_{j+3}^{\pm }=E_{j}^{\mp }$ for $j=1,2,3$ since $q_{j+3}=-q_j\left( j=1,2,3 \right)$.


\section{Bifurcation equation of surface instability based on the surface impedance method} \label{AppeB}


Assume that the SMA half-space in the reference and current configurations occupies the region ${X}_{2}\ge 0$ and ${{x}_{2}}\ge 0$, respectively. To satisfy the decay condition at ${{x}_{2}}\to +\infty$ in the half-space, we only keep the eigenvalues in Eq.~\eqref{39} with positive imaginary parts. Thus, we take the three eigenvalues $q_1,q_2,q_3$ according to Eqs.~\eqref{44} and \eqref{45}. The general solution to the Stroh formulation \eqref{33} is written as
\begin{equation} \label{B1}
\bm{\overline{\eta }}\left( k{{x}_{2}} \right)=\left[ \begin{matrix}
\mathbf{\overline{U}}\left( k{{x}_{2}} \right)  \\
\mathbf{\overline{S}}\left( k{{x}_{2}} \right)  \\
\end{matrix} \right]=\sum\limits_{j=1}^{3}{{{A}_{j}}{{{\bm{\overline{\eta }}}}^{\left( j \right)}}{\text{e}^{\text{i}{{q}_{j}}k{{x}_{2}}}}}=\sum\limits_{j=1}^{3}{{{A}_{j}}{{{\bm{\overline{\eta }}}}^{\left( j \right)}}{\text{e}^{-{{p}_{j}}k{{x}_{2}}}}},
\end{equation}
where $q_j=\text{i}p_j\,(j=1,2,3)$ with $p_j>0$. In matrix form, Eq.~\eqref{B1} is expressed as
\begin{equation} \label{B2}
\bm{\overline{\eta}}\left(k{{x}_{2}}\right)=\left[ \begin{matrix}
{{\mathbf{P}}_{1}}  \\
{{\mathbf{P}}_{2}}  \\
\end{matrix} \right]\left[ \begin{matrix}
{\text{e}^{-{{p}_{1}}k{{x}_{2}}}} & 0 & 0  \\
0 & {\text{e}^{-{{p}_{2}}k{{x}_{2}}}} & 0  \\
0 & 0 & {\text{e}^{-{{p}_{3}}k{{x}_{2}}}}  \\
\end{matrix} \right]\left[ \begin{matrix}
{{A}_{1}}  \\
{{A}_{2}}  \\
{{A}_{3}}  \\
\end{matrix} \right],
\end{equation}
where
\renewcommand{\arraystretch}{1.4}
\begin{equation} \label{B3}
{{\mathbf{P}}_{1}}=\left[ \begin{matrix}
\overline{\eta }_{1}^{\left( 1 \right)} & \overline{\eta }_{1}^{\left( 2 \right)} & \overline{\eta }_{1}^{\left( 3 \right)}  \\
\overline{\eta }_{2}^{\left( 1 \right)} & \overline{\eta }_{2}^{\left( 2 \right)} & \overline{\eta }_{2}^{\left( 3 \right)}  \\
\overline{\eta }_{3}^{\left( 1 \right)} & \overline{\eta }_{3}^{\left( 2 \right)} & \overline{\eta }_{3}^{\left( 3 \right)}  \\
\end{matrix} \right],  \qquad {{\mathbf{P}}_{2}}=\left[ \begin{matrix}
\overline{\eta }_{4}^{\left( 1 \right)} & \overline{\eta }_{4}^{\left( 2 \right)} & \overline{\eta }_{4}^{\left( 3 \right)}  \\
\overline{\eta }_{5}^{\left( 1 \right)} & \overline{\eta }_{5}^{\left( 2 \right)} & \overline{\eta }_{5}^{\left( 3 \right)}  \\
\overline{\eta }_{6}^{\left( 1 \right)} & \overline{\eta }_{6}^{\left( 2 \right)} & \overline{\eta }_{6}^{\left( 3 \right)}  \\
\end{matrix} \right],
\end{equation}

Setting ${{x}_{2}}=0$, we obtain from Eqs.~\eqref{B1} and \eqref{B2} that
\begin{equation} \label{B4}
\mathbf{\overline{S}}\left( 0 \right)= \text{i}\mathbf{\overline{Z}\overline{U}}\left( 0 \right),
\end{equation}
where $\mathbf{\overline{Z}}= -\text{i} {{\mathbf{P}}_{2}}\mathbf{P}_{1}^{-1}$ is the surface impedance matrix of the half-space, through which the generalized traction and displacement vectors at the face $x_2=0$ are connected. 

In view of Eqs.~\eqref{60} and \eqref{61}, the surface impedance matrix exterior to the material is
\begin{equation}  \label{B5}
{{\mathbf{\overline{Z}}}^\star}=\left[ \begin{matrix}
0 & \text{i}\overline{B}_{2}^{2}/2 & \text{i}{{{\overline{B}}}_{2}}  \\
-\text{i}\overline{B}_{2}^{2}/2 & \overline{B}_{2}^{2} & {{{\overline{B}}}_{2}}  \\
-\text{i}{{{\overline{B}}}_{2}} & {{{\overline{B}}}_{2}} & 1  \\
\end{matrix} \right],
\end{equation}
which, at the face $x_2=0$, satisfies
\begin{equation} \label{B6}
\mathbf{\overline{S}}\left( 0 \right)=\text{i}{{\mathbf{\overline{Z}}}^\star}\mathbf{\overline{U}}\left( 0 \right).
\end{equation}

From Eqs.~\eqref{B4} and \eqref{B6}, we get the bifurcation equation governing the surface wrinkling instability, as

\begin{equation}  \label{B7}
\text{det}\left(\mathbf{\overline{Z}}- {{{\mathbf{\overline{Z}}}}^\star} \right)=0.
\end{equation}

Substituting Eqs.~\eqref{tau22}, \eqref{41}, \eqref{45}$_{1,2}$, \eqref{46}$_{1,2,3}$ and \eqref{B5} into Eq.~\eqref{B7} and with the help of Mathematica, we obtain the explicit bifurcation equation for the surface wrinkles as
\begin{multline}
\left[ 1+\beta \left( \lambda _{3}^{2}-1 \right) \right]\left( 1+2{\overline{F}_5}{{p}_{3}} \right)\left[ {{\left( \lambda _{1}^{2}{{\lambda }_{3}} \right)}^{3}}+{{\left( \lambda _{1}^{2}{{\lambda }_{3}} \right)}^{2}}+3\lambda _{1}^{2}{{\lambda }_{3}}-1 \right]
\\
-\overline{B}_{2}^{2}{{\left( 1-2{\overline{F}_5} \right)}^{2}}\lambda _{1}^{2}\lambda _{3}^{2}\left( \lambda _{1}^{2}{{\lambda }_{3}}+1 \right)=0,
\end{multline}
which is the same as Eq.~\eqref{70}.


\section{Thin-plate buckling approximations} \label{AppeC}


For the neo-Hookean ideal magneto-elastic model \eqref{72}, this appendix makes use of the exact bifurcation equation \eqref{82} to establish thin-plate approximations to antisymmetric wrinkling modes, which always occur first.


\subsection{Plane-strain loading}\label{AppeC1}


For plane-strain loading, the exact bifurcation equation \eqref{82} becomes
\begin{multline} \label{C1}
\left[ 1-\chi +\tanh \left( kh \right) \right]\left[ {{\left( 1+{{\lambda }^{4}} \right)}^{2}}\tanh \left( kh \right)-4{{\lambda }^{2}}\tanh \left( {{\lambda }^{2}}kh \right) \right] \\
=\overline{B}_{2}^{2}{{\chi }^{2}}{{\lambda }^{2}}\left( {{\lambda }^{4}}-1 \right)\tanh \left( kh \right),
\end{multline}
where $kh={{\lambda }^{-1}}kH$. At the zero-th order in $kH$, the thin-plate equation \eqref{84} gives
\begin{equation} \label{C2}
\overline{B}_{2}^{2}={{\chi }^{-2}}\left( 1-\chi  \right)\left( {{\lambda }^{2}}-{{\lambda }^{-2}} \right).
\end{equation}
%


\subsubsection{Critical stretch}\label{AppeC1.1}


For a fixed magnetic induction field $\overline{B}_{2}$, we first derive the approximations of the critical stretch $\lambda^{\text{cr}}$. In this case, we denote the root of Eq.~\eqref{C2} as $\lambda_0$. In the thin-plate limit ($kH\to 0$), we may expand the tanh functions in Eq.~\eqref{C1} in power series as follows:
\begin{equation} \label{C3}
\tanh \left( kh \right) \approx {{\lambda }^{-1}}kH-{{\left( {{\lambda }^{-1}}kH \right)}^{3}}/3, \qquad 
\tanh \left( {{\lambda }^{2}}kh\right) \approx \lambda kH-{{\left( \lambda kH \right)}^{3}}/3.
\end{equation}
Inserting Eq.~\eqref{C3} into Eq.~\eqref{C1} and retaining only terms of first order in $kH$, we obtain
\begin{equation} \label{C4}
(1-\lambda^2) (1+\lambda^2)kH +\lambda 
\left[ (1-\chi)(1-\lambda^2) (1+\lambda^2)+\overline{B}_{2}^{2}\chi^2\lambda^2 \right]=0.
\end{equation}
As expected, when $kH=0$, $\lambda$ is equal to
$\lambda_0$. Hence, substituting the first-order expansion $\lambda^\text{cr} ={{\lambda }_{0}}+{{\varepsilon }_{1}}kH$ in Eq.~\eqref{C4} and retaining terms to the first order in $kH$, we find the equation governing $\varepsilon_1$ as
\begin{equation} \label{C5}
1-\lambda_0^4+\varepsilon_1 \left[ (1-\chi)(5-9\lambda_0^4)+7\overline{B}_{2}^{2}\chi^2\lambda_0^2 \right]=0.
\end{equation}
Using Eq.~\eqref{C2} in Eq.~\eqref{C5}, we thus obtain the first-order correction of the critical stretch
\begin{equation} \label{C6}
\lambda^\text{cr} ={{\lambda }_{0}}+\frac{1-\lambda _{0}^{4}}{2\left( 1+\lambda _{0}^{4} \right)\left( 1-\chi  \right)}kH.
\end{equation}

Similarly, we insert the power series expansion \eqref{C3} into Eq.~\eqref{C1} and retain only terms of second order in $kH$. Then, we introduce the second-order expansion of stretch $\lambda^\text{cr} ={{\lambda }_{0}}+{{\varepsilon }_{1}}kH+{{\varepsilon }_{2}}{{\left( kH \right)}^{2}}$ in the resultant equation, keep terms to the second order in $kH$, and thereby get the second-order correction of the critical stretch
\begin{equation} \label{C7}
\lambda^\text{cr} ={{\lambda }_{0}}+\frac{1-\lambda _{0}^{4}}{2\left( 1+\lambda _{0}^{4} \right)\left( 1-\chi  \right)}kH+{{\varepsilon }_{2}}{{\left( kH \right)}^{2}},
\end{equation}
where
\begin{equation} \label{C8}
{{\varepsilon }_{2}}=-\frac{2\lambda _{0}^{3}}{3\left( 1+\lambda _{0}^{4} \right)}+\frac{\lambda _{0}^{12}+11\lambda _{0}^{8}-9\lambda _{0}^{4}-3}{8{{\lambda }_{0}}{{\left( 1+\lambda _{0}^{4} \right)}^{3}}{{\left( 1-\chi  \right)}^{2}}}.
\end{equation}

When $\overline{B}_{2}=0$, we have $\lambda_0=1$ and from Eq.~\eqref{C7} the first correction for stretch is of order two: $\lambda^\text{cr} =1-{{\left( kH \right)}^{2}}/3$. This is equivalent to the classical Euler solution for the buckling of a slender or a thin plate under plane-strain loading in the purely elastic case \citep{beatty1998stability}.


\subsubsection{Critical magnetic induction field} \label{AppeC1.2}


Next, we derive the approximations of the critical magnetic field $\overline{B}_2^\text{cr}$ for a given pre-stretch $\lambda$. In this case, the root of Eq.~\eqref{C2} is represented by $\overline{B}_{20}$. Analogous to the derivations of the critical stretch described in \ref{AppeC1.1}, we obtain
\begin{equation} \label{C9}
\overline{B}_2^\text{cr}={{\overline{B}}_{20}}\left[1+\frac{1}{2\lambda \left( 1-\chi \right)} kH \right]
\end{equation}
for the first-order correction, and
\begin{equation} \label{C10}
\overline{B}_2^\text{cr}={{\overline{B}}_{20}}\left[1+\frac{1}{2\lambda \left( 1-\chi \right)}kH +{{\gamma }_{2}}{{\left( kH \right)}^{2}} \right],  \qquad
{{\gamma }_{2}}=\frac{3+{{\lambda }^{4}}\left[ 13+16\chi \left( \chi -2 \right) \right]}{24{{\lambda }^{2}}\left( {{\lambda }^{4}}-1 \right){{\left( 1-\chi \right)}^{2}}}
\end{equation}
for the second-order correction.

However, in the special case where there is no pre-stretch ($\lambda=1$), we see that $\overline{B}_{20}=0$ from Eq.~\eqref{C2}. In that case, Eqs.~\eqref{C9} and \eqref{C10} give $\overline{B}_2^\text{cr}\equiv0$, which is independent of $kH$ and unphysical. That case thus requires a separate treatment. 
Expanding the bifurcation equation \eqref{C1} in power series in $kH$, keeping terms up to the third order in $kH$, and setting $\lambda=1$, we have
\begin{equation} \label{C11}
-4(kH)^3+\left[4(\chi-1)-\overline{B}_{2}^{2}\chi^2\right] (kH)^2 +3\overline{B}_{2}^{2}\chi^2=0. 
\end{equation}
Then, we introduce the first-order expansion of the critical magnetic field $\overline{B}_2^\text{cr}={{\gamma }_{10}}kH$ in Eq.~\eqref{C11}, keep terms to the second order in $kH$, and obtain the first-order correction as
\begin{equation} \label{C12}
\overline{B}_2^\text{cr}=\frac{2\sqrt{1-\chi }}{\sqrt{3}\chi } kH.
\end{equation}
Further, the second-order correction of the critical magnetic field is found as
\begin{equation} \label{C13}
\overline{B}_2^\text{cr}=\frac{2\sqrt{1-\chi }}{\sqrt{3}\chi } kH+\frac{1}{\chi \sqrt{3\left( 1-\chi  \right)}}{{\left( kH \right)}^{2}}.
\end{equation}

Note that the thin-plate buckling approximation \eqref{C13} of $\overline{B}_2^\text{cr}$ agrees  with the classical asymptotic formula obtained by \citet{yih1973linear} (see their Eq.~(8.13) and let the Poisson ratio in their formula tend to 1/2 for incompressible materials).


\subsection{Uni-axial loading}\label{AppeC2}


For uni-axial loading, the exact bifurcation equation for antisymmetric wrinkles is governed by Eq.~\eqref{82}, which is reproduced here, as
\begin{multline} \label{C14}
\left[ 1-\chi +\tanh \left( kh \right) \right]\left[ {{\left( 1+\lambda _{1}^{4}\lambda _{3}^{2} \right)}^{2}}\tanh \left( kh \right)-4\lambda _{1}^{2}{{\lambda }_{3}}\tanh \left( \lambda _{1}^{2}{{\lambda }_{3}}kh \right) \right] \\ =\overline{B}_{2}^{2}{{\chi }^{2}}\lambda _{1}^{2}\lambda _{3}^{2}\left( \lambda _{1}^{4}\lambda _{3}^{2}-1 \right)\tanh \left( kh \right),
\end{multline}
where $kh=\lambda _{1}^{-1}\lambda _{3}^{-1}kH$, and from Eq.~\eqref{80}$_2$ we obtain the nonlinear mechanical response (determining $\lambda_3$) for the ideal magneto-elastic model as
\begin{equation} \label{C15}
\lambda _{3}^{2}-\lambda _{1}^{-2}\lambda _{3}^{-2}+\chi \overline{B}_{2}^{2}=0.
\end{equation}
At the zero-th order in $kH$, the thin-plate equation \eqref{84} gives
\begin{equation} \label{C16}
\overline{B}_{2}^{2}={{\chi }^{-2}}\left( 1-\chi  \right)\left( \lambda _{1}^{2}-\lambda _{1}^{-2}\lambda _{3}^{-2} \right).
\end{equation}
%


\subsubsection{Critical stretch}\label{AppeC2.1}


First, the approximations of the critical stretches $\lambda_1^{\text{cr}}$ and $\lambda_3^{\text{cr}}$ are derived according to Eqs.~\eqref{C14} and \eqref{C15} for a fixed $\overline{B}_{2}$. The root of the zero-order equations \eqref{C15} and \eqref{C16} is denoted by $\lambda_{10}$ and $\lambda_{30}$. For $kH\to 0$, we expand Eq.~\eqref{C14} up to the first order in $kH$, as
\begin{equation} \label{C17}
(1-\lambda_1^{4} \lambda_3^2) kH +\lambda_1 \lambda_3 \left[ (1-\chi)(1-\lambda_1^{4} \lambda_3^2)+\overline{B}_{2}^{2}\chi^2 \lambda_1^{2} \lambda_3^2 \right]=0.
\end{equation}

We then introduce the first-order corrections ${{\lambda }_1^\text{cr}}={{\lambda }_{10}}+{{\varepsilon }_{1}}kH$ and ${{\lambda }_3^\text{cr}}={{\lambda }_{30}}+{{\varepsilon }_{2}}kH$ into Eqs.~\eqref{C15} and \eqref{C17}, and retain terms to the first order in $kH$. The resultant zero-order terms satisfy the zero-order thin-plate equations \eqref{C15} and \eqref{C16}, while the first-order terms constitute a set of two linear algebraic equations for $\varepsilon_1$ and $\varepsilon_2$, which are solved as
\begin{equation} \label{C18}
{{\varepsilon }_{1}}=\frac{\left( 1-\lambda _{10}^{4}\lambda _{30}^{2} \right)\left( 1+\lambda _{10}^{2}\lambda _{30}^{4} \right)}{2\lambda _{10}^{2}\lambda _{30}^{3}\left( \lambda _{10}^{2}+\lambda _{30}^{2}+\lambda _{10}^{4}\lambda _{30}^{4} \right)\left( 1-\chi  \right)},
\qquad
{{\varepsilon }_{2}}=-\frac{\lambda_{30}} {\lambda_{10} \left( 1+\lambda _{10}^{2}\lambda _{30}^{4} \right)} {{\varepsilon }_{1}}.
\end{equation}

The next order in $kH$ is order two. With similar manipulations of Eqs.~\eqref{C14} and \eqref{C15}, we find the second-order correction of the critical stretches, as
\begin{equation} \label{C19}
{{\lambda }_1^\text{cr}}={{\lambda }_{10}}+{{\varepsilon }_{1}}kH+{{\varepsilon }_{3}}{{\left( kH \right)}^{2}}, \qquad
{{\lambda }_3^\text{cr}}={{\lambda }_{30}}+{{\varepsilon }_{2}}kH+{{\varepsilon }_{4}}{{\left( kH \right)}^{2}},
\end{equation}
where $\varepsilon_1$ and $\varepsilon_2$ are determined by Eq.~\eqref{C18}, and 
\begin{equation} \label{C20}
\left[{{\varepsilon }_{3}}, \, {{\varepsilon }_{4}} \right]^\text{T}=2\mathbf{Q}_{1}^{-1} \left[t_1, \, t_2 \right]^\text{T},
\end{equation}
in which we have
\begin{align} \label{C21}
{{\mathbf{Q}}_{1}}= & \left( \begin{matrix}
9\lambda _{10}^{4}\lambda _{30}^{5}\left( 1+\lambda _{10}^{4}\lambda _{30}^{2} \right)\left( \chi -1 \right) & 9\lambda _{10}^{5}\lambda _{30}^{4}\left( \chi -1 \right)  \\
\lambda _{10}^{-1} & \lambda _{30}^{-1}\left( 1+\lambda _{10}^{2}\lambda _{30}^{4} \right)  \\
\end{matrix} \right), \notag \\[4pt]
{{t}_{1}}= & 12\lambda _{10}^{7}\lambda _{30}^{5}\left( 1-\chi  \right)+\frac{9{{k}_{1}}\left( 1-\lambda _{10}^{4}\lambda _{30}^{2} \right)}{4{{\lambda }_{10}}{{\lambda }_{30}}{{\left( \lambda _{10}^{2}+\lambda _{30}^{2}+\lambda _{10}^{4}\lambda _{30}^{4} \right)}^{2}}\left( 1-\chi  \right)}, \notag \\[4pt]
{{t}_{2}}= & \frac{{{\left( 1-\lambda _{10}^{4}\lambda _{30}^{2} \right)}^{2}}\left( 2+\lambda _{10}^{2}\lambda _{30}^{4}+3\lambda _{10}^{4}\lambda _{30}^{8} \right)}{4\lambda _{10}^{6}\lambda _{30}^{6}{{\left( \lambda _{10}^{2}+\lambda _{30}^{2}+\lambda _{10}^{4}\lambda _{30}^{4} \right)}^{2}}{{\left( 1-\chi  \right)}^{2}}}, 
\notag \\[4pt]
{{k}_{1}}= & \lambda _{10}^{4}\lambda _{30}^{2}\left[ \lambda _{10}^{2}\lambda _{30}^{4}\left( 14+\lambda _{10}^{6}\lambda _{30}^{6} \right)+4\lambda _{10}^{4}\lambda _{30}^{6}\left( \lambda _{10}^{2}+3\lambda _{30}^{2} \right) \right. 
\notag \\[2pt]
& \left. +3\left( 1+\lambda _{30}^{6}+\lambda _{10}^{4}\lambda _{30}^{2} \right) \right]-2\left( 1+\lambda _{10}^{2}\lambda _{30}^{4} \right).
\end{align}

Again we verify that for $\overline{B}_{2}=0$, we have $\lambda_{10}=\lambda_{30}=1$ and from Eqs.~\eqref{C18}-\eqref{C21} we recover the purely elastic result: ${{\lambda }_{1}^\text{cr}}= 1-4{{\left( kH \right)}^{2}}/9$ and ${{\lambda }_{3}^\text{cr}}=1+2{{\left( kH \right)}^{2}}/9$. This is the classical Euler solution for the buckling of a slender or a thin plate under uni-axial loading \citep{beatty1998stability}.


\subsubsection{Critical magnetic induction field} \label{AppeC2.2}


We now derive the approximations of the critical magnetic field $\overline{B}_2^\text{cr}$ for a given pre-stretch $\lambda_1$. We call the root of Eqs.~\eqref{C15} and \eqref{C16} as $\overline{B}_{20}$ and $\lambda_{30}$. The derivation is essentially the same as the one of the critical stretch given in \ref{AppeC2.1}. Hence, we get the first-order corrections
\begin{equation} \label{C22}
{{\overline{B}}_2^\text{cr}}={{\overline{B}}_{20}}+{{\gamma }_{1}}kH,   \qquad 
{{\lambda }_3^\text{cr}}={{\lambda }_{30}}+{{\gamma }_{2}}kH,
\end{equation}
with
\begin{align}  \label{C23}
& {{\gamma }_{1}}=\frac{\left( 1-\lambda _{1}^{4}\lambda _{30}^{2} \right)\left( 2\lambda _{30}^{2}+\chi \overline{B}_{20}^{2} \right)}{\chi {{{\overline{B}}}_{20}}{{\lambda }_{1}}{{\lambda }_{30}}\left\{ 5\left( 1-\chi  \right)+\lambda _{1}^{2}\lambda _{30}^{2}\left[ 7\lambda _{1}^{2}\left( \chi -1 \right)+\chi \left( 5\chi \overline{B}_{20}^{2}-4\lambda _{30}^{2} \right) \right] \right\}}, \notag \\[4pt] 
& {{\gamma }_{2}}=\frac{\lambda _{1}^{4}\lambda _{30}^{2}-1}{{{\lambda }_{1}}\left\{ 5\left( 1-\chi  \right)+\lambda _{1}^{2}\lambda _{30}^{2}\left[ 7\lambda _{1}^{2}\left( \chi -1 \right)+\chi \left( 5\chi \overline{B}_{20}^{2}-4\lambda _{30}^{2} \right) \right] \right\}},
\end{align}
and the second-order corrections
\begin{equation}  \label{C24}
{{\overline{B}}_2^\text{cr}}={{\overline{B}}_{20}}+{{\gamma }_{1}}kH+{{\gamma }_{3}}{{\left( kH \right)}^{2}},  \qquad 
{{\lambda }_3^\text{cr}}={{\lambda }_{30}}+{{\gamma }_{2}}kH+{{\gamma }_{4}}{{\left( kH \right)}^{2}},
\end{equation}
with
\begin{equation} \label{C25}
\left[{{\gamma }_{3}}, \, {{\gamma}_{4}} \right]^\text{T} = -\mathbf{Q}_{2}^{-1} \left[t_3, \, t_4 \right]^\text{T},
\end{equation}
where
\begin{align} \label{C26}
{{\mathbf{Q}}_{2}} = & \left( \begin{matrix}
6\lambda _{1}^{4}\lambda _{30}^{4}{{\chi }^{2}}{{{\overline{B}}}_{20}} & 3\lambda _{1}^{2}{{\lambda }_{30}}\left\{ 5\left( 1-\chi  \right)+7\lambda _{1}^{2}\lambda _{30}^{2}\left[ \lambda _{1}^{2}\left( \chi -1 \right)+{{\chi }^{2}}\overline{B}_{20}^{2} \right] \right\}  \\
2\lambda _{30}^{2}\chi {{{\overline{B}}}_{20}} & 2\left( 2\lambda _{30}^{3}+{{\lambda }_{30}}\chi \overline{B}_{20}^{2} \right)  \\
\end{matrix} \right), 
\notag \\[4pt] 
{{t}_{3}} = & \left( 1+3\lambda _{1}^{4}\lambda _{30}^{2} \right)\left( \chi -1 \right)+\lambda _{1}^{2}\lambda _{30}^{2}{{\chi }^{2}}\left( 3\lambda _{1}^{2}\lambda _{30}^{2}\gamma _{1}^{2}-\overline{B}_{20}^{2} \right)+ 
\notag \\[2pt] 
& +3\lambda _{1}^{2}\gamma _{2}^{2}\left\{ 10\left( 1-\chi  \right)+21\lambda _{1}^{2}\lambda _{30}^{2}\left[ \lambda _{1}^{2}\left( \chi -1 \right)+\overline{B}_{20}^{2}{{\chi }^{2}} \right] \right\} 
\notag \\[2pt]
& +6{{\lambda }_{1}}{{\gamma }_{2}}\left[ 2+\lambda _{1}^{3}\lambda _{30}^{2}\left( 7{{\lambda }_{30}}{{\chi }^{2}}{{{\overline{B}}}_{20}}{{\gamma }_{1}}-3{{\lambda }_{1}} \right) \right], 
\notag \\[4pt]
{{t}_{4}} = & 4{{\lambda }_{30}}\chi {{{\overline{B}}}_{20}}{{\gamma }_{1}}{{\gamma }_{2}}+\lambda _{30}^{2}\chi \gamma _{1}^{2}+\left( 6\lambda _{30}^{2}+\chi \overline{B}_{20}^{2} \right)\gamma _{2}^{2}.
\end{align}

Again, when there is no pre-stretch ($\lambda_1=1$), the zero-order thin-plate equations \eqref{C15} and \eqref{C16} yield one root  $\overline{B}_{20}=0$ and $\lambda_{30}=1$. Hence, the corrections \eqref{C22} and \eqref{C24} give $\overline{B}_2^\text{cr}\equiv0$ independent of $kH$, are not applicable in this case and we need to re-do the expansion. Specifically, expanding the bifurcation equation \eqref{C14} to the third order in $kH$, and setting $\lambda_1=1$, we obtain
\begin{multline} \label{C27}
-2(1+\lambda_3^2)(kH)^3 + \lambda_3 \left[(\chi-1)(1+3\lambda_3^2) -\overline{B}_{2}^{2}\chi^2 \lambda_3^2\right] (kH)^2 
\\
+3\lambda_3^2(1-\lambda_3^2)kH
+3\lambda_3^3 \left[(1-\chi)(1-\lambda_3^2) +\overline{B}_{2}^{2}\chi^2 \lambda_3^2\right]=0. 
\end{multline}

Conducting the same operations of Eqs.~\eqref{C15} and \eqref{C27} as those in \ref{AppeC1.2}, we finally get
\begin{equation} \label{C28}
{{\overline{B}}_2^\text{cr}}=\frac{2\sqrt{2\left( 1-\chi  \right)}}{\sqrt{3\chi \left( 1+\chi  \right)}} kH, \qquad
{{\lambda }_3^\text{cr}}=1+\frac{2\left( \chi -1 \right)}{3\left( 1+\chi  \right)}{{\left( kH \right)}^{2}}
\end{equation}
for the first-order correction of ${{\overline{B}}_2^\text{cr}}$, and
\begin{align} \label{C29}
{{\overline{B}}_2^\text{cr}} & =\frac{2\sqrt{2\left( 1-\chi  \right)}}{\sqrt{3\chi \left( 1+\chi  \right)}} kH +\frac{2\sqrt{2\chi \left( 1+\chi  \right)}}{{{\left( 1+\chi  \right)}^{2}}\sqrt{3\left( 1-\chi \right)}} {{\left( kH \right)}^{2}},
\notag \\
{{\lambda }_3^\text{cr}} & =1+\frac{2\left( \chi -1 \right)}{3\left( 1+\chi  \right)}{{\left( kH \right)}^{2}}-\frac{4\chi }{3{{\left( 1+\chi  \right)}^{2}}}{{\left( kH \right)}^{3}}
\end{align}
for the second-order correction of ${{\overline{B}}_2^\text{cr}}$.


\section*{References}

\bibliographystyle{elsarticle-harv.bst}
\nocite{*}
\bibliography{Stability-in-SMA-plates-R2.bib}

\begin{thebibliography}{51}
\expandafter\ifx\csname natexlab\endcsname\relax\def\natexlab#1{#1}\fi
\expandafter\ifx\csname url\endcsname\relax
  \def\url#1{\texttt{#1}}\fi
\expandafter\ifx\csname urlprefix\endcsname\relax\def\urlprefix{URL }\fi

\bibitem[{Beatty and Pan(1998)}]{beatty1998stability}
Beatty, M.~F., Pan, F., 1998. Stability of an internally constrained,
  hyperelastic slab. International Journal of Non-Linear Mechanics 33~(5),
  867--906.

\bibitem[{Brigadnov and Dorfmann(2003)}]{brigadnov2003mathematical}
Brigadnov, I., Dorfmann, A., 2003. Mathematical modeling of magneto-sensitive
  elastomers. International Journal of Solids and Structures 40~(18),
  4659--4674.

\bibitem[{Brown(1966)}]{brown1966magnetoelastic}
Brown, W.~F., 1966. Magnetoelastic Interactions. Springer, New York.

\bibitem[{Bustamante(2010)}]{bustamante2010transversely}
Bustamante, R., 2010. Transversely isotropic nonlinear magneto-active
  elastomers. Acta Mechanica 210~(3-4), 183--214.

\bibitem[{Casta{\~n}eda and Galipeau(2011)}]{castaneda2011homogenization}
Casta{\~n}eda, P.~P., Galipeau, E., 2011. Homogenization-based constitutive
  models for magnetorheological elastomers at finite strain. Journal of the
  Mechanics and Physics of Solids 59~(2), 194--215.

\bibitem[{Dalrymple et~al.(1974)Dalrymple, Peach, and
  Viegelahn}]{dalrymple1974magnetoelastic}
Dalrymple, J.~M., Peach, M.~O., Viegelahn, G.~L., 1974. Magnetoelastic buckling
  of thin magnetically soft plates in cylindrical mode. Journal of Applied
  Mechanics 41~(1), 145--150.

\bibitem[{Danas and Triantafyllidis(2014)}]{danas2014instability}
Danas, K., Triantafyllidis, N., 2014. Instability of a magnetoelastic layer
  resting on a non-magnetic substrate. Journal of the Mechanics and Physics of
  Solids 69, 67--83.

\bibitem[{Destrade and Ogden(2011)}]{destrade2011magneto}
Destrade, M., Ogden, R.~W., 2011. On magneto-acoustic waves in finitely
  deformed elastic solids. Mathematics and Mechanics of Solids 16~(6),
  594--604.

\bibitem[{Dorfmann and Ogden(2004)}]{dorfmann2004nonlinear}
Dorfmann, A., Ogden, R., 2004. Nonlinear magnetoelastic deformations. Quarterly
  Journal of Mechanics and Applied Mathematics 57~(4), 599--622.

\bibitem[{Dorfmann and Ogden(2014)}]{dorfmann2014nonlinear}
Dorfmann, L., Ogden, R.~W., 2014. Nonlinear Theory of Electroelastic and
  Magnetoelastic Interactions. Springer, New York.

\bibitem[{Flavin(1963)}]{flavin1963surface}
Flavin, J., 1963. Surface waves in pre-stressed {M}ooney material. The
  Quarterly Journal of Mechanics and Applied Mathematics 16~(4), 441--449.

\bibitem[{Galipeau et~al.(2014)Galipeau, Rudykh, deBotton, and
  Casta{\~n}eda}]{galipeau2014magnetoactive}
Galipeau, E., Rudykh, S., deBotton, G., Casta{\~n}eda, P.~P., 2014.
  Magnetoactive elastomers with periodic and random microstructures.
  International Journal of Solids and Structures 51~(18), 3012--3024.

\bibitem[{Gerbal et~al.(2015)Gerbal, Wang, Lyonnet, Bacri, Hocquet, and
  Devaud}]{gerbal2015refined}
Gerbal, F., Wang, Y., Lyonnet, F., Bacri, J.-C., Hocquet, T., Devaud, M., 2015.
  A refined theory of magnetoelastic buckling matches experiments with
  ferromagnetic and superparamagnetic rods. Proceedings of the National Academy
  of Sciences 112~(23), 7135--7140.

\bibitem[{Ginder et~al.(2002)Ginder, Clark, Schlotter, and
  Nichols}]{ginder2002magnetostrictive}
Ginder, J., Clark, S., Schlotter, W., Nichols, M., 2002. Magnetostrictive
  phenomena in magnetorheological elastomers. International Journal of Modern
  Physics B 16, 2412--2418.

\bibitem[{Ginder et~al.(2001)Ginder, Schlotter, and
  Nichols}]{ginder2001magnetorheological}
Ginder, J.~M., Schlotter, W.~F., Nichols, M.~E., 2001. Magnetorheological
  elastomers in tunable vibration absorbers. Proceedings SPIE 4331, 103--110.

\bibitem[{Goshkoderia et~al.(2020)Goshkoderia, Chen, Li, Juhl, Buskohl, and
  Rudykh}]{goshkoderia2020instability}
Goshkoderia, A., Chen, V., Li, J., Juhl, A., Buskohl, P., Rudykh, S., 2020.
  Instability-induced pattern formations in soft magnetoactive composites.
  Physical Review Letters 124~(15), 158002.

\bibitem[{Hoang et~al.(2010)Hoang, Zhang, and Du}]{hoang2010adaptive}
Hoang, N., Zhang, N., Du, H., 2010. An adaptive tunable vibration absorber
  using a new magnetorheological elastomer for vehicular powertrain transient
  vibration reduction. Smart Materials and Structures 20~(1), 015019.

\bibitem[{Kankanala and Triantafyllidis(2004)}]{kankanala2004finitely}
Kankanala, S., Triantafyllidis, N., 2004. On finitely strained
  magnetorheological elastomers. Journal of the Mechanics and Physics of Solids
  52~(12), 2869--2908.

\bibitem[{Kankanala and Triantafyllidis(2008)}]{kankanala2008magnetoelastic}
Kankanala, S., Triantafyllidis, N., 2008. Magnetoelastic buckling of a
  rectangular block in plane strain. Journal of the Mechanics and Physics of
  Solids 56~(4), 1147--1169.

\bibitem[{Karami~Mohammadi et~al.(2019)Karami~Mohammadi, Galich, Krushynska,
  and Rudykh}]{karami2019soft}
Karami~Mohammadi, N., Galich, P.~I., Krushynska, A.~O., Rudykh, S., 2019. Soft
  magnetoactive laminates: {L}arge deformations, transverse elastic waves and
  band gaps tunability by a magnetic field. Journal of Applied Mechanics
  86~(11).

\bibitem[{Kim et~al.(2018)Kim, Yuk, Zhao, Chester, and Zhao}]{kim2018printing}
Kim, Y., Yuk, H., Zhao, R., Chester, S.~A., Zhao, X., 2018. Printing
  ferromagnetic domains for untethered fast-transforming soft materials. Nature
  558~(7709), 274--279.

\bibitem[{Lanotte et~al.(2003)Lanotte, Ausanio, Hison, Iannotti, and
  Luponio}]{lanotte2003potentiality}
Lanotte, L., Ausanio, G., Hison, C., Iannotti, V., Luponio, C., 2003. The
  potentiality of composite elastic magnets as novel materials for sensors and
  actuators. Sensors and Actuators A: Physical 106~(1-3), 56--60.

\bibitem[{Lef{\`e}vre et~al.(2017)Lef{\`e}vre, Danas, and
  Lopez-Pamies}]{lefevre2017general}
Lef{\`e}vre, V., Danas, K., Lopez-Pamies, O., 2017. A general result for the
  magnetoelastic response of isotropic suspensions of iron and ferrofluid
  particles in rubber, with applications to spherical and cylindrical
  specimens. Journal of the Mechanics and Physics of Solids 107, 343--364.

\bibitem[{Lef{\`e}vre et~al.(2020)Lef{\`e}vre, Danas, and
  Lopez-Pamies}]{lefevre2020two}
Lef{\`e}vre, V., Danas, K., Lopez-Pamies, O., 2020. Two families of explicit
  models constructed from a homogenization solution for the magnetoelastic
  response of {MRE}s containing iron and ferrofluid particles. International
  Journal of Non-Linear Mechanics 119, 103362.

\bibitem[{Luo et~al.(2019)Luo, Evans, and Chang}]{luo2019magnetically}
Luo, Z., Evans, B.~A., Chang, C.-H., 2019. Magnetically actuated dynamic
  iridescence inspired by the neon tetra. ACS Nano 13~(4), 4657--4666.

\bibitem[{Makarova et~al.(2016)Makarova, Alekhina, Rusakova, and
  Perov}]{makarova2016tunable}
Makarova, L.~A., Alekhina, Y.~A., Rusakova, T.~S., Perov, N.~S., 2016. Tunable
  properties of magnetoactive elastomers for biomedical applications. Physics
  Procedia 82, 38--45.

\bibitem[{Martin et~al.(2006)Martin, Anderson, Read, and
  Gulley}]{martin2006magnetostriction}
Martin, J.~E., Anderson, R.~A., Read, D., Gulley, G., 2006. Magnetostriction of
  field-structured magnetoelastomers. Physical Review E 74~(5), 051507.

\bibitem[{Maugin(1988)}]{maugin2013continuum}
Maugin, G.~A., 1988. Continuum Mechanics of Electromagnetic Solids.
  North-Holland, Amsterdam.

\bibitem[{Miya et~al.(1978)Miya, Hara, and Someya}]{miya1978experimental}
Miya, K., Hara, K., Someya, K., 1978. Experimental and theoretical study on
  magnetoelastic buckling of a ferromagnetic cantilevered beam-plate. Journal
  of Applied Mechanics 45, 355–--360.

\bibitem[{Moon and Pao(1968)}]{moon1968magnetoelastic}
Moon, F.~C., Pao, Y.-H., 1968. Magnetoelastic buckling of a thin plate. Journal
  of Applied Mechanics 35, 53–--58.

\bibitem[{Mukherjee et~al.(2020)Mukherjee, Bodelot, and
  Danas}]{mukherjee2020microstructurally}
Mukherjee, D., Bodelot, L., Danas, K., 2020. Microstructurally-guided explicit
  continuum models for isotropic magnetorheological elastomers with iron
  particles. International Journal of Non-Linear Mechanics 120, 103380.

\bibitem[{Ott{\'e}nio et~al.(2008)Ott{\'e}nio, Destrade, and
  Ogden}]{ottenio2008incremental}
Ott{\'e}nio, M., Destrade, M., Ogden, R.~W., 2008. Incremental magnetoelastic
  deformations, with application to surface instability. Journal of Elasticity
  90~(1), 19--42.

\bibitem[{Pao and Yeh(1973)}]{yih1973linear}
Pao, Y.-H., Yeh, C.-S., 1973. A linear theory for soft ferromagnetic elastic
  solids. International Journal of Engineering Science 11~(4), 415--436.

\bibitem[{Popelar(1972)}]{popelar1972postbuckling}
Popelar, C.~H., 1972. Postbuckling analysis of a magnetoelastic beam. Journal
  of Applied Mechanics 39, 207–--212.

\bibitem[{Psarra et~al.(2017)Psarra, Bodelot, and Danas}]{psarra2017two}
Psarra, E., Bodelot, L., Danas, K., 2017. Two-field surface pattern control via
  marginally stable magnetorheological elastomers. Soft Matter 13~(37),
  6576--6584.

\bibitem[{Psarra et~al.(2019)Psarra, Bodelot, and Danas}]{psarra2019wrinkling}
Psarra, E., Bodelot, L., Danas, K., 2019. Wrinkling to crinkling transitions
  and curvature localization in a magnetoelastic film bonded to a non-magnetic
  substrate. Journal of the Mechanics and Physics of Solids 133, 103734.

\bibitem[{Rambausek and Keip(2018)}]{rambausek2018analytical}
Rambausek, M., Keip, M.-A., 2018. Analytical estimation of non-local
  deformation-mediated magneto-electric coupling in soft composites.
  Proceedings of the Royal Society A: Mathematical, Physical and Engineering
  Sciences 474~(2216), 20170803.

\bibitem[{Reddy and Saxena(2018)}]{reddy2018instabilities}
Reddy, N.~H., Saxena, P., 2018. Instabilities in the axisymmetric
  magnetoelastic deformation of a cylindrical membrane. International Journal
  of Solids and Structures 136, 203--219.

\bibitem[{Rigbi and Jilken(1983)}]{rigbi1983response}
Rigbi, Z., Jilken, L., 1983. The response of an elastomer filled with soft
  ferrite to mechanical and magnetic influences. Journal of Magnetism and
  Magnetic Materials 37~(3), 267--276.

\bibitem[{Rudykh and Bertoldi(2013)}]{rudykh2013stability}
Rudykh, S., Bertoldi, K., 2013. Stability of anisotropic magnetorheological
  elastomers in finite deformations: {A} micromechanical approach. Journal of
  the Mechanics and Physics of Solids 61~(4), 949--967.

\bibitem[{Saxena et~al.(2013)Saxena, Hossain, and Steinmann}]{saxena2013theory}
Saxena, P., Hossain, M., Steinmann, P., 2013. A theory of finite deformation
  magneto-viscoelasticity. International Journal of Solids and Structures
  50~(24), 3886--3897.

\bibitem[{Shuvalov et~al.(2004)Shuvalov, Poncelet, and
  Deschamps}]{shuvalov2004general}
Shuvalov, A., Poncelet, O., Deschamps, M., 2004. General formalism for plane
  guided waves in transversely inhomogeneous anisotropic plates. Wave Motion
  40~(4), 413--426.

\bibitem[{Singh and Onck(2018)}]{singh2018magnetic}
Singh, R., Onck, P., 2018. Magnetic field induced deformation and buckling of
  slender bodies. International Journal of Solids and Structures 143, 29--58.

\bibitem[{Su et~al.(2018)Su, Broderick, Chen, and Destrade}]{su2018wrinkles}
Su, Y., Broderick, H.~C., Chen, W., Destrade, M., 2018. Wrinkles in soft
  dielectric plates. Journal of the Mechanics and Physics of Solids 119,
  298--318.

\bibitem[{Su et~al.(2020)Su, Chen, Dorfmann, and Destrade}]{su2020effect}
Su, Y., Chen, W., Dorfmann, L., Destrade, M., 2020. The effect of an exterior
  electric field on the instability of dielectric plates. Proceedings of the
  Royal Society A 476~(2239), 20200267.

\bibitem[{Su et~al.(2019)Su, Wu, Chen, and Destrade}]{su2019finite}
Su, Y., Wu, B., Chen, W., Destrade, M., 2019. Finite bending and pattern
  evolution of the associated instability for a dielectric elastomer slab.
  International Journal of Solids and Structures 158, 191--209.

\bibitem[{Tang et~al.(2019)Tang, Qiao, Chu, Tong, Zhou, Zhang, Xie, Hu, and
  Wang}]{tang2019magnetic}
Tang, J., Qiao, Y., Chu, Y., Tong, Z., Zhou, Y., Zhang, W., Xie, S., Hu, J.,
  Wang, T., 2019. Magnetic double-network hydrogels for tissue hyperthermia and
  drug release. Journal of Materials Chemistry B 7~(8), 1311--1321.

\bibitem[{Tian et~al.(2011)Tian, Li, and Deng}]{tian2011sensing}
Tian, T., Li, W., Deng, Y., 2011. Sensing capabilities of graphite based {MR}
  elastomers. Smart materials and structures 20~(2), 025022.

\bibitem[{Tiersten(1964)}]{tiersten1964coupled}
Tiersten, H., 1964. Coupled magnetomechanical equations for magnetically
  saturated insulators. Journal of Mathematical Physics 5~(9), 1298--1318.

\bibitem[{Wallerstein and Peach(1972)}]{wallerstein1972magnetoelastic}
Wallerstein, D.~V., Peach, M.~O., 1972. Magnetoelastic buckling of beams and
  thin plates of magnetically soft material. Journal of Applied Mechanics 39,
  451–--455.

\bibitem[{Yu et~al.(2018)Yu, Fang, Huang, and Wang}]{yu2018magnetoactive}
Yu, K., Fang, N.~X., Huang, G., Wang, Q., 2018. Magnetoactive acoustic
  metamaterials. Advanced Materials 30~(21), 1706348.

\end{thebibliography}







\end{document}